\documentclass[aps,prd,12pt,notitlepage,nofootinbib,tightenlines]{revtex4-1}
\usepackage{amsmath}
\usepackage{bm}
\usepackage{times}
\usepackage{braket}
\usepackage{float}
\usepackage{color}
\usepackage{epsfig}
\usepackage{slashed}
\usepackage{hyperref}
\usepackage{subfigure}
\usepackage{graphicx}
\usepackage{multirow}
\usepackage{longtable}
\usepackage{epstopdf}
\usepackage{rotating}
\allowdisplaybreaks[1]

\newcommand{\non}{\nonumber\\ }

\newcommand{\calb}{ {\cal B} }

\newcommand{\xipp}{\Xi_{cc}^{++}}
\newcommand{\xip}{\Xi_{cc}^+}
\newcommand{\opp}{\Omega_{cc}^+}

\definecolor{Red}{rgb}{1.,0.,0.}

\definecolor{Blue}{rgb}{0.,0.,1.}

\definecolor{nicered}{rgb}{0.7,0.1,0.1}
\definecolor{nicegreen}{rgb}{0.1,0.5,0.1}
\definecolor{diff}{rgb}{1,0,0}

\hypersetup{colorlinks,citecolor=nicegreen,linkcolor=nicered}

\begin{document}
\title{Rescattering mechanism of weak decays of double-charm baryons}
\author{Jia-Jie Han$^{1,2}$}\email{hanjiajie1020@163.com}
\author{Hua-Yu Jiang$^{1,3}$} \email{jianghy15@lzu.edu.cn, corresponding author}
\author{Wei Liu$^{1}$}\email{liuw18@lzu.edu.cn}
\author{Zhen-Jun Xiao$^{2}$}\email{xiaozhenjun@njnu.edu.cn, corresponding author}
\author{Fu-Sheng Yu$^{1,4,5,6}$}  \email{yufsh@lzu.edu.cn, corresponding author}

\affiliation{$^1$ School of Nuclear Science and Technology, Lanzhou University, Lanzhou 730000, China}
\affiliation{$^2$ Department of Physics and Institute of Theoretical Physics, Nanjing Normal University, Nanjing, Jiangsu 210023, China}
\affiliation{$^3$ Theoretische Physik 1, Naturwissenschaftlich-Technische Fakult\"at,
Universit\"at Siegen, 57068 Siegen, Germany}
\affiliation{$^4$ Lanzhou Center for Theoretical Physics, Lanzhou University, Lanzhou 730000, China}
\affiliation{$^5$ Frontier Science Center for Rare Isotopes, Lanzhou University,  Lanzhou 730000, China}
\affiliation{$^6$ Center for High Energy Physics, Peking University, Beijing 100871, China}

\begin{abstract}
The doubly charmed baryon $\Xi_{cc}^{++}$ was recently observed by LHCb via the decay processes of  $\Xi_{cc}^{++}\to \Lambda_c^+ K^-\pi^+\pi^+$ and $\Xi_c^+\pi^+$.
These discovery channels were successfully predicted in the framework that the short-distance contributions are calculated under the factorization hypothesis
and the long-distance contributions are estimated using the rescattering mechanism for the final-state-interaction effects.
In this paper, we illustrate the above framework in details by systematic studies on the two-body baryonic decays $\mathcal{B}_{cc}\to\mathcal{B}_{c}P$ involving
the doubly charmed baryons $\mathcal{B}_{cc}=(\Xi_{cc}^{++} , \Xi_{cc}^+,\Omega_{cc}^+)$, the singly charmed baryons  $\mathcal{B}_{c}=(\mathcal{B}_{\bar{3}}, \mathcal{B}_{6})$
and the light pseudoscalar mesons $P=(\pi,K,\eta_{1,8})$.
\end{abstract}


\pacs{13.30.-a, 13.30.Eg, 14.20.Mr}

\vspace{1cm}
\maketitle

\section{Introduction}\label{sec:intro}

The doubly heavy baryons with two heavy flavor quarks (the $b$ or $c$ quark) were predicted by the quark model and the quantum chromodynamics (QCD) 
several decades ago \cite{DeRujula:1975qlm,Jaffe:1975us,Ponce:1978gk}. Their structure analogous to a heavy double-star system with an attached light 
planet \cite{PDG}, is very different from single-heavy-flavor baryons and light baryons. Besides, the research to the doubly heavy baryons is being a 
powerful tool to the investigations on the doubly and fully heavy tetraquark states \cite{Karliner:2017qjm,Eichten:2017ffp}. Therefore, the doubly heavy 
baryons open a new window for QCD properties \cite{Chen:2016spr}.

However, the experimental searches for doubly heavy baryons went through a flexuous exploration process. In 2002, $\Xi_{cc}^+$ was firstly reported as
observed by the SELEX collaboration via the mode of  $\Xi_{cc}^+\to\Lambda_c^+K^-\pi^+$ \cite{Mattson:2002vu}. All the following measurements 
by FOCUS \cite{Ratti:2003ez}, BarBar \cite{Aubert:2006qw}, Belle \cite{Chistov:2006zj}, and LHCb \cite{Aaij:2013voa}, did not find signatures using the 
same decay mode. Actually, the production rate of the doubly charmed baryons is large enough at the beginning running of LHCb \cite{Chang:2006eu,Wu:2019gta}. 
The left problem is the decay properties, i.e. which decaying processes have largest branching fractions and final particles easily detected in the experiments 
\cite{Yu:2019lxw}. In 2017, a theoretical analysis on all the decaying processes pointed out that $\Xi_{cc}^{++}\to\Lambda_c^+K^-\pi^+\pi^+$ and $\Xi_{c}^+\pi^+$ 
are the most favorable ones for the discovery of doubly charmed baryons \cite{Yu:2017zst}. Subsequently, the LHCb collaboration observed the doubly charmed baryon 
for the first time via $\Xi_{cc}^{++}\to\Lambda_c^+K^-\pi^+\pi^+$ \cite{Aaij:2017ueg} under the theoretical suggestions, and confirmed the discovery via 
$\Xi_{cc}^{++}\to\Xi_{c}^+\pi^+$ in 2018 \cite{Aaij:2018gfl}. It is clear that the theoretical studies on the decay properties play an important role in the 
experimental searches for the doubly heavy baryons.

The dynamics of the doubly charmed baryon decays has two difficulties in the theoretical calculations: the charm decay with large non-perturbative contributions 
and the baryon decay with a three-body problem. For charm decays, the QCD-inspired methods do not work well at the scale of around 1 GeV. In the charmed 
meson decays, the significant non-perturbative contribution are involved in the topological amplitudes which are extracted from the experimental data of the 
decaying branching fractions \cite{Cheng:2010ry,Li:2012cfa,Li:2013xsa,Bhattacharya:2009ps,Wang:2020gmn}. However, the topological diagrammatic approach 
cannot be directly used in the doubly charmed baryon decays since there are no available data. The charmed baryon decays are even more complicated \cite{Lu:2016ogy,Geng:2017esc,Geng:2017mxn,Wang:2017gxe,Geng:2018plk,Cheng:2018hwl,Jiang:2018iqa,Zhao:2018zcb,Geng:2018bow,Geng:2018upx,He:2018joe,Zhao:2018mov,Grossman:2018ptn,Geng:2018rse,Wang:2019dls,Geng:2019xbo,Hsiao:2019yur,Geng:2019awr,Jia:2019zxi,Zou:2019kzq,Geng:2020zgr,Niu:2020gjw,Hsiao:2020iwc,Pan:2020qqo,Meng:2020euv,Hu:2020nkg,Cheng:2018rkz}. 
As a first attempt to study the non-leptonic doubly charmed baryon decays, Ref. \cite{Yu:2017zst}, calculate the factorizable contributions under the factorization hypothesis 
and the non-factorizable ones considering the rescattering mechanism of the final-state-interaction(FSI) effects. In this work, we will systematically illustrate the theoretical 
framework in the decays of doubly charmed baryons into a charmed baryon and a light pseudoscalar meson.

The success of the factorization and rescattering mechanism on the suggestion of the discovery channels of the doubly charmed baryons, manifests that the above framework 
roughly describes the correct dynamics of doubly charmed baryon decays. There is no doubt that the factorization approach works well for the short-distance tree-emitted 
diagrams \cite{Bauer:1986bm,Ali:1998eb,Cheng:2010ry,Li:2012cfa,Li:2013xsa}. The problem is how to calculate the long-distance contributions, which are usually considered 
as the FSIs effects. Many works have been done to calculate the FSIs effects of weak decays of heavy-flavor meson 
\cite{Li:2002pj,Ablikim:2002ep,Li:1996cj,Dai:1999cs,Locher:1993cc,Cheng:FSIB,Lu:2005mx,Chen:2002jr}. 
Before \cite{Yu:2017zst}, there was no any work for the long-distance contributions of doubly heavy baryon decay, but only a few works for the short-distance factorizable 
ones \cite{Onishchenko:1999yu,Kiselev:2001fw}. As pointed out before, the non-perturbative contributions are very important in charm decays. The rescattering mechanism 
of the FSIs effect was firstly investigated for $\Lambda_c^+$ decays in \cite{Chen:2002jr}, and firstly studied for the doubly charmed baryon decays in \cite{Yu:2017zst}.

The theoretical framework of the rescattering mechanism is as follows. The doubly charmed baryon decays via a short-distance tree emitted process 
into one baryon and one meson, which scatter with each other by exchanging one particle as a long-distance effect into the final states. It forms a triangle diagram at 
the hadron level. The short-distance and long-distance contributions are separated to avoid the double-counting problem.  In this work, the calculating techniques are 
followed by \cite{Cheng:FSIB}, in which the cutting rules are used to compute the imaginary part of the triangle diagram. There is a basic difference of our framework 
from those most known works in \cite{Ablikim:2002ep,Cheng:FSIB}. The triangle diagrams of the rescattering mechanism are taken as an independent method by the 
calculation of hadron-level Feynman diagrams, while in \cite{Ablikim:2002ep,Cheng:FSIB} the triangle diagrams are calculated corresponding to the quark topological 
diagrams. The problem of the latter method will be discussed elsewhere \cite{topoFSI}.

It is well known that the FSIs calculations suffer large theoretical uncertainties. The branching fractions could be changed by one order of magnitude with the variation 
of the non-perturbative parameters like $\eta$ in the form factor of the cutting rules or the cut-off $\Lambda$ in the loop calculation. The parameters are always 
determined by the measured results of the branching fractions \cite{Ablikim:2002ep,Cheng:FSIB}. However, in the case of doubly charmed baryons, there is no available 
data to determine the non-perturbative parameters. Therefore, the biggest problem in the FSIs calculations is how to control the theoretical uncertainties. The 
innovation of our method is to calculate the ratios of branching fractions, which are however not sensitive to the non-perturbative parameters. The uncertainties 
of the ratios are thus well under control. That is why Ref. \cite{Yu:2017zst} could correctly and reliably predict the modes with the largest branching fractions.
		
The knowledge of the relative sizes of the topological diagrams of heavy baryon decays provided important implications on the predictions of the most favorable 
modes to discover the doubly charmed baryons. In the soft-colliner effective theory \cite{Leibovich:2003tw,Mantry:2003uz}, the power counting rules of 
${|C|\over |T|}\sim{|C^\prime|\over |C|}\sim{|E_1|\over |C|}\sim{|E_2|\over |C|}\sim {O({\Lambda^h_{\text{QCD}}\over m_c}})$ are obtained. These relations are manifested 
by the most precise measurements of the $\Lambda_c^+$ decays performed by the BESIII collaboration \cite{Ablikim:2015flg}. In addition, the discovery channel 
of $\Xi_{cc}^{++}\to \Lambda_c^+K^-\pi^+\pi^+$ dominated by $\Xi_{cc}^{++}\to \Sigma_c^{++}\overline K^{*0}$ can be directly related to the result of 
$\Lambda_c^+\to p\phi$ \cite{Ablikim:2016tze}  by exactly the same topological diagram with an interchange of a spectator quark. In this work, we will calculate 
the topological diagrams by the rescattering mechanism to test the above relations. Besides, the flavor $SU(3)$ symmetry and its breaking effects will also be discussed. 
It has to be stressed that, the weak decays of doubly charmed baryons have been widely studied  
\cite{Onishchenko:1999yu,Kiselev:2001fw,Onishchenko:2000yp,Egolf:2002nk,Li:2017ndo,Jiang:2018oak,Wang:2017mqp,Wang:2017azm,Shi:2019hbf,Gutsche:2017hux,Hu:2017dzi,Shi:2017dto,Zhao:2018mrg,Xing:2018lre,Dhir:2018twm,Zhang:2018llc,Shi:2019fph,Xing:2019hjg,Gutsche:2019wgu,Hu:2019bqj,Gutsche:2019iac,Ke:2019lcf,Cheng:2020wmk,Hu:2020mxk,Shi:2020qde,Ivanov:2020xmw,Li:2020qrh,Cheng:2019sxr,Cheng:2018mwu}, especially after the work of Ref. \cite{Yu:2017zst} and the experimental observation of $\Xi_{cc}^{++}$. 
The clarification of our framework will be helpful to understand the dynamics and nature of doubly charmed baryons.

This paper is arranged as follows. In section \ref{sec:frame}, we introduce the theoretical framework of the rescattering mechanism and demonstrate the calculation 
details with $\xipp\to\Xi_c^+\pi^+$ as an example.
Then the parameter inputs, numerical results of branching fractions and relevant discussions are presented in Sec. \ref{sec:inputs}.
In the end, we give a brief summary.
The effective hadron-strong-interaction Lagrangians and corresponding strong coupling constants are collected in Appendix \ref{app:lag}.
The expressions of the decay amplitudes for all considered decay modes are gathered in Appendix \ref{app:amp}.

\section{Theoretical framework}\label{sec:frame}

\subsection{Effective Hamiltonian and topological diagrams}\label{sec:hamil}\label{sec:topo}

The exclusive non-leptonic weak decays of doubly charmed baryons are induced by the charge currents of charm decays at the tree level. The penguin contributions are safely neglected in the branching fractions of charm decays, due to the smallness of the corresponding CKM matrix elements. The effective Hamiltonian is given by
\begin{align}
\mathcal{H}_{eff}=\frac{G_F}{\sqrt{2}}\sum_{q^{\prime}=d,s}^{}V_{cq^{\prime}}^\ast V_{uq}[C_1(\mu)O_1(\mu)+C_2(\mu)O_2(\mu)]+h.c..
\end{align}
with the  four-fermion operators of
\begin{align}
O_1=(\bar{u}_\alpha q_\beta)_{V-A}(\bar{q^{\prime}}_\beta c_\alpha)_{V-A},\quad
O_2=(\bar{u}_\alpha q_\alpha)_{V-A}(\bar{q^{\prime}}_\beta c_\beta)_{V-A}.
\end{align}
where $q^{(\prime)}=(s,d)$, $\alpha$ and $\beta$ as color indices, $V_{cq^{\prime}}$ and $V_{uq}$ are the Cabibbo-Kobayashi-Maskawa (CKM) matrix elements, and $G_F=1.166\times 10^{-5}\mbox{ GeV}^{-2}$ is the Fermi constant.
$C_{1,2}(\mu)$ denote the Wilson coefficients which include the short-distance QCD dynamics scaling from $\mu=M_W$ to $\mu=m_c$.
To obtain the amplitudes of $\mathcal{B}_{cc}\to\mathcal{B}_cP$ decays, one needs to evaluate the next hadronic matrix element of the effective Hamiltonian:
	\begin{align}\label{eq:hme} \langle\mathcal{B}_cP|\mathcal{H}_{eff}|\mathcal{B}_{cc}\rangle=\frac{G_F}{\sqrt{2}}V^{\ast}_{cq^{\prime}}V_{uq}\sum_{i=1,2}C_i\langle\mathcal{B}_cP|O_i|\mathcal{B}_{cc}\rangle.
	\end{align}
		
\begin{figure}[htp]
		\includegraphics[scale=0.2]{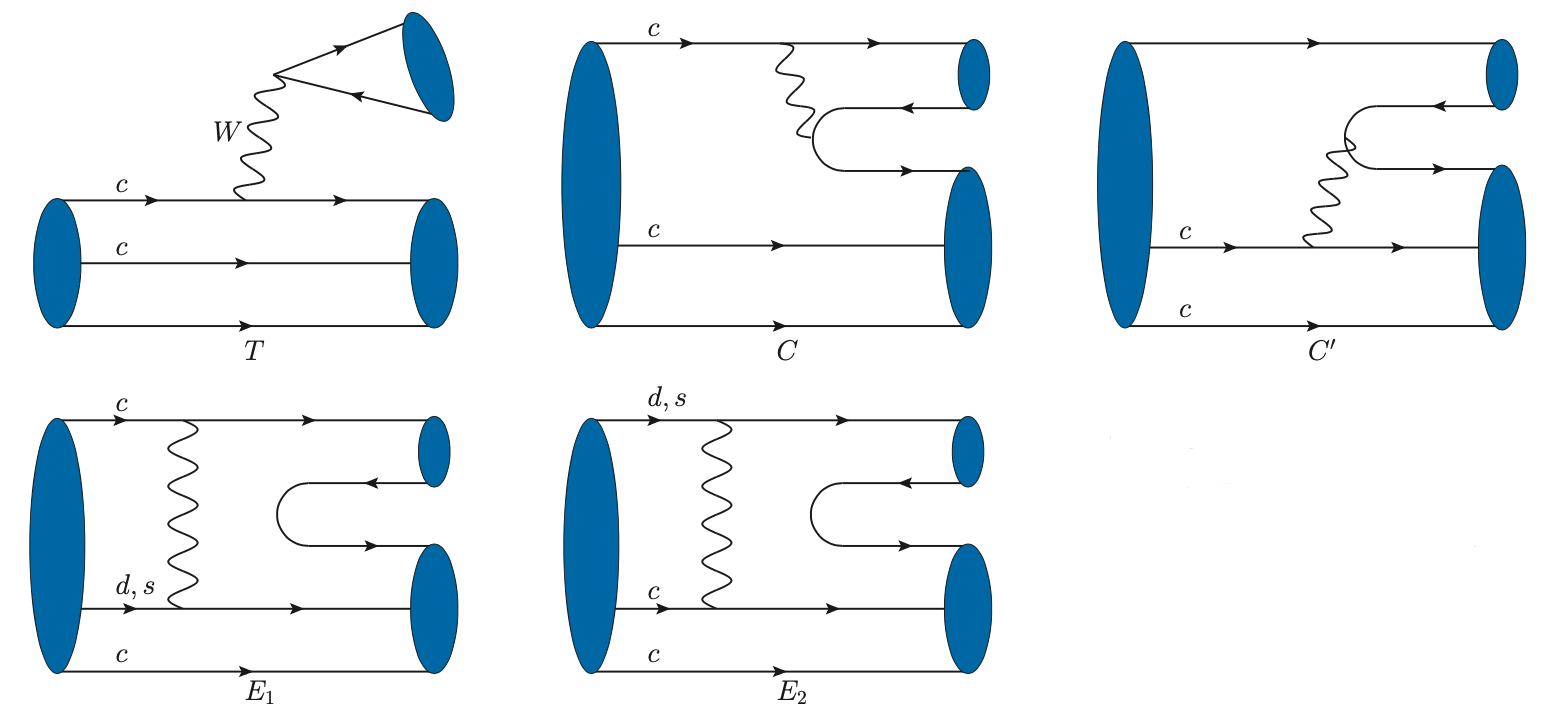}
		\caption{Tree-level topological diagrams for two-body non-leptonic decays of the doubly charmed baryons $\calb_{cc} =(\xipp,\xip,\opp)$ to a singly heavy baryon $\calb_c$ with a light pseudoscalar meson 
$P=(\pi, K, \eta_1,\eta_8)$. }		\label{fig:fig1}
\end{figure}


The general tree level topological diagrams of doubly charmed baryons decaying into a singly charmed baryon and a light meson are displayed in Fig.~\ref{fig:fig1}.
These diagrams can  be sorted by their different topologies. 
Each diagram has contained both short-distance and long-distance contributions.
$T$ describes the color-allowed external $W$-emission diagram, while $C$ and $C^\prime$ can be both used to represent color-suppressed internal $W$-emission diagrams.
$C$ diagram is the one with both the quark and anti-quark of the final light meson state coming from the weak vertex, while for $C^\prime$ only the antiquark is generated 
from the weak vertex and the quark is directly transferred from the light spectator of doubly charmed baryons.
There are also two different types of $W$-exchange diagrams which labeled by $E_1$ and $E_2$. In $E_1$, the light quark decayed from the charmed quark absorbed into the 
light mesons; and in $E_2$, the light quark decayed from the charmed quark absorbed into singly heavy baryons.
On the other hand, the possible quark loop (penguin) diagrams relevant with the tree level topological diagrams also have non-ignorable impact on the estimation of the 
long-distance contributions in our framework.

In the calculation of these topological diagrams, it has been demonstrated that $T$ is dominated by factorizable contributions \cite{Lu:2009cm} and can be calculated under the factorization hypothesis.
However, this factorizable contribution of the $C$ diagram is deeply suppressed by the color factor at charm scale with the effective Wilson coefficient $a_2(m_c)=C_1(m_c)+C_2(m_c)/N_c$.
So that the factorizable short-distance contributions are negligible, but the long-distance dynamics of $C$ can play an important role \cite{Yu:2017zst}.
The short-distance amplitudes of topological diagrams $C^\prime$, $E_1$ and $E_2$ are also expected to be suppressed at least by one order \cite{Lu:2009cm}, 
while the long distance dynamics is more important at the scale of charm quark mass.

In the following sections, we take the second discovery channel $\Xi_{cc}^{++}\to\Xi_c^+\pi^+$ \cite{Yu:2017zst,Aaij:2018gfl} as an example to introduce our framework in detail.
This decay contains two contributions of topological amplitude, i.e. $T+C^\prime$, in which the $T$ is dominated by the short-distance dynamics, while as 
our discussion before, $C^\prime$ is dominated by the non-factorizable long-distance dynamics, and $T$ usually plays the central role comparing with $C^\prime$ (mainly due to the colour suppression).
However, from our calculation, it can be seen that the non-factorizable contributions of $C^\prime$ may impose significant impact on the total amplitude.

\subsection{Short-distance amplitudes under the factorization hypothesis}\label{sec:short}

In this section, we discuss how to calculate the factorizable short-distance contributions of the topological amplitudes $T$ and $C$.
The feasible approach is the factorization approach with the matrix elements $\langle\mathcal{B}_cM|O_i|\mathcal{B}_{cc}\rangle$ in Eq.~(\ref{eq:hme}) factorized
into the product of two parts. One is parameterized as the decay constant of the emitted mesons and the other is expressed as the transition form factors.
The factorizable contribution of the  $T$ diagram is expressed as:
\begin{align} 
\langle\mathcal{B}_{c}M|\mathcal{H}_{eff}|\mathcal{B}_{cc}\rangle_{SD}^T=\frac{G_F}{\sqrt{2}}V^\ast_{cq^\prime}V_{uq}a_1(\mu)\langle M|\bar{u}\gamma^\mu(1-\gamma_5)q|0\rangle\langle
\mathcal{B}_c|\bar{q}^\prime\gamma_\mu(1-\gamma_5)c|\mathcal{B}_{cc} \rangle , \label{eq:hadronic1}
\end{align}
while the factorizable $C$ diagram is given by
\begin{align}
\langle\mathcal{B}_{c}M|\mathcal{H}_{eff}|\mathcal{B}_{cc}\rangle_{SD}^C=
\frac{G_F}{\sqrt{2}}V^\ast_{cq^\prime}V_{uq}a_2(\mu)\langle M|\bar{q}^\prime\gamma^\mu(1-\gamma_5)q|0\rangle\langle\mathcal{B}_c|\bar{u}\gamma_\mu(1-\gamma_5)c|\mathcal{B}_{cc}\rangle. \label{eq:hadronic2}
\end{align}
where $a_1(a_2)$ represents the effective Wilson coefficients, $a_1(\mu)=C_1(\mu)+C_2(\mu)/3$
and $a_2(\mu)=C_2(\mu)+C_1(\mu)/3$, with the Wilson coefficients $C_1(\mu)=1.21$ and $C_2(\mu)=-0.42$ at the scale of charm decays $\mu=m_c$ \cite{Li:2012cfa}. The meson $M$ represents both the pseudoscalar and vector mesons, since as will be seen in the next subsections, the vector meson will contribute to the long-distance dynamics as an intermediate states. 
		
In both Eqs.~(\ref{eq:hadronic1}) and (\ref{eq:hadronic2}), the first hadronic matrix element are parameterized in  the same way as:
\begin{align}
\langle P(p)|\bar{u}\gamma^\mu(1-\gamma_5)q|0\rangle&=-if_Pp^\mu,\label{eq:Pdecay}\\
\langle V(p)|\bar{u}\gamma^\mu(1-\gamma_5)q|0\rangle&=m_Vf_V\epsilon^{\ast \mu}.\label{eq:Vdecay}
\end{align}
where $f_P$ and $f_V$ are the corresponding decay constants of the pseudoscalar and vector mesons,  and $\epsilon^\mu$ denotes the polarization of the vector meson. The second matrix element is usually defined as:
	\begin{align}
		\begin{aligned}
			\langle {\cal B}_c(p^\prime,s_z^\prime)| \bar{q}^{\prime}\gamma_\mu(1-\gamma_5)c |{\cal B}_{cc}(p,s_z)\rangle&=\\
			\bar u(p^\prime,s^\prime_z)\Big [ \gamma_\mu f_1(q^2) &+ i\sigma_{\mu\nu}\frac{q^\nu}{M_{\mathcal{B}_{cc}}} f_2(q^2)+\frac{q^\mu}{M_{\mathcal{B}_{cc}}} f_3(q^2) \Big ] u(p,s_z)\\
			-\bar u(p^\prime,s^\prime_z)\Big [ \gamma_\mu g_1(q^2) &+ i\sigma_{\mu\nu}\frac{q^\nu}{M_{\mathcal{B}_{cc}}}g_2(q^2) +\frac{q^\mu}{M_{\mathcal{B}_{cc}}} g_3(q^2) \Big ] \gamma_5 u(p,s_z).
		\label{eq:ff}
		\end{aligned}
	\end{align}
with $q=p-p^\prime$, $M_{\mathcal{B}_{cc}}$ is the mass of doubly charmed baryons, and $f_i$, $g_i$ denote the heavy-light transition form factors which can only be extracted from 
non-perturbative approaches.
		
In general, the weak decay amplitudes of ${\cal B}_{cc}\to{\cal B}_c P$ and $\mathcal{B}_c V$ have the following parametrization form
\begin{align}
		\label{eq:B2BP} \mathcal{A}({\cal B}_{cc}\to{\cal B}_c P)&=i\bar{u}_{\mathcal{B}_c}(A + B\gamma_5) u_{\mathcal{B}_{cc}},\\
		\label{eq:B2BV} \mathcal{A}({\cal B}_{cc}\to{\cal B}_c V)&=\epsilon^{\ast \mu}\bar{u}_{\mathcal{B}_c}
		\left [ A_1\gamma_\mu\gamma_5 + A_2\frac{p_\mu(\mathcal{B}_c)}{M_{\mathcal{B}_{cc}}}\gamma_5 +B_1\gamma_\mu+B_2\frac{p_\mu(\mathcal{B}_c)}{M_{\mathcal{B}_{cc}}}
		\right ] u_{\mathcal{B}_{cc}}.
	\end{align}
The formulas of the parameters $A,B$ and $A_{1,2},B_{1,2}$ under the factorization approach are
	\begin{align}
		A&=\lambda f_P(M_{\mathcal{B}_{cc}}-M_{\mathcal{B}_{c}})f_1(m^2), \hspace{3cm}   B=\lambda f_P(M_{\mathcal{B}_{cc}}+M_{\mathcal{B}_{c}})g_1(m^2),\\
		A_1&=-\lambda f_Vm \left ( g_1(m^2)+g_2(m^2)\frac{M_{\mathcal{B}_{cc}}-M_{\mathcal{B}_{c}}}{M_{\mathcal{B}_{cc}}} \right ),
		\ \ \ \  A_2=-2\lambda f_Vmg_2(m^2), \\
		B_1&=\lambda f_V m \left ( f_1(m^2)-f_2(m^2)\frac{M_{\mathcal{B}_{cc}}+M_{\mathcal{B}_{c}}}{M_{\mathcal{B}_{cc}}} \right ),
		\ \ \ \ \ \ \ \ B_2=2\lambda f_Vmf_2(m^2).
	\end{align}
where $\lambda=\frac{G_F}{\sqrt{2}}V_{CKM}a_{1,2}(\mu)$, $m$ is the mass of pseudoscalar or vector meson.

\subsection{Long-distance contributions from the rescattering mechanism}\label{sec:long}

\begin{figure}[t]
\centering\includegraphics[width=0.4\textwidth]{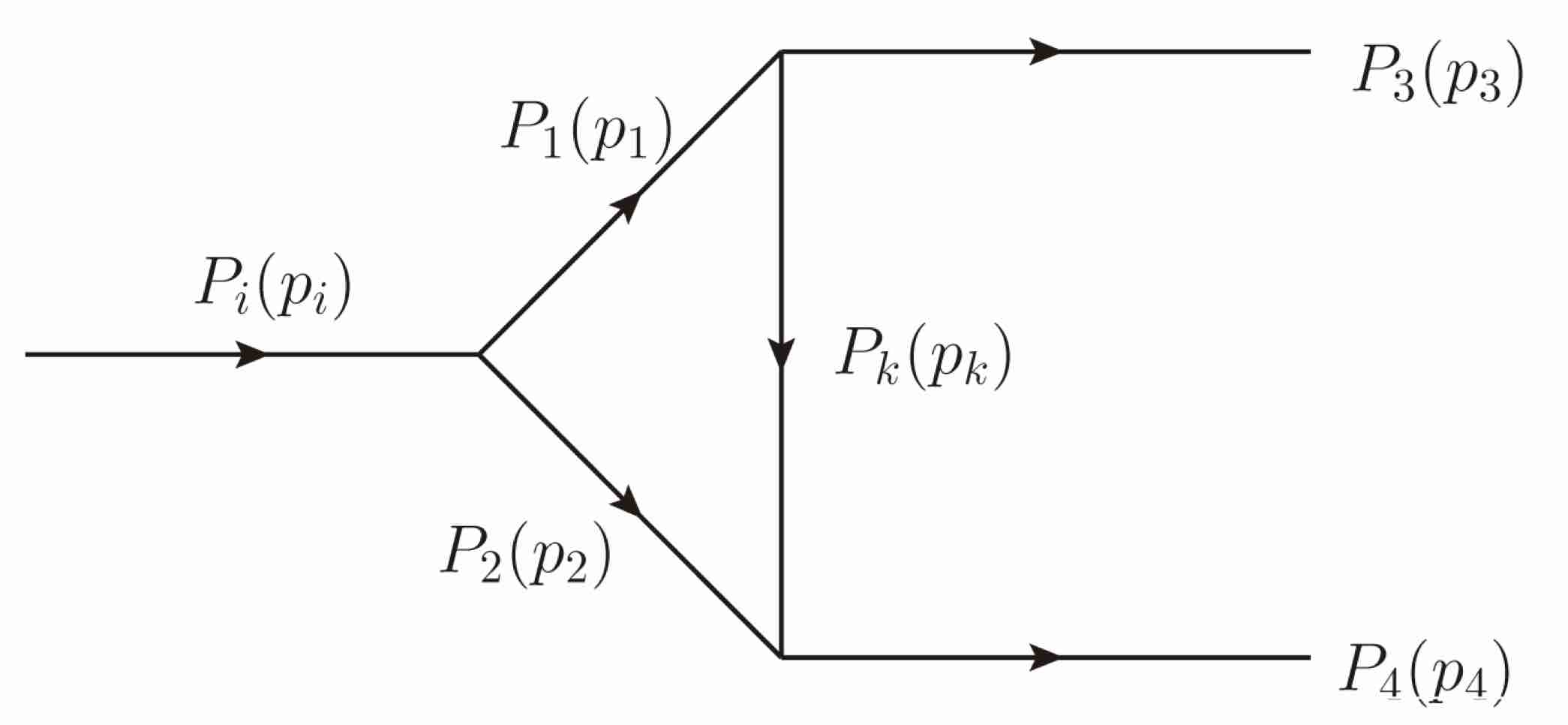}
\caption{The diagram description of the rescattering mechanism at the hadron level.}\label{fig:fig2}
\end{figure}

The long-distance contributions are important but not easy to evaluate. As is done in Ref.\cite{Yu:2017zst}, we calculate the FSIs effects by the rescattering of two intermediate particles at the hadron level using the hadronic strong-interaction effective Lagrangian.
The diagram description of the rescattering mechanism is shown in Fig.~\ref{fig:fig2}.
The weak vertex displayed in hadron-level diagrams only involves the short-distance contributions and thus can be evaluated under the factorization hypothesis.
The subsequent scattering process could be in principle either a $s$-channel resonant-state process or a $t/u$-channel one.
The dominant contribution of the $s$-channel diagram comes from the momentum region $k^2\sim m_{\xip}^2$, which demand the mass of the exchanged particle (a singly charmed baryon in this case) approaching to $m_{\xip}\sim 3.6$GeV. However, until now the heaviest observed singly charmed baryons are much lighter than $m_{\xip}$.
Therefore, the $s$-channel diagram is usually highly suppressed by the off-shell effect and can be safely neglected.
In our calculations, the main contribution will be the $t/u$-channel triangle diagram as shown in Fig.~\ref{fig:fig2}.
	
The particles in a triangle diagram are labeled as $P_n$, where the subscripts $n=i,1,2,3,4,k$ represent the parent doubly charmed particle $(i)$, 
two intermediate hadrons$(1,2)$, two final hadron states$(3,4)$ and the exchanged hadron$(k)$, respectively, as seen in Fig.~\ref{fig:fig2}. The corresponding momenta are assigned as $p_n$.
In general, there are several methods to calculate the amplitude of triangle diagram \cite{Li:2002pj,Ablikim:2002ep,Li:1996cj,Dai:1999cs,Locher:1993cc,Cheng:FSIB,Lu:2005mx}. Their main difference is how to deal with the hadronic loop integration.

{We adopt the optical theorem and Cutkosky cutting rule as in Ref.~\cite{Cheng:FSIB}.
The absorptive part of the amplitude of $P_i\to P_3P_4$ is a product of two distinct parts, the decay of $P_i\to \{P_1P_2\}$ and the rescattering of $\{P_1P_2\}\to P_3P_4$, with the internal particles ($P_1$ and $P_2$) being on-shell.  According to the optical theorem, the absorptive amplitude should sum over all possible on-shell intermediate states $\{P_1P_2\}$ with a phase integration.} It can be expressed as
\begin{align}
{\cal A}bs [\mathcal{M}(P_i\to P_3P_4)] &=\frac{1}{2}\sum_{\{P_1P_2\}} \int\frac{\mbox{d}^3 p_1}{(2\pi)^3 2E_1}\int\frac{\mbox{d}^3 p_2}{(2\pi)^3 2E_2}(2\pi)^4
\delta^4(p_3 + p_4 - p_1-p_2)\non
& \cdot M(P_i\to \{P_1P_2\})T^*(P_3P_4\to \{P_1P_2\}).\label{eq:optical}
\end{align}
Based on the argument in Ref.\cite{FSI:2arguement,Buras:1985xv}, the $2$-body $\rightleftharpoons$ $n$-body rescattering is negligible.
In this approach, the loop integration is transferred into the dispersive part which can be calculated via the dispersion relation
\begin{align}\label{eq:dispersion}
{\cal D}is[ \mathcal{M}(m_1^2)] =\frac{1}{\pi}\int_s^\infty \frac{{\cal A}bs[\mathcal{M}(s^\prime)]}{s^\prime - m_1^2} \mbox{d}s^\prime.
\end{align}
But it suffers from large ambiguities since we cannot reliably describe $\mathcal{M}(s')$ for the whole region. On the other hand, in the charmed meson decays, the large strong phases of the topological diagrams \cite{Cheng:2010ry,Li:2012cfa} indicate that the absorptive (imaginary) part is a dominant one.  Therefore, we will only calculate the absorptive part but neglecting the dispersive one, as done in \cite{Cheng:FSIB}.
In a phenomenological analysis, the non-negligible dispersive contributions can be effectively absorbed into the varying of the parameter $\eta$ which will  be introduced in the following. 

\begin{figure}[tbp]
\centering
\includegraphics[width=0.9\textwidth]{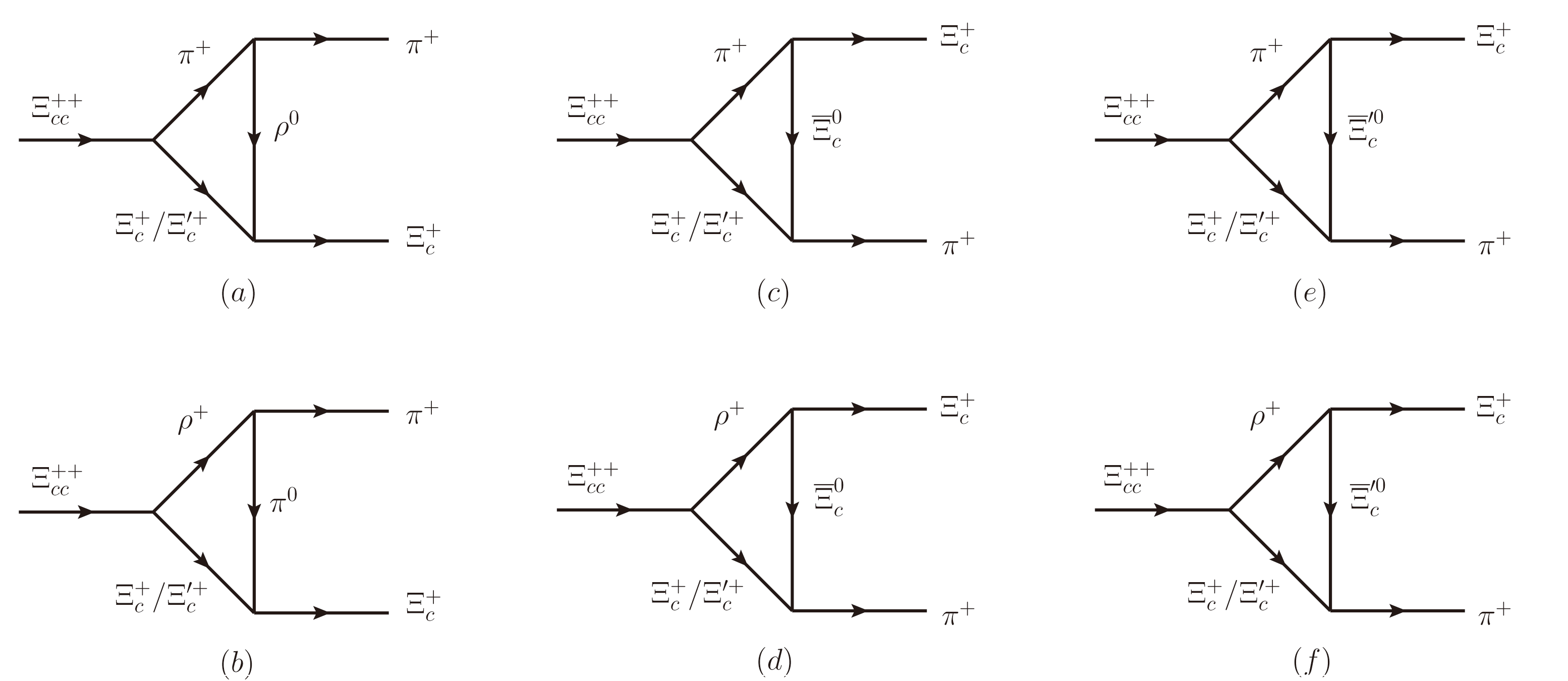}
\caption{The long-distance rescattering contributions to $\xipp\to\Xi_c^+\pi^+$ manifested at hadron level. }\label{fig:fig3}
\end{figure}

Next we express the amplitude of the decay mode $\xipp\to\Xi_c^+\pi^+$ as an example. All the rescattering diagrams are represented in Fig.~\ref{fig:fig3}, which can be summarized as 
$\xipp\to\Xi_c^+(\Xi_c^{\prime+}) \pi^+(\rho^+)\to\Xi_c^+\pi^+$. The intermediate particles can be either light pseudoscalar or vector mesons, or anti-triplet or sextet singly charmed baryons. We use the symbol $\mathcal{M}(P_1,P_2;P_k)$ to denote a triangle amplitude with the intermediate states of $P_1$, $P_2$ and $P_k$.

The Fig.~\ref{fig:fig3}(a) contains two different triangle diagrams, 
with the intermediate two-particle states of $\{\Xi_c^+\pi^+\}$ and $\{\Xi_c^{\prime+}\pi^+\}$, respectively.  The amplitudes of the weak vertex $\xipp\to\Xi_c^+\pi^+$ 
and $\xipp\to\Xi_c^{'\prime}\pi^+$ are taken from Eq.~(\ref{eq:B2BP}) in the factorization approach for the short-distance contributions. The rescattering amplitudes of $\Xi_c^+\pi^+\to\Xi_c^{+}\pi^+$ and $\Xi_c^{\prime+}\pi^+\to\Xi_c^+\pi^+$ 
can be simply carried out from the hadronic strong Lagrangian in Appendix A. Then the absorptive part is written as
\\
{\small
\begin{align}
{\cal A}bs[\mathcal{M}(\pi^+,\Xi_c^+/\Xi_c^{\prime +};\rho^0)] &=\frac{1}{2}\sum_{s_1,\lambda_k}\int\frac{d^3\vec{p}_1}{(2\pi)^32E_1}\frac{d^3\vec{p}_2}{(2\pi)^32
E_2}(2\pi)^4\delta^4(p_i-p_1-p_2) \non
& \hspace{-2.5cm} \cdot i \bar{u}(p_4,s_4) \left(f_{1\rho^0(\Xi_c^+/\Xi_c^{\prime +})\to\Xi_c^+}\gamma_\nu-\frac{if_{2\rho^0(\Xi_c^+/\Xi_c^{\prime +})\to\Xi_c^+}}{2m_{\Xi_c^+}}\sigma_{\mu\nu}k^\mu\right)
u(p_2,s_2)\epsilon^\nu(k,\lambda_k) \non
& \hspace{-2.5cm} \cdot \frac{F^2(t,m_{\rho})}{t-m_{\rho}^2+im_{\rho}\Gamma_{\rho}}(-ig_{\pi^+\to\rho^0\pi^+})\epsilon^{\ast \alpha}(k,\lambda_k)
(p_1+p_3)_\alpha \; i\bar{u}(p_2,s_2)(A+B\gamma_5)u(p_i,s_i)\non
& \hspace{-3.5cm}=\int\frac{|\vec{p}_1|sin\theta d\theta d\phi}{32\pi^2m_{\Xi_{cc}^{++}}}\; i^2(-ig_{\pi^+\to\rho^0\pi^+})
\frac{F^2(t,m_{\rho})}{t-m_{\rho}^2+im_{\rho}\Gamma_{\rho}}  \bar{u}(p_4,s_4) \non
& \hspace{-3cm}\cdot \left[ f_{1\rho^0(\Xi_c^+/\Xi_c^{\prime +})\to\Xi_c^+}\left (-p\llap{/}_1-p\llap{/}_3+\frac{k\cdot (p_1+p_3)\slashed{k}}{m_{\rho}^2} \right )\right.\non
& \hspace{-2.5cm}+\left. \frac{f_{2\rho^0(\Xi_c^+/\Xi_c^{\prime +})\to\Xi_c^+}}{2m_{\Xi_c^+}} \left(-\slashed{k}(p\llap{/}_1+p\llap{/}_3)+k\cdot (p_1+p_3) \right)\right]  \cdot (p\llap{/}_2+m_2)(A+B\gamma_5)u(p_i,s_i).
\label{eq11}
\end{align} }
\\
where $t=p_k^2=(p_3-p_1)^2$. In the center of mass frame of $\xipp$, the $3$-momentum of final-state baryon $\vec{p}_3$ is defined at the plus direction of $z$ axis, with two angles $\theta$ and $\phi$ standing for the polar and azimuthal angles of $\vec{p}_1$ in spherical coordinate system.
$g_{\rho\pi\pi}$, $f_{1\Xi_c^+\Xi_c^+\rho^0}$ and $f_{2\Xi_c^+\Xi_c^+\rho^0}$  are relevant strong coupling constants, which are usually extracted or calculated in the on-shell condition.
However, the exchanged $\rho^0$ is generally off-shell, so that the strong coupling constants are not exactly correct.
To include the off-shell effect of $\rho^0$, a form factor $F(t,m_\rho)$ \cite{Cheng:FSIB} is introduced as
\begin{align}
F(t,m_\rho)=\left(\frac{\Lambda^2-m_\rho^2}{\Lambda^2-t}\right)^n. \label{eq:Ffactor}
\end{align}
This form factor normalize to unity at the on-shell situation $t=p_k^2=m_\rho^2$. The cutoff $\Lambda$ has the form of
\begin{align}
\Lambda= m_\rho + \eta \Lambda_{\rm QCD}.
\end{align}
with $\Lambda_{\rm QCD}=330\text{MeV}$ for the charm decays. The parameter $\eta$ can not be calculated from the first-principle QCD method, and usually needs to be phenomenologically determined by the experimental data. The results are always sensitive to the value of $\eta$. 
More discussions about $\eta$ can be found in Section \ref{sec:inputs}.
{Usually, the exponential factor $n$ in Eq.~(\ref{eq:Ffactor}) can be taken as $1$ or $2$ as a monopole or dipole behavior. For $B$ meson decays in Ref.~\cite{Cheng:FSIB}, the resultant branching ratios are almost the same for both choices. Hence in our work, we choose $n=1$ for convenience following by Ref.~\cite{Cheng:FSIB}.}

By the same way, the absorptive amplitude of the rest triangle diagrams in Fig.~\ref{fig:fig3} are
{\small
\begin{align}
Abs[\mathcal{M}(\rho^+,\Xi_c^+/\Xi_c^{\prime+};\pi^0)]&=\int\frac{|\vec{p_1}|sin\theta d\theta d\phi}{32\pi^2m_{\Xi_{cc}^{++}}}
(-ig_{\rho^+\to\pi^0\pi^+})(ig_{\pi^0(\Xi_c^+/\Xi_c^{\prime+})\to\Xi_c^+})\bar{u}(p_4,s_4)i\gamma_5(\slashed{p_2}+m_2)\non
&\cdot\frac{F^2(t,m_{\pi^0})}{t-m_{\pi^0}^2+im_{\pi^0}\Gamma_{\pi^0}}((-2\slashed{p_3}+\frac{2p_3\cdot p_1\slashed{p_1}}{m_1^2})(A_1\gamma_5+B_1)\non
& +\frac{-2m_1^2p_3\cdot p_2+2p_3\cdot p_1p_1\cdot p_2}{m_1^2m_{\Xi_{cc}^{++}}}(A_2\gamma_5+B_2))u(p_i,s_i),
\end{align}}
\begin{align}
Abs[\mathcal{M}(\pi^+,\Xi_c^+/\Xi_c^{\prime+};\Xi_c^0)]&=\int\frac{|\vec{p_1}|sin\theta d\theta d\phi}{32\pi^2m_{\Xi_{cc}^{++}}}
ig_{\pi^+\Xi_c^0\to\Xi_c^+}g_{(\Xi_c^+/\Xi_c^{\prime+})\to\Xi_c^0\pi^+}\bar{u}(p_3,s_3)\gamma_5(\slashed{k}+m_k)\non
&\cdot\gamma_5(\slashed{p_2}+m_2)(A+B\gamma_5)u(p_i,s_i)\frac{F^2(t,m_{\Xi_c^0})}{t-m_{\Xi_c^0}^2+im_{\Xi_c^0}\Gamma_{\Xi_c^0}},
\end{align}
\begin{align}
Abs[\mathcal{M}(\rho^+,\Xi_c^+/\Xi_c^{\prime+},\Xi_c^0)]&=\int\frac{|\vec{p_1}|sin\theta d\theta d\phi}{32\pi^2m_{\Xi_{cc}^{++}}}i^3\bar{u}(p_3,s_3)
\left(f_{\rho^+\Xi_c^0\to\Xi_c^+}\gamma_\nu-\frac{if_{2\rho^+\Xi_c^0\to\Xi_c^+}}{m_k+m_3}\sigma_{\mu\nu}p_1^\mu\right)\non
&\cdot g_{(\Xi_c^+/\Xi_c^{\prime+})\to\Xi_c^0\pi^+}(\slashed{k}+m_k) \frac{F^2(t,m_{\Xi_c^0})}{t-m_{\Xi_c^0}^2+im_{\Xi_c^0}\Gamma_{\Xi_c^0}}\gamma_5(-g^{\nu\alpha}+\frac{p_1^\nu p_1^\alpha}{m_1^2})
\non
&\cdot(\slashed{p_2}+m_2)(A_1\gamma_\alpha\gamma_5+A_2\frac{p_{2\alpha}}{m_{\xip}}\gamma_5+B_1\gamma_\alpha+B_2\frac{p_{2\alpha}}{m_{\xip}})u(p_i,s_i),
\end{align}
\begin{align}
Abs[\mathcal{M}(\pi^+,\Xi_c^+/\Xi_c^{\prime+};\Xi_c^{\prime 0})]&=\int\frac{|\vec{p_1}|sin\theta d\theta d\phi}{32\pi^2m_{\Xi_{cc}^{++}}}
ig_{\pi^+\Xi_c^{\prime 0}\to\Xi_c^+}g_{(\Xi_c^+/\Xi_c^{\prime+})\to\Xi_c^{\prime 0}\pi^+}\bar{u}(p_3,s_3)\gamma_5(\slashed{k}+m_k)\non
&\cdot\gamma_5(\slashed{p_2}+m_2)(A+B\gamma_5)u(p_i,s_i)\frac{F^2(t,m_{\Xi_c^0})}{t-m_{\Xi_c^0}^2+im_{\Xi_c^0}\Gamma_{\Xi_c^0}},
\end{align}
\begin{align}
Abs[\mathcal{M}(\rho^+,\Xi_c^+/\Xi_c^{\prime+},\Xi_c^{\prime 0})]&=\int\frac{|\vec{p_1}|sin\theta d\theta d\phi}{32\pi^2m_{\Xi_{cc}^{++}}}i^3\bar{u}(p_3,s_3)
\left(f_{\rho^+\Xi_c^{\prime 0}\to\Xi_c^+}\gamma_\nu-\frac{if_{2\rho^+\Xi_c^{\prime 0}\to\Xi_c^+}}{m_k+m_3}\sigma_{\mu\nu}p_1^\mu\right)\non
&\cdot g_{(\Xi_c^+/\Xi_c^{\prime+})\to\Xi_c^{\prime 0}\pi^+}(\slashed{k}+m_k) \frac{F^2(t,m_{\Xi_c^0})}{t-m_{\Xi_c^0}^2+im_{\Xi_c^0}\Gamma_{\Xi_c^0}}\gamma_5(-g^{\nu\alpha}+\frac{p_1^\nu p_1^\alpha}{m_1^2})
\non
&\cdot(\slashed{p_2}+m_2)(A_1\gamma_\alpha\gamma_5+A_2\frac{p_{2\alpha}}{m_{\xip}}\gamma_5+B_1\gamma_\alpha+B_2\frac{p_{2\alpha}}{m_{\xip}})u(p_i,s_i).
\end{align}
		
Collecting all the pieces together, the amplitude of  the decay $\xipp \to \Xi_c^+ \pi^+$ can be written in the following form 
\begin{align}
\mathcal{A}(\Xi_{cc}^{++}\to\Xi_c^+\pi^+) & = T_{SD}(\Xi_{cc}^{++}\to\Xi_c^+\pi^+) \non
& \hspace{-2cm}+ i Abs \Big[\mathcal{M}(\pi^+,\Xi_c^+/\Xi_c^{\prime +};\rho^0) + \mathcal{M}(\rho^+,\Xi_c^+/\Xi_c^{\prime+};\pi^0) + \mathcal{M}(\pi^+,\Xi_c^+/\Xi_c^{\prime+};\Xi_c^0) \non
& \hspace{-2cm}+ \mathcal{M}(\rho^+,\Xi_c^+/\Xi_c^{\prime+},\Xi_c^0) 
+ \mathcal{M}(\pi^+,\Xi_c^+/\Xi_c^{\prime+};\Xi_c^{\prime 0}) + \mathcal{M}(\rho^+,\Xi_c^+/\Xi_c^{\prime+},\Xi_c^{\prime 0}) \Big].
\end{align}
The short-distance contributions of this decay mode is labeled by $T_{SD}$. The analytical expression for all the other channels are given in Appendix \ref{app:amp}.

\section{Numerical results and discussions}\label{sec:inputs}

The width of a two-body decay $\mathcal{B}_{cc}\to \mathcal{B}_cP$ is
\begin{align}
\Gamma(\mathcal{B}_{cc}(\lambda_i)\to \mathcal{B}_c(\lambda_f)P) =  {|\vec{p}|\over 8\pi m_{\mathcal{B}_{cc}}} {1\over2} \sum_{\lambda_i,\lambda_f}
 \Big\vert\mathcal{A}(\mathcal{B}_{cc}(\lambda_i)\to \mathcal{B}_c(\lambda_f)P)\Big\vert^2.
\end{align}
where $|\vec{p}|=\sqrt{[m_{\mathcal{B}_{cc}}^2-(m_{\mathcal{B}_c}+m_P)^2][m_{\mathcal{B}_{cc}}^2 -(m_{\mathcal{B}_c}-m_P)^2]}/(2m_{\mathcal{B}_{cc}})$, and $\lambda_i$, $\lambda_f$ are the corresponding spin polarizations.

\subsection{Inputs}

\begin{table}[b]
\centering
\caption{Masses (in units of GeV) and lifetime (in units of fs) of the doubly charmed baryons used in this work.} \label{table:mass}
\begin{tabular}{cccc}	\hline 				\hline
~~baryons~~~~~&~~~~~~~~~~$\Xi_{cc}^{++}$ ~~~~~~~~~~&~~~~~~~~~~$\Xi_{cc}^{+}$ ~~~~~~~~~~&~~~~~~~~~~$\Omega_{cc}^{+}$~~~~~\\
\hline
masses&3.621\cite{Aaij:2017ueg,Aaij:2018gfl}&3.621\cite{Yu:2018com}&3.738\cite{Yu:2018com}\\
\hline
lifetime&256\cite{Aaij:2018wzf}&140\cite{Berezhnoy:2018bde}&180\cite{Berezhnoy:2018bde}\\
\hline
\hline
\end{tabular}
\end{table}

All the inputs used in this work are clarified here. The mass of $\Xi_{cc}^{++}$ is well measured by LHCb \cite{Aaij:2017ueg,Aaij:2018gfl}. A lot of theoretical papers have studied the masses of the doubly charmed baryons \cite{PDG,Chen:2016spr}. As to the ground states, there would be not much difference between the theoretical predictions on the masses, benefit from the measurement on the mass of $\Xi_{cc}^{++}$. We adopt the results from Ref.\cite{Yu:2018com}, as shown in Table \ref{table:mass}. 

The lifetimes of the doubly charmed baryons play an essential role in the theoretical predictions and experimental searches \cite{Yu:2019lxw}. The branching fractions are proportional to the lifetimes, $\mathcal{BR}_i=\Gamma_i\cdot\tau$. Besides, the longer lifetime will be helpful for the experiments to reject the large backgrounds at the decaying vertex. However, the theoretical predictions on the lifetimes of the doubly charmed baryons are of large ambiguities due to the non-perturbative contributions at the charm scale \cite{Onishchenko:2000yp,Kiselev:2001fw,Chang:2007xa,Karliner:2014gca}. Currently, the lifetime of $\xipp$ has been well measured by LHCb \cite{Aaij:2018wzf}. After this measurement, a phenomenological analysis considering the contributions of dimension-7 operators gives $\tau(\xipp)=298$fs,  $\tau(\xip)=44$fs and  $\tau(\opp)=206$fs \cite{Cheng:2018mwu}. Another work predicts $\tau(\xip)=140$fs and  $\tau(\opp)=180$fs \cite{Berezhnoy:2018bde}. It can be seen that there still exists large uncertainties in understanding the lifetimes of $\xip$ and $\opp$. Since it is just an overall factor of the branching fractions of individual initial particles, the values of the lifetimes could be easily changed in our results. Not any result is preferred, but we have to use one of them, as shown in Table \ref{table:mass}.

The masses and decay constants of all the final state hadrons come from Refs. \cite{PDG,Choi:2015ywa,Feldmann:1998vh}. The heavy-light transition form factors of $\mathcal{B}_{cc}\to \mathcal{B}_c$ have been calculated in several methods. The results of form factors from the light-front quark model \cite{Wang:2017mqp} have been successfully used in the prediction on the discovery channels in \cite{Yu:2017zst}, which will be used in this work as well. 
Besides, the strong coupling constants of the various hadrons are also important inputs, some of them can be extracted from the experimental decay widthes, such as from $\Gamma(\rho^0\to\pi^+\pi^-)$ and $\Gamma(K^{\ast+}\to K^0\pi^+)$, $g_{\rho\pi\pi}$($g_{\rho^0\pi^+\pi^-}$, $g_{\rho^\pm\pi^\pm\pi^0}$) and $g_{K^{\ast+} K^0\pi^+}$ are respectively determined as 6.05 and 4.6 \cite{Cheng:FSIB}.
Under the flavor $SU(3)$ symmetry, $g_{K^{\ast+} K^0\pi^+}$ can be related to $g_{\rho\pi\pi}$, with $g_{K^{\ast+} K^0\pi^+}^{s} = {1\over\sqrt{2}} g_{\rho\pi\pi} = 4.28$. It deviates from the extracted value 4.6 about $7\%\sim{m_s-m_{u,d}\over \Lambda_{QCD}}$, which is the flavor $SU(3)$ breaking effects mainly caused by the mass difference of $s$ and $u,d$ quark.
In our calculation, we take the extracted one $g_{K^{\ast+} K^0\pi^+}=4.6$ and relate any other $VPP$ coupling with strange mesons participating to this value, e.g. $g_{K^{\ast+} K^+\eta_8}={3\over\sqrt{6}}g_{K^{\ast+} K^0\pi^+}=5.63$.
For coupling constants of the singly heavy baryons coupled with light mesons, we adopt the theoretical calculation results from the literatures \cite{Aliev:2010yx,Yan:1992gz,Casalbuoni:1996pg,Meissner:1987ge,Li:2012bt,Aliev:2010nh}. A simple counting the uncertainty of the strong coupling constants is about $30\%$ caused by the QCD sum rules\cite{Aliev:2010yx,Aliev:2010nh}.
All the strong coupling constants appeared are gathered in the Appendix \ref{app:lag}.

\subsection{Dependence on $\eta$ and its cancellation by ratio of branching fractions}

\begin{figure}[t]
\centering
\subfigure[]{\begin{minipage}[]{0.45\linewidth}	 \includegraphics[width=3.0in]{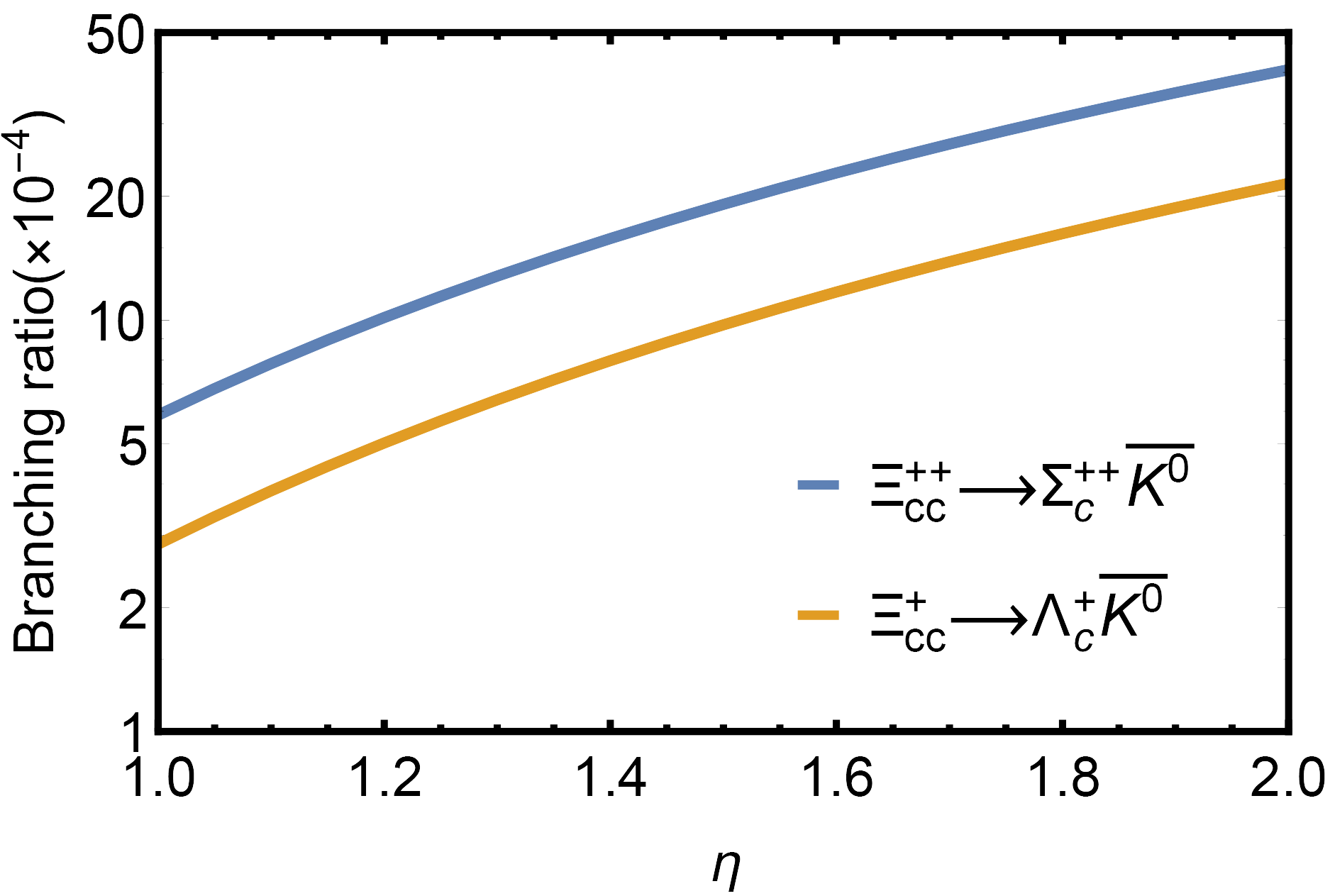} \end{minipage} 	}\hspace{1cm}
\subfigure[]{	\begin{minipage}[]{0.45\linewidth} \includegraphics[width=3.0in]{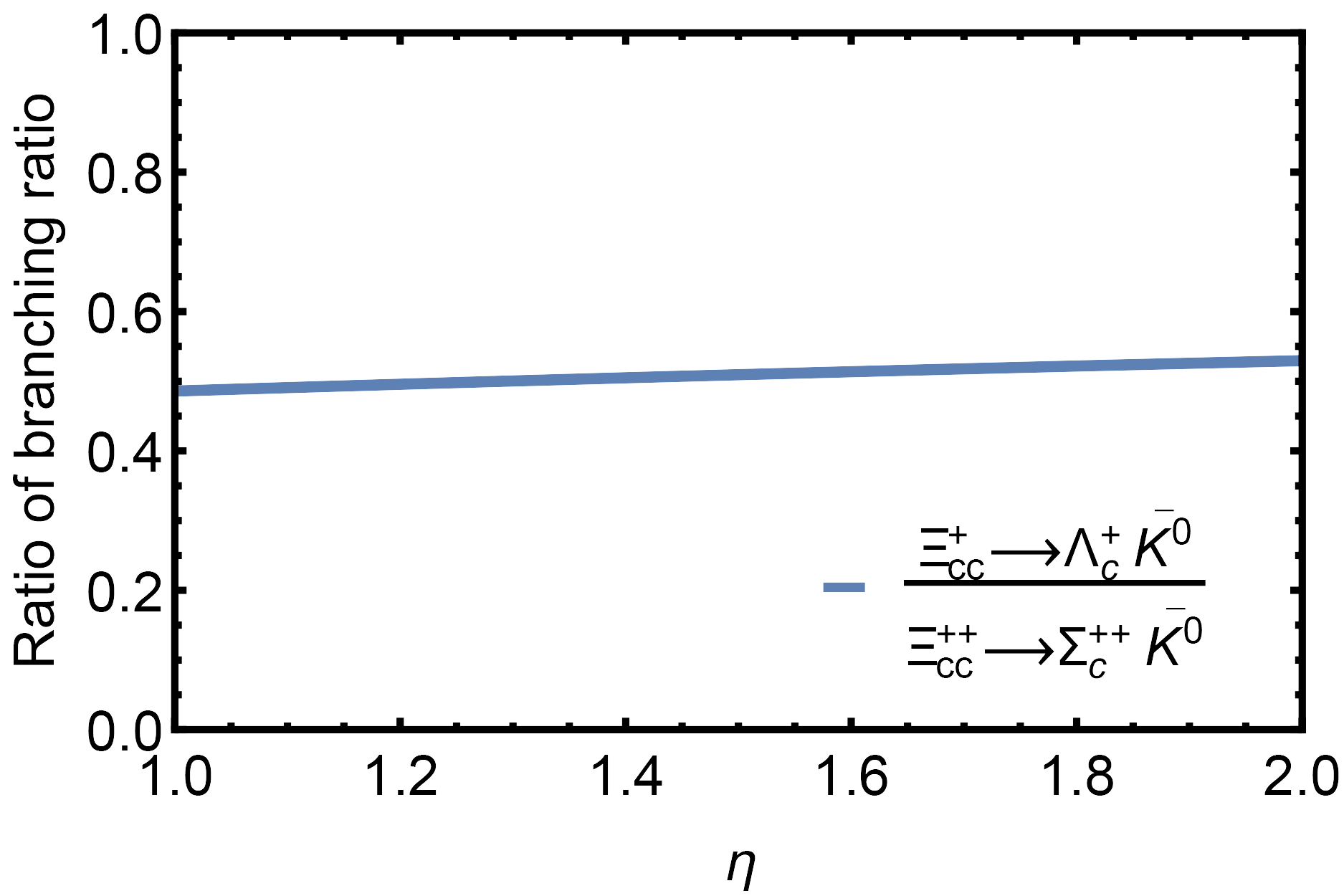}			
\end{minipage} }
\caption{ (a): The theoretical predictions for the branching ratios of $\Xi_{cc}^{++}\to\Sigma_c^{++}\overline{K}^{0}$ and $\Xi_{cc}^{+}\to\Lambda_c^+\overline{K}^{0}$ in logarithmic coordinates; 
(b): Ratio of branching fractions with $\eta$ varying from 1.0 to 2.0.}
\label{fig:fig4}
\end{figure}

A form factor in Eq.~(\ref{eq:Ffactor}) is introduced into the amplitude of triangle diagrams, to describe the off-shell effect of the exchanged particles. 
When the exponential factor and $\Lambda_{QCD}$ are fixed with specified values (in our case, we take $n=1$ and $\Lambda_{QCD}=330$MeV), the form factor is a function of three variables, i.e. the momentum square, the mass of the exchanged particle and the parameter $\eta$, apparently $F(k^2, m_{exe}^2, \eta)$.
The $k^2$ (varying in a range) and $m_{exe}$ are definite for any individual diagram.
The left parameter, $\eta$, is however a {process-dependent} parameter which cannot be calculated from the first-principle QCD methods. 
The value of $\eta$ is usually determined by the experimental data as in \cite{Cheng:FSIB}. As in the case of the doubly charmed baryon decays without any available data, it can be expected that $\eta$ will cause large uncertainties in the theoretical predictions on the branching fractions.

In Fig.~\ref{fig:fig4}(a), we display the dependence of the branching fractions on the parameter $\eta$, taking $\Xi_{cc}^{++} \to \Sigma_c^{++} \overline{K}^0$ and $\Xi_{cc}^+ \to \Lambda_c^+ \overline{K}^0$ as examples, which both are dominated by the long-distance dynamics. It is clear that the branching fractions are very sensitive to the value of $\eta$. With $\eta$ varying in the range between $1.0$ and $2.0$, the branching fractions could be changed by nearly one order of magnitude. 
This is the well-known feature of the FSIs effects which have very large theoretical uncertainties. 

The problem of large uncertainty induced by varying the value of $\eta$ can be solved by the ratio of branching fractions, firstly proposed in \cite{Yu:2017zst}. 
From Fig.~\ref{fig:fig4}(a), we can easily find that the dependence of the branching ratios of $\Xi_{cc}^{++} \to \Sigma_c^{++} \overline{K}^0$ and $\Xi_{cc}^+ \to \Lambda_c^+ \overline{K}^0$ on $\eta$ have similar line shapes.
That means that the dependence of the ratio of the two branching fractions on $\eta$ can be mostly canceled. Therefore, the ratio of branching fractions will be insensitive to the value of $\eta$, as apparently shown in Fig.~\ref{fig:fig4}(b).
{From Fig.\ref{fig:fig4}(a), the branching fraction of $\Xi_{cc}^{++}\to \Sigma_c^{++}\overline K^0$ changes from $0.59\times 10^{-3}$ to $4.06\times 10^{-3}$ with the parameter $\eta$ varying from 1.0 to 2.0. Similarly for $\Xi_{cc}^{+}\to \Lambda_c^{+}\overline K^0$, it changes from $0.29\times 10^{-3}$ to $2.15\times 10^{-3}$. Both the branching fractions of both processes are varied by nearly one order of magnitude. Taking the central values of the above results, they are $(2.33\pm1.74)\times10^{-3}$ and $(1.22\pm0.93)\times10^{-3}$, respectively, with the uncertainties around $75\%$. However, the ratio of the above branching fractions is $0.48\sim 0.53$, seen in Fig.\ref{fig:fig4}(b). Written as $0.51\pm0.03$, the uncertainty of the ratio is $5\%$. Therefore, almost $70\%$ of the uncertainties of the absolute branching fractions are cancelled by the ratio of branching fractions.
 Similar cancelation behavior also happens between all other decay channels.}
The reason of this cancelation is as follows. The difference of the triangle diagrams between the two modes of $\Xi_{cc}^{++} \to \Sigma_c^{++} \overline{K}^0$ and $\Xi_{cc}^+ \to \Lambda_c^+ \overline{K}^0$ on $\eta$ mainly stems from the strong coupling constants and the masses of the particles in the rescattering diagrams. 
The corresponding quantities for different decay modes can be related under the flavor $SU(3)$ symmetry. The dependence on $\eta$ would be similar for these both modes, thus can be mostly cancelled in the ratio of branching fractions. In this way, the theoretical uncertainties could be under control. That is why we can successfully predict the discovery channels of $\xipp$ in \cite{Yu:2017zst}. In this work, we will present the results of branching fractions with different values of $\eta$, to show the absolute uncertainty of any individual mode. The readers can take a ratio between any two processes, to get a more reliable result. 

Note that the above cancellation can be broken down by the so-called triangle singularity, in which all the intermediate particles are on-shell. We will not tackle this issue in this work. Besides, as discussed in Ref.~\cite{Cheng:FSIB}, the value of $\eta$ are not identical for different classes of decay modes. For example, the values of $\eta$ are different for $B\to D\pi$ and $B\to \pi\pi$ decays, to satisfy the constraints from the  experimental data. This can be easily understood that the exchanged particles are different in different decay modes. Therefore, we should be careful to consider the values of $\eta$'s among the classes of decaying processes.

\subsection{Numerical results of branching fractions}\label{sec:results}

\begin{sidewaystable}
\centering
\caption{Branching ratios for the short-distance dynamics dominated modes. The ``CF", ``SCS" and ``DCS" represent CKM favored, singly CKM suppressed and doubly CKM suppressed processes, respectively.}
\label{result1}
\begin{tabular}{clc|cccc|c}	\hline\hline
\rule{0pt}{12pt}Particles&Decay modes&Topology&\ \ \ $\mathcal{BR}_{T_{SD}}(\%)$\ \ \ &$\mathcal{BR}_{\eta=1.0}(\%)$&$\mathcal{BR}_{\eta=1.5}(\%)$&$\mathcal{BR}_{\eta=2.0}(\%)$&\ \ \ CKM\ \ \ \\ \hline 				\hline
\rule{0pt}{10pt}$\Xi_{cc}^{++}$&$\to\Xi_c^+\pi^+$ &$\lambda_{sd} (T+C^\prime)$&6.76&7.11&8.48&10.75&CF\\ 				\hline
\rule{0pt}{10pt}&$\to\Xi_c^{\prime +}\pi^+$&${1\over\sqrt{2}}\lambda_{sd}\big(\tilde{T}+\tilde{C}^{\prime}\big)$&4.71&4.72&4.72&4.74&CF\\ 				\hline
\rule{0pt}{10pt}&$\to\Sigma_c^+\pi^+$ &${1\over\sqrt{2}}\lambda_{d}\big(\tilde{T}+\tilde{C}^{\prime}\big)$&0.248&0.251&0.255&0.261&SCS\\ 				\hline
\rule{0pt}{10pt}&$\to\Lambda_c^+\pi^+$&$\lambda_d(T+C^{\prime})$&0.386&0.390&0.393&0.396&SCS\\ 				\hline
\rule{0pt}{10pt}&$\to\Xi_c^{\prime +}K^+$ &${1\over\sqrt{2}}\lambda_s \big(\tilde{T}+\tilde{C}^{\prime}\big)$&0.303&0.304&0.304&0.305&SCS\\
				\hline
				\rule{0pt}{10pt}&$\to\Xi_c^+K^+$&$\lambda_s(T+C^{\prime})$&0.538&0.538&0.538&0.538&SCS\\
				\hline
				\rule{0pt}{10pt}&$\to\Lambda_c^+K^+$&$\lambda_{ds}(T+C^\prime)$&0.028&0.029&0.030&0.031&DCS\\
				\hline
				\rule{0pt}{10pt}&$\to\Sigma_c^+K^+$&${1\over\sqrt{2}}\lambda_{ds}\big(\tilde{T}+ \tilde{C}^{\prime}\big)$&0.016&0.016&0.018&0.021&DCS\\
				\hline
				\hline
				\rule{0pt}{10pt}$\Xi_{cc}^{+}$&$\to\Xi_c^0\pi^+$&$\lambda_{sd}(T-E_2)$&4.08&4.34&4.58&4.74&CF\\
				\hline
				\rule{0pt}{10pt}&$\to\Xi_c^{\prime 0}\pi^+$&${1\over \sqrt{2}} \lambda_{sd} \big(\tilde{T}+\tilde{E}_2\big)$ &2.84&2.84&2.84&2.84&CF\\
				\hline
				\rule{0pt}{10pt}&$\to\Sigma_c^0\pi^+$&$\lambda_d(\tilde{T}+\tilde{E}_2)$&0.31&0.32&0.33&0.35&SCS\\
				\hline
				\rule{0pt}{10pt}&$\to\Xi_c^{\prime 0}K^+$&${1\over\sqrt{2}}\big(\lambda_s \tilde{T} + \lambda_d \tilde{E}_2\big)$ &0.18&0.18&0.18&0.18&SCS\\
				\hline
				\rule{0pt}{10pt}&$\to\Xi_c^{0}K^+$&$\lambda_s T + \lambda_d E_2$ &0.32&0.32&0.33&0.33&SCS\\
				\hline
				\rule{0pt}{10pt}&$\to\Sigma_c^0K^+$&$\lambda_{ds}\tilde{T}$&0.02&&&&DCS\\
				\hline
				\hline
				\rule{0pt}{10pt}$\Omega_{cc}^{+}$&$\to\Omega_c^{0}\pi^+$&$\lambda_{sd}\tilde{T}$&6.09&&&&CF\\
				\hline
				\rule{0pt}{10pt}&$\to\Xi_c^{0}\pi^+$&$-\lambda_d T - \lambda_s E_2 $ &0.23&0.23&0.23&0.23&SCS\\
				\hline
				\rule{0pt}{10pt}&$\to\Xi_c^{\prime 0}\pi^+$&${1\over\sqrt{2}}\big(\lambda_d \tilde{T} + \lambda_s \tilde{E}_2\big)$ &0.16&0.16&0.17&0.17&SCS\\
				\hline
				\rule{0pt}{10pt}&$\to\Omega_c^{0}K^+$&$\lambda_s\big(\tilde{T}+\tilde{E}_2\big)$&0.38&0.40&0.40&0.40&SCS\\
				\hline
				\rule{0pt}{10pt}&$\to\Xi_c^0K^+$&$\lambda_{ds}(-T+E_2)$ &0.019&0.091&0.091&0.091&DCS\\
				\hline
				\rule{0pt}{10pt}&$\to\Xi_c^{\prime 0}K^+$&${1\over \sqrt{2}}\lambda_{ds} \big(\tilde{T}+\tilde{E}_2\big)$&0.011&0.011&0.011&0.011&DCS\\
				\hline
				\hline
			\end{tabular}
		\end{sidewaystable}

\begin{sidewaystable}
\centering
\caption{Branching ratios for the long-distance dominated Cabibbo-favored ($\lambda_{sd}$) modes. For the channels
involving internal W-emission contributions, the short-distance factorizable contributions are also listed at the fourth column for comparison.}
			\label{result2}
			\begin{tabular}{clc|cccc}
				\hline
				\hline
				\rule{0pt}{12pt}Particles&Decay modes&Topology&\ \ \ $\mathcal{BR}_{T_{SD}}(\times 10^{-3})$\ \ \ &$\mathcal{BR}_{\eta=1.0}(\times 10^{-3})$&
$\mathcal{BR}_{\eta=1.5}(\times 10^{-3})$&$\mathcal{BR}_{\eta=2.0}(\times 10^{-3})$\\
				\hline
				\hline
				\rule{0pt}{10pt}$\Xi_{cc}^{++}$&$\to\Sigma_c^{++}\overline{K}^0$&$\tilde{C}$&0.015&0.59&1.91&4.06\\
				\hline
				\hline
				\rule{0pt}{10pt}$\Xi_{cc}^{+}$&$\to\Omega_c^0K^+$&$\tilde{E}_2$&&0.049&0.15&0.29\\
				\hline
				\rule{0pt}{10pt}&$\to\Sigma_c^+\overline{K}^0$&${1\over\sqrt{2}}\big(\tilde{C}+\tilde{E}_1\big)$&0.009&1.55&4.72&9.95\\
				\hline
				\rule{0pt}{10pt}&$\to\Lambda_c^+\overline{K}^0$&$-C+E_1$&0.017&0.29&0.98&2.15\\
				\hline
				\rule{0pt}{10pt}&$\to\Sigma_c^{++}K^-$&$\tilde{E}_1$&&0.075&0.26&0.55\\
				\hline
				\rule{0pt}{10pt}&$\to\Xi_c^{\prime +}\pi^0$&${1\over2}(-\tilde{C}^{\prime}+\tilde{E}_2)$&&0.33&0.98&1.97\\
				\hline
				\rule{0pt}{10pt}&$\to\Xi_c^{\prime +}\eta_1$ &${1\over\sqrt{6}}(\tilde{C}^{\prime}+\tilde{E}_1+\tilde{E}_2)$&&0.57&1.73&3.50\\
				\hline
				\rule{0pt}{10pt}&$\to\Xi_c^{\prime +}\eta_8$ &${1\over2\sqrt{3}}(\tilde{C}^{\prime}-2\tilde{E}_1+\tilde{E}_2)$&&0.22&0.66&1.37\\
				\hline
				\rule{0pt}{10pt}&$\to\Xi_c^+\pi^0$&$-{1\over\sqrt{2}}(C^{\prime}+E_2)$&&3.24&10.2&21.0\\
				\hline
				\rule{0pt}{10pt}&$\to\Xi_c^+\eta_1$&${1\over\sqrt{3}}(C^{\prime}+E_1-E_2)$&&0.18&0.57&1.20\\
				\hline
				\rule{0pt}{10pt}&$\to\Xi_c^+\eta_8$&${1\over\sqrt{6}}(C^{\prime}-2E_1-E_2)$&&0.11&0.35&0.66\\
				\hline
				\hline
				\rule{0pt}{10pt}$\Omega_{cc}^{+}$&$\to\Xi_c^{\prime +}\overline{K}^0$&${1\over\sqrt{2}}(\tilde{C}+\tilde{C}^{\prime})$&0.010&1.10&3.38&6.84\\
				\hline
				\rule{0pt}{10pt}&$\to\Xi_c^+\overline{K}^0$&$-C+C^{\prime}$&0.017&0.73&2.30&4.38\\
				\hline
				\hline
			\end{tabular}
		\end{sidewaystable}

\begin{sidewaystable}
\centering
			\caption{Same as Table.\ref{result2} but for the long-distance dominated singly Cabibbo-suppressed modes.}
			\label{result3}
			\begin{tabular}{clc|cccc}
				\hline
				\hline
				\rule{0pt}{12pt}Particles&Decay modes&Topology&\ \ \ $\mathcal{BR}_{T_{SD}}(\times 10^{-5})$\ \ \ &$\mathcal{BR}_{\eta=1.0}(\times 10^{-5})$&$\mathcal{BR}_{\eta=1.5}(\times 10^{-5})$&
$\mathcal{BR}_{\eta=2.0}(\times 10^{-5})$\\
				\hline
				\hline
				\rule{0pt}{10pt}$\Xi_{cc}^{++}$&$\to\Sigma_c^{++}\pi^0$&$-{1\over\sqrt{2}}\tilde{C}$ &0.062&3.39&11.5&25.4\\
				\hline
				\rule{0pt}{10pt}&$\to\Sigma_c^{++}\eta_1$&${1\over\sqrt{3}}(\lambda_d+\lambda_s) \tilde{C}$&0.022&1.18&2.43&4.35\\
				\hline
				\rule{0pt}{10pt}&$\to\Sigma_c^{++}\eta_8$&${1\over\sqrt{6}}(\lambda_d-2\lambda_s) \tilde{C}$&0.059&3.24&8.62&15.3\\
				\hline
				\hline
				\rule{0pt}{10pt}$\Xi_{cc}^{+}$&$\to\Sigma_c^{+}\pi^0$ &${1\over2}\lambda_d\big(-\tilde{C}-\tilde{C}^{\prime}+\tilde{E}_1+\tilde{E}_2\big)$&0.038&3.85&12.1&25.0\\
				\hline
				\rule{0pt}{10pt}&$\to\Sigma_c^{+}\eta_1$ &${1\over\sqrt{6}}\big[\lambda_d\big(\tilde{C}+\tilde{C}^{\prime}+\tilde{E}_1+\tilde{E}_2\big) + \lambda_s\tilde{C}\big]$&0.026&2.52&7.69&16.4\\
				\hline
				\rule{0pt}{10pt}&$\to\Sigma_c^{+}\eta_8$ &${1\over2\sqrt{3}}\big[\lambda_d\big(\tilde{C}+\tilde{C}^{\prime}+\tilde{E}_1+\tilde{E}_2\big) - 2\lambda_s\tilde{C}\big]$&0.016&1.64&4.52&8.70\\
				\hline
				\rule{0pt}{10pt}&$\to\Lambda_c^{+}\pi^0$&${1\over\sqrt{2}}\lambda_d(C-C^{\prime}-E_1-E_2)$ &0.057&1.53&4.76&9.79\\
				\hline
				\rule{0pt}{10pt}&$\to\Lambda_c^{+}\eta_1$ &${1\over\sqrt{3}}\big[\lambda_d\big(C^{\prime}-C+E_1-E_2\big) - \lambda_sC\big]$ &0.031&0.84&1.90&4.07\\
				\hline
				\rule{0pt}{10pt}&$\to\Lambda_c^{+}\eta_8$ &${1\over\sqrt{6}}\big[\lambda_d\big(C^{\prime}-C+E_1-E_2\big) + 2\lambda_sC\big]$ &0.020&0.55&0.93&1.59\\
				\hline
				\rule{0pt}{10pt}&$\to\Sigma_c^{++}\pi^-$&$\lambda_d \tilde{E}_1$ &&0.38&1.18&2.46\\
				\hline
				\rule{0pt}{10pt}&$\to\Xi_c^{\prime +}K^0$&${1\over\sqrt{2}} \big(\lambda_s\tilde{C}^{\prime}+\lambda_d\tilde{E}_1\big)$ &&1.17&3.79&8.00\\
				\hline
				\rule{0pt}{10pt}&$\to\Xi_c^+K^0$&$\lambda_sC^{\prime}+\lambda_dE_1$&&2.77&8.75&18.1\\
				\hline
				\hline
				\rule{0pt}{10pt}$\Omega_{cc}^{+}$&$\to\Sigma_c^+\overline{K}^0$&${1\over\sqrt{2}}\big(\lambda_d \tilde{C}^{\prime} + \lambda_s \tilde{E}_1\big)$&&0.26&0.75&1.48\\
				\hline
				\rule{0pt}{10pt}&$\to\Lambda_c^+\overline{K}^0$&$\lambda_dC^{\prime}+\lambda_sE_1$&&0.52&1.66&3.49\\
				\hline
				\rule{0pt}{10pt}&$\to\Sigma_c^{++}K^-$&$\lambda_s \tilde{E}_1$ &&0.23&0.77&1.69\\
				\hline
				\rule{0pt}{10pt}&$\to\Xi_c^{\prime +}\pi^0$&${1\over2}\big(-\lambda_d \tilde{C} + \lambda_s \tilde{E}_2\big)$ &&0.38&1.48&3.61\\
				\hline
				\rule{0pt}{10pt}&$\to\Xi_c^{\prime +}\eta_1$ &${1\over\sqrt{6}}\big[\lambda_s\big(\tilde{C}+\tilde{C}^{\prime}+\tilde{E}_1+\tilde{E}_2\big) + \lambda_d\tilde{C}\big]$&&0.49&1.71&3.45\\
				\hline
				\rule{0pt}{10pt}&$\to\Xi_c^{\prime +}\eta_8$ &${1\over2\sqrt{3}}\big[\lambda_s\big(\tilde{E}_2-2\tilde{C}-2\tilde{C}^{\prime}-2\tilde{E}_1\big) +\lambda_d\tilde{C}\big]$&&2.04&6.26&13.0\\
				\hline
				\rule{0pt}{10pt}&$\to\Xi_c^+\pi^0$ &${1\over\sqrt{2}}(\lambda_dC-\lambda_sE_2)$ &&4.50&13.2&26.0\\
				\hline
				\rule{0pt}{10pt}&$\to\Xi_c^+\eta_1$ &${1\over\sqrt{3}}\big[\lambda_s\big(C^{\prime}-C+E_1-E_2\big) - \lambda_dC\big] $&&11.3&37.3&77.3\\
				\hline
				\rule{0pt}{10pt}&$\to\Xi_c^+\eta_8$&${1\over\sqrt{6}}\big[\lambda_s\big(2C-2C^{\prime}+2E_1-E_2\big) -\lambda_dC\big]$&&6.66&21.9&45.5\\
				\hline
				\hline
			\end{tabular}
		\end{sidewaystable}
	
		\begin{sidewaystable}
			\centering
			\caption{Same as Table.\ref{result2} but for the long-distance dominated doubly Cabibbo-suppressed modes.}
			\label{result4}
			\begin{tabular}{clc|cccc}
				\hline
				\hline
				\rule{0pt}{12pt}Particles&Decay modes&Topology&\ \ \ $\mathcal{BR}_{T_{SD}}(\times 10^{-6})$\ \ \ &$\mathcal{BR}_{\eta=1.0}(\times 10^{-6})$&
$\mathcal{BR}_{\eta=1.5}(\times 10^{-6})$&$\mathcal{BR}_{\eta=2.0}(\times 10^{-6})$\\
				\hline
				\hline
				\rule{0pt}{10pt}$\Xi_{cc}^{++}$&$\to\Sigma_c^{++}K^0$ &$\tilde{C}$ &0.043&1.31&4.69&10.75\\
				\hline
				\hline
				\rule{0pt}{10pt}$\Xi_{cc}^{+}$&$\to\Sigma_c^+K^0$&${1\over\sqrt{2}}\big(\tilde{C} + \tilde{C}^\prime\big)$ &0.035&4.52&16.0&36.4\\
				\hline
				\rule{0pt}{10pt}&$\to\Lambda_c^+K^0$&$-C+C^\prime$&0.05&2.39&8.88&21.0\\
				\hline
				\hline
				\rule{0pt}{10pt}$\Omega_{cc}^{+}$&$\to\Sigma_c^+\pi^0$&${1\over2}(-\tilde{E}_1+\tilde{E}_2)$&&0.07&0.23&0.52\\
				\hline
				\rule{0pt}{10pt}&$\to\Sigma_c^+\eta_1$&${1\over\sqrt{6}}\big(\tilde{C}^\prime+\tilde{E}_1+\tilde{E}_2\big)$ &&0.14&0.45&0.89\\
				\hline
				\rule{0pt}{10pt}&$\to\Sigma_c^+\eta_8$ &${1\over2\sqrt{3}}\big(-2\tilde{C}^\prime+\tilde{E}_1+\tilde{E}_2\big)$ &&0.09&0.30&0.66\\
				\hline
				\rule{0pt}{10pt}&$\to\Lambda_c^+\pi^0$&$-{1\over\sqrt{2}}(E_1+E_2)$&&0.19&0.61&1.34\\
				\hline
				\rule{0pt}{10pt}&$\to\Lambda_c^+\eta_1$&${1\over\sqrt{3}}(C^\prime+E_1-E_2)$&&0.59&1.89&4.13\\
				\hline
				\rule{0pt}{10pt}&$\to\Lambda_c^+\eta_8$&$-{1\over\sqrt{6}}(2C^\prime-E_1+E_2)$&&0.28&0.89&1.97\\
				\hline
				\rule{0pt}{10pt}&$\to\Sigma_c^0\pi^+$&$\tilde{E}_2$ &&0.14&0.45&0.91\\
				\hline
				\rule{0pt}{10pt}&$\to\Sigma_c^{++}\pi^-$&$\tilde{E}_1$ &&0.16&0.59&1.31\\
				\hline
				\rule{0pt}{10pt}&$\to\Xi_c^{\prime +}K^0$ &${1\over\sqrt{2}}\big(\tilde{C}+\tilde{E}_1\big)$ &0.028&9.68&31.9&68.0\\
				\hline
				\rule{0pt}{10pt}&$\to\Xi_c^+K^0$&$-C+E_1$&0.045&0.82&2.84&6.54\\
				\hline
				\hline
			\end{tabular}
		\end{sidewaystable}

We list all the branching ratios of the decay modes $\mathcal{B}_{cc}\to\mathcal{B}_cP$ in Tables \ref{result1}-\ref{result4}.
The short-distance dynamics dominant channels (with $T$ topology) are collected into Table \ref{result1}.
For the long-distance contribution dominated processes, the numerical results are classified into three groups according to the CKM matrix elements:
(a) the Cabibbo-favored (CF) decays induced by $c\to su\bar{d}$ (with the CKM element $V_{cs}^\ast V_{ud}$) are listed in Table \ref{result2};
(b) the  singly Cabibbo-suppressed (SCS) ones induced by $c\to du\bar{d}$ or $c\to su\bar{s}$ (with the CKM element $V_{cd}^\ast V_{ud}$ or $V_{cs}^\ast V_{us}$) are listed in Table \ref{result3} ;
and (c) the doubly Cabibdbo-suppressed (DCS) ones induced by $c\to du\bar{s}$ (with the CKM element $V_{cd}^\ast V_{us}$) are given in Table \ref{result4}.
The topological amplitudes for the channels with the sextet single charmed baryons are distinguished from the anti-triplet baryons by adding a $tilde$, e.g. $\tilde{T}$.

In Table \ref{result1}, it can be obviously read out that the factorizable short-distance contributions of diagram amplitude $T$ are dominant relative to the long-distance contributions of $C^\prime$ and $E_2$. When the parameter $\eta$ tends to $2.0$, the long-distance contributions of $C^\prime$ and $E_2$ also have a visible or comparable effect. On the other hand, from Table \ref{result2}-\ref{result4}, the long-distance dynamics dominate the decay modes in Table \ref{result2} to \ref{result4}, since the only calculable short-distance amplitude $C_{SD}$ is deeply suppressed by the effective Wilson coefficient $a_2(\mu)$ at the charm mass scale, i.e. $a_2(m_c)=-0.017$ is much smaller than $a_1(m_c)=1.07$. The latter is used at the weak decay vertex of the triangle diagram in our calculations.

\subsection{Discussions on the topological diagrams}
		
The topological diagrams have the relations of ${|C|\over |T|}\sim{|C^\prime|\over |C|}\sim{|E_1|\over |C|}\sim{|E_2|\over |C|}\sim O({\Lambda^h_{\text{QCD}}\over m_Q})$ in the heavy baryon decays, manifested by the soft-colliner effective theory \cite{Leibovich:2003tw,Mantry:2003uz}. These relations are important in the phenomenological studies on the searches for the double-heavy-flavor baryons, and give us more hints on the dynamics of heavy baryon decays. From the above relations, all the tree-level topological diagrams are at the same order in charmed baryon decays due to $\Lambda^h_{\text{QCD}}/ m_c\sim1$. It would be very useful to numerically test these relations in our framework. 

From the amplitudes of $\Xi_{cc}^{++}\to \Sigma_c^{++}\overline{K}^{0}$ \big($\tilde{C}$\big) and $\Xi_{cc}^{++}\to \Xi_c^{\prime +}\pi^+$ \big(${1\over\sqrt{2}}\big(\tilde{T}+\tilde{C}^\prime\big)$\big), 
we obtain the ratios $\tilde{C}_{LD}/\tilde{T}_{SD}$ and $\tilde{C}^\prime_{LD}/\tilde{C}_{LD}$ as,
\begin{align}
	\frac{|\tilde{C}_{LD}|}{|\tilde{T}_{SD}|} & = \frac{|\mathcal{A}(\Xi_{cc}^{++}\to \Sigma_c^{++}\overline{K}^{0})|_{LD}}{\sqrt{2}|\mathcal{A}(\Xi_{cc}^{++}\to \Xi_c^{\prime +}\pi^+)|_{SD}}= 0.21\sim 0.55,\label{eq33}\\
    \frac{|\tilde{C}^\prime_{LD}|}{|\tilde{C}_{LD}|} & =\frac{\sqrt{2}|\mathcal{A}(\Xi_{cc}^{++}\to \Xi_c^{\prime +}\pi^+)|_{LD}}{|\mathcal{A}(\Xi_{cc}^{++}\to \Sigma_c^{++}\overline{K}^{0})|_{LD}}= 1.33\sim 1.45.
\end{align}

The ratio $\tilde{C}^\prime_{LD}/\tilde{C}_{LD}$ can also be calculated by the decay channels $\Xi_{cc}^{++} \to \Sigma_c^{++}\pi^0$\big($-{1\over\sqrt{2}}\tilde{C}$\big), $\Xi_{cc}^{++} 
\to \Sigma_c^+\pi^+$\big(${1\over\sqrt{2}}\big(\tilde{T}+\tilde{C}^\prime\big)$\big), $\Xi_{cc}^{++} \to \Xi_c^{\prime +}K^+$\big(${1\over\sqrt{2}}\big(\tilde{T}+\tilde{C}^\prime\big)$\big) 
and $\Xi_{cc}^{++}\to\Sigma_c^{++}K^0$\big($\tilde{C}$\big), $\Xi_{cc}^{++}\to\Sigma_c^+K^+$\big(${1\over\sqrt{2}}\big(\tilde{T}+\tilde{C}^\prime\big)$\big) as
\begin{align}
	\frac{|\tilde{C}^\prime_{LD}|}{|\tilde{C}_{LD}|} & =\frac{|\mathcal{A}(\Xi_{cc}^{++}\to\Sigma_c^+\pi^+)|_{LD}}{|\mathcal{A}(\Xi_{cc}^{++}\to\Sigma_c^{++}\pi^0)|_{LD}}= 0.72\sim 0.88,\\
	\frac{|\tilde{C}^\prime_{LD}|}{|\tilde{C}_{LD}|} & =\frac{|\mathcal{A}(\Xi_{cc}^{++}\to\Xi_c^{\prime+}K^+)|_{LD}}{|\mathcal{A}(\Xi_{cc}^{++}\to\Sigma_c^{++}\pi^0)|_{LD}}= 0.69\sim 0.85,\\
	\frac{|\tilde{C}^\prime_{LD}|}{|\tilde{C}_{LD}|} & =\frac{\sqrt{2}|\mathcal{A}(\Xi_{cc}^{++}\to\Sigma_c^+K^+)|_{LD}}{|\mathcal{A}(\Xi_{cc}^{++}\to\Sigma_c^{++}K^0)|_{LD}}= 0.98\sim 1.24.
\end{align}

In the following to consider the relations between $C$, $E_1$ and $E_2$, it would be more convenient to study the processes with a pure topological diagram. The advantage of avoiding the interference between diagrams is that they could be directly determined by the experimental data in the future. 

In Table \ref{result2}, by the three single amplitude channels $\Xi_{cc}^{+}\to\Sigma_c^{++}K^-$ \big($\tilde{E}_1$\big), $\Xi_{cc}^{+}
\to\Omega_c^0K^+$ \big($\tilde{E}_2$\big) and $\Xi_{cc}^{++}\to\Sigma_c^{++}\overline{K}^{0}$ \big($\tilde{C}$\big), the ratios among 
the amplitudes $\tilde{E}_{1LD}$, $\tilde{E}_{2LD}$ and $\tilde{C}_{LD}$ can be easily carried out, with the next results
		\begin{align}
			\frac{|\tilde{E}_{1LD}|}{|\tilde{C}_{LD}|} &= \frac{|\mathcal{A}(\Xi_{cc}^{+}\to\Sigma_c^{++}K^-)|_{LD}}{|\mathcal{A}(\Xi_{cc}^{++}\to\Sigma_c^{++}\overline{K}^{0})|_{LD}}=0.45\sim 0.46,\\
			\frac{|\tilde{E}_{2LD}|}{|\tilde{C}_{LD}|} &= \frac{|\mathcal{A}(\Xi_{cc}^{+}\to\Omega_c^0K^+)|_{LD}}{|\mathcal{A}(\Xi_{cc}^{++}\to\Sigma_c^{++}\overline{K}^{0})|_{LD}}=0.38\sim 0.40, \\
            \frac{|\tilde{E}_{1LD}|}{|\tilde{E}_{2LD}|} &=\frac{|\mathcal{A}(\Xi_{cc}^{+}\to\Sigma_c^{++}K^-)|_{LD}}{|\mathcal{A}(\Xi_{cc}^{+}\to\Omega_c^0K^+)|_{LD}}=1.12\sim 1.24.
		\end{align}
From above, the ratio $\tilde{E}_{1LD}/\tilde{C}_{SD}$ and $\tilde{E}_{2LD}/\tilde{C}_{SD}$ have the similar values at the order one.
The ratios between $\tilde{E}_{1LD}$ and $\tilde{C}_{SD}$ can also be calculated from Table.\ref{result3},
i.e. from the modes $\Xi_{cc}^{++}\to\Sigma_c^{++}\pi^0$\big($-\frac{1}{\sqrt{2}}\tilde{C}$\big), $\Xi_{cc}^{+}\to\Sigma_c^{++}\pi^-$\big($\tilde{E}_1$\big) and $\Omega_{cc}^{+}\to\Sigma_c^{++}K^-$\big($\tilde{E}_1$\big),
		\begin{align}
			\frac{|\tilde{E}_{1LD}|}{|\tilde{C}_{LD}|}&=\frac{|\mathcal{A}(\Xi_{cc}^{+}\to\Sigma_c^{++}\pi^-)|_{LD}}{\sqrt{2}|\mathcal{A}(\Xi_{cc}^{++}\to\Sigma_c^{++}\pi^0)|_{LD}}=0.28\sim 0.26,\\
			\frac{|\tilde{E}_{1LD}|}{|\tilde{C}_{LD}|}&=\frac{|\mathcal{A}(\Omega_{cc}^{+}\to\Sigma_c^{++}K^-)|_{LD}}{\sqrt{2}|\mathcal{A}(\Xi_{cc}^{++}\to\Sigma_c^{++}\pi^0)|_{LD}}=0.29\sim 0.30.
		\end{align}
We can testify the results using the data from Table.\ref{result4}.
By the three channels $\Xi_{cc}^{++}\to\Sigma_c^{++}K^0$\big($\tilde{C}$\big), $\Omega_{cc}^{+}\to\Sigma_c^0\pi^+$\big($\tilde{E}_2$\big) 
and $\Omega_{cc}^{+}\to\Sigma_c^{++}\pi^-$\big($\tilde{E}_1$\big), we get the ratios as
		\begin{align}
			\frac{|\tilde{E}_{2LD}|}{|\tilde{C}_{LD}|}&=\frac{|\mathcal{A}(\Omega_{cc}^{+}\to\Sigma_c^0\pi^+)|_{LD}}{|\mathcal{A}(\Xi_{cc}^{++}\to\Sigma_c^{++}K^0)|_{LD}}=0.34\sim 0.37,\\
			\frac{|\tilde{E}_{1LD}|}{|\tilde{C}_{LD}|}&=\frac{|\mathcal{A}(\Omega_{cc}^{+}\to\Sigma_c^{++}\pi^-)|_{LD}}{|\mathcal{A}(\Xi_{cc}^{++}\to\Sigma_c^{++}K^0)|_{LD}}=0.40\sim 0.41.
			\label{eq39}
		\end{align}
From Eqs.~(\ref{eq33}-\ref{eq39}), considering the relatively large parameter uncertainty, all these results are consistent with the relations found in \cite{Leibovich:2003tw,Mantry:2003uz},
\begin{align}
\frac{|C|}{|T|}\sim\frac{|C^\prime|}{|C|}\sim\frac{|E_1|}{|C|}\sim\frac{|E_2|}{|C|}\sim \mathcal{O}\left({\textcolor{diff}{\Lambda^h_{QCD}\over m_c}}\right)\sim \mathcal{O}(1).
\end{align}
Some results between Eqs.~(\ref{eq33}-\ref{eq39}) are different with each other. This can be understood by the flavor $SU(3)$ breaking effects, shown in the following subsection. 

\begin{figure}[tbp]
	\centering
	\includegraphics[width=0.6\textwidth]{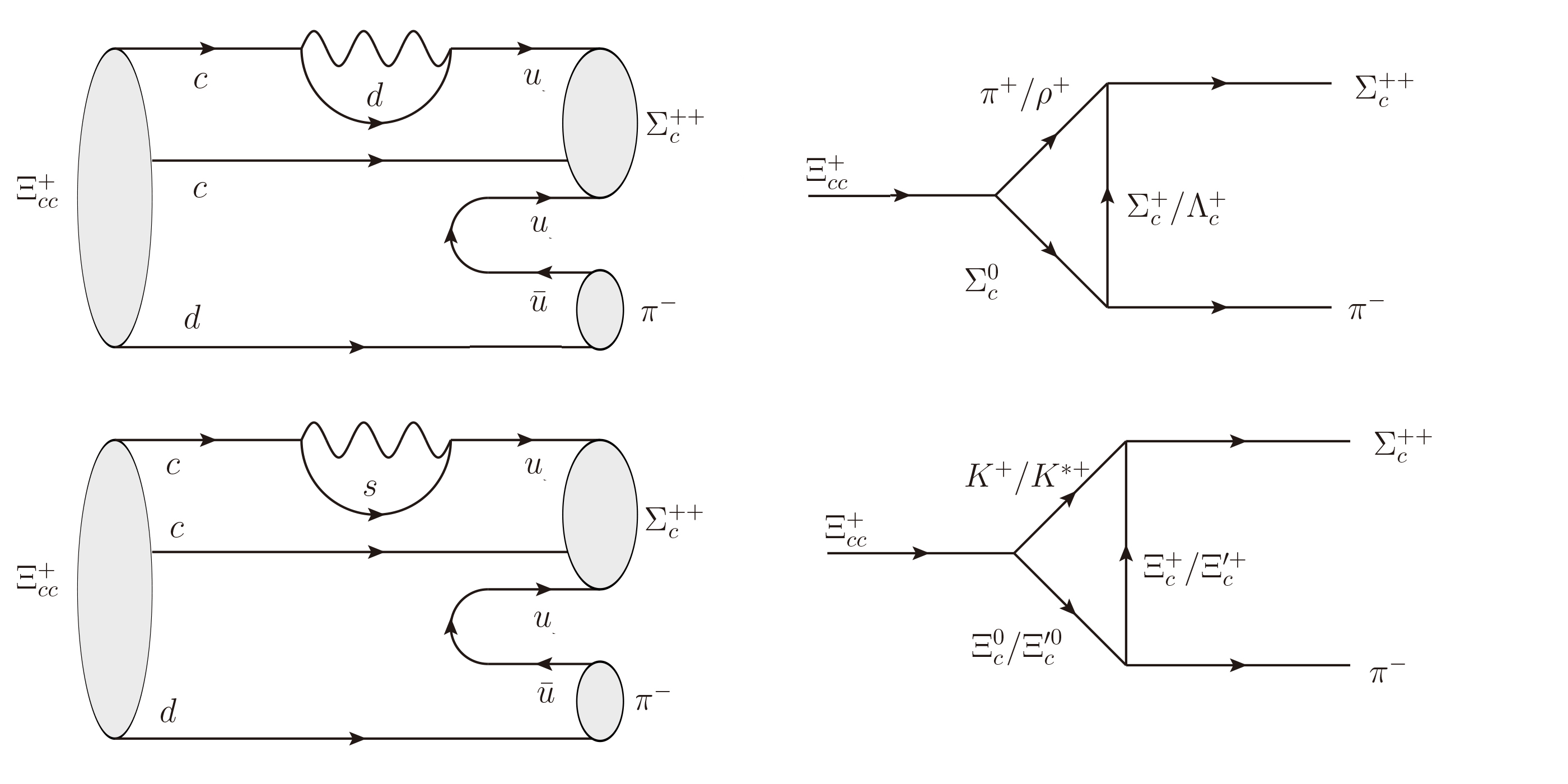}
	\caption{Quark-loop topological diagrams and their corresponding hadronic triangle diagrams of $\Xi_{cc}^+\to\Sigma_c^{++}\pi^-$.}\label{fig:fig5}
\end{figure}

{As mentioned in Sec.\ref{sec:topo}, the long-distance quark-loop diagrams are also taken into account in our calculations. Taking $\Xi_{cc}^+\to\Sigma_c^{++}\pi^-$ as an example, the quark-loop topological diagrams and the corresponding hadronic triangle diagrams are shown in Fig.\ref{fig:fig5}. It can be expected that any individual loop diagram is as large as the tree diagrams. However, the $d$-quark and $s$-quark loop diagrams are mostly cancelled due to the GIM mechanism. Therefore, the total contribution of quark loop diagrams comes from the flavor $SU(3)$ symmetry breaking effects.  Numerically, the magnitudes of $d$-quark loop, $s$-quark loop and the sum of the both loop diagrams  are $|\mathcal{A}_{d}|=2.6\times 10^{-7}$ GeV, $|\mathcal{A}_{s}|=1.0\times 10^{-7}$ GeV, $|\mathcal{A}_{d+s}|=1.5\times 10^{-7}$ GeV, respectively. It can be seen that the long-distance dynamics contributes the relatively large $SU(3)$ breaking effect. It is difficult to test this effect in the doubly charmed baryons in experiments. We will investigate the long-distance quark-loop contributions in the $D$ meson and $\Lambda_c$ decays in a further study. }

\subsection{Discussions on the flavor $SU(3)$ symmetry and its breaking}

The flavor $SU(3)$ symmetry is of great significance in the weak decays of heavy hadrons. In terms of a few SU(3) irreducible amplitudes, 
a number of relations between the widths of doubly charmed baryon decays are obtained \cite{Wang:2017azm}. It is important to numerically test the flavor $SU(3)$ symmetry and its breaking effects. 

In the following, we show our numerical results on the ratios of decay widths. In the $SU(3)$ limit, they should be unity. Any deviation would indicate the $SU(3)$ breaking effects.
\begin{align}
	\Gamma\big( \Xi_{cc}^{++}\to\Lambda_c^+\pi^+ \big)/\Gamma\big( \Xi_{cc}^{++}\to\Xi_c^+K^+ \big)\quad&=|\lambda_d(T+C^\prime)|^2/|\lambda_s(T+C^\prime)|^2\nonumber\\
	&=0.73\sim 0.74,\label{eq:SU3-1}\\
	\Gamma\big( \Xi_{cc}^{+}\to\Xi_c^+K^0\big) / \Gamma\big( \Omega_{cc}^{+}\to\Lambda_c^+\overline{K}^0 \big)\ \ \quad&=|\lambda_s C^\prime+\lambda_d E_1|^2/|\lambda_d C^\prime+\lambda_s E_1|^2\nonumber\\
	&=6.67\sim 6.85,\label{eq:SU3-2}\\
	 \Gamma\big( \Omega_{cc}^{+}\to\Xi_c^0\pi^+ \big)/ \Gamma\big( \Xi_{cc}^{+}\to\Xi_c^0K^+ \big)\ \ \quad&=|-\lambda_d T-\lambda_s E_2 |^2/| \lambda_s T+\lambda_d E_2|^2\nonumber\\
	 &= 0.54\sim 0.56,\label{eq:SU3-3}\\
	 \Gamma\big( \Xi_{cc}^{++}\to\Sigma_c^{++}\pi^0 \big)/\frac{1}{3}\Gamma\big( \Xi_{cc}^{++}\to\Sigma_c^{++}\eta_8 \big) &=|-\frac{1}{\sqrt{2}}\tilde{C} |^2/|-\frac{1}{\sqrt{6}}(\lambda_d-2\lambda_s)\tilde{C} |^2\nonumber\\
	 &= 3.14\sim 4.98,\label{eq:SU3-4}\\
	 \Gamma\big( \Xi_{cc}^{++}\to\Sigma_c^{+}\pi^+ \big)/ \Gamma\big( \Xi_{cc}^{++}\to\Xi_c^{\prime+}K^+ \big)\ \ &=| \frac{1}{\sqrt{2}}\lambda_d(\tilde{T}+\tilde{C}^\prime)|^2/| \frac{1}{\sqrt{2}}\lambda_s(\tilde{T}+\tilde{C}^\prime)|^2\nonumber\\
	 &= 1.28\sim 1.30,\label{eq:SU3-5}\\
	 \Gamma\big( \Xi_{cc}^{+}\to\Sigma_c^{++}\pi^- \big)/\Gamma\big( \Omega_{cc}^{+}\to\Sigma_c^{++}K^- \big)\ \ &=|\lambda_d \tilde{E}_1|^2/|\lambda_s \tilde{E}_1 |^2\nonumber\\
	 &= 1.87\sim 2.12,\label{eq:SU3-6}\\
	\Gamma\big( \Xi_{cc}^{+}\to\Sigma_c^{0}\pi^+ \big) / \Gamma\big( \Omega_{cc}^{+}\to\Omega_c^{0}K^+ \big)\ \quad&=| \lambda_d(\tilde{T}+\tilde{E}_2)|^2/|\lambda_s(\tilde{T}+\tilde{E}_2) |^2\nonumber\\
	&= 1.03\sim 1.13,\label{eq:SU3-7}\\
	 	\Gamma\big( \Xi_{cc}^{+}\to\Xi_c^{\prime+}K^0 \big)/\Gamma\big( \Omega_{cc}^{+}\to\Sigma_c^{+}\overline{K}^0 \big)\ \quad&=| \frac{1}{\sqrt{2}}(\lambda_s\tilde{C}^\prime+\lambda_d\tilde{E}_1)|^2/|\frac{1}{\sqrt{2}}(\lambda_d\tilde{C}^\prime+\lambda_s\tilde{E}_1) |^2\nonumber\\
	 	&= 5.79\sim 6.95,\label{eq:SU3-8}\\
	 \Gamma\big( \Omega_{cc}^{+}\to\Xi_c^{\prime 0}\pi^+ \big)/ \Gamma\big( \Xi_{cc}^{+}\to\Xi_c^{\prime 0}K^+ \big)\ \quad&=| \frac{1}{\sqrt{2}}(\lambda_d\tilde{T}^\prime+\lambda_s\tilde{E}_2|^2/|\frac{1}{\sqrt{2}}(\lambda_s\tilde{T}^\prime+\lambda_d\tilde{E}_2) |^2\nonumber\\
	 &= 0.69\sim 0.74.\label{eq:SU3-9}
\end{align}

From the above numerical results, it can be found that the long-distance final-state interactions can contribute to large $SU(3)$ breaking effect. It stems from the exchanged particles, the hadronic strong coupling constants, the transition form factors, the decay constants, and the interference between different diagrams. 
{In our calculation of Eq.(40) and Eq.(46), the large values mainly stem from the strong coupling constants. Taking $\Gamma\big( \Xi_{cc}^{+}\to\Xi_c^+K^0\big) / \Gamma\big( \Omega_{cc}^{+}\to\Lambda_c^+\overline{K}^0 \big)=6.67\sim 6.85$ as an example, a factor of 2.5 in the amplitude could account for the large ratio. The decay mode $\Xi_{cc}^{+}\to\Xi_c^+K^0$ is dominated by the triangle diagram $\mathcal{M}(K^+,\Xi_c^0;\rho^+)$, while $\Omega_{cc}^{+}\to\Lambda_c^+\overline{K}^0$ dominated by $\mathcal{M}(\pi^+,\Xi_c^0;K^{\ast +})$. To evaluate these two triangle diagrams, we need strong coupling constants $f_1(\Xi_c^+\to\Xi_c^0\rho^+)=8.5$, $f_2(\Xi_c^+\to\Xi_c^0\rho^+)=10.6$\cite{Aliev:2010nh} and $f_1(\Xi_c^0\to\Lambda_c^+K^{\ast -})=-4.6$, $f_2(\Xi_c^0\to\Lambda_c^+K^{\ast -})=-6$\cite{Aliev:2010nh}, respectively. Therefore, $\Gamma\big( \Xi_{cc}^{+}\to\Xi_c^+K^0\big)$ is much larger than $\Gamma\big( \Omega_{cc}^{+}\to\Lambda_c^+\overline{K}^0 \big)$. As for $\Gamma\big( \Xi_{cc}^{+}\to\Xi_c^{\prime+}K^0 \big)/\Gamma\big( \Omega_{cc}^{+}\to\Sigma_c^{+}\overline{K}^0 \big)=5.79\sim 6.95$ with the dominant triangle diagrams $\mathcal{M}(K^+,\Xi_c^0;\rho^+)$ and $\mathcal{M}(\pi^+,\Xi_c^0;K^{\ast +})$, the relevant strong coupling constants are $f_1(\Xi_c^{\prime +}\to\Xi_c^0\rho^+)=2.1$, $f_2(\Xi_c^{\prime +}\to\Xi_c^0\rho^+)=116$\cite{Aliev:2010nh} and $f_1(\Sigma_c^{+}\to\Xi_c^0K^{\ast +})=-2.2$, $f_2(\Sigma_c^{+}\to\Xi_c^0K^{\ast +})=-13$\cite{Aliev:2010nh}, respectively, which lead to the large value as Eq.(46). All the values of strong couplings are taken from the literature. These non-perturbative quantities actually have large theoretical uncertainties, and need to be carefully studied in the future. }

\section{Summary}\label{sec:summary}

In this work, we introduced the whole theoretical framework of the rescattering mechanism by investigating the forty-nine  two-body baryon decays $\mathcal{B}_{cc}\to\mathcal{B}_{c}P$ , 
where $\mathcal{B}_{cc}=(\Xi_{cc}^{++} , \Xi_{cc}^+,\Omega_{cc}^+)$ are the doubly charmed baryons, $\mathcal{B}_{c}=(\mathcal{B}_{\bar{3}}, \mathcal{B}_{6})$ are the singly charmed baryons  
and $P=(\pi,K,\eta_{1,8})$ are the light pseudoscalar mesons.
It was interpreted in detail for the physical foundation of the rescattering mechanism at the hadron level.
On the other hand, as a self-consistent test to the rescattering mechanism, the relations of topological diagrams and flavor $SU(3)$ symmetry have been discussed. The main points are the following:
\begin{enumerate}
\item[(1)]
As the standard programme, we provided the theoretical predictions for the branching ratios of all considered $\mathcal{B}_{cc}\to\mathcal{B}_{c}P$  and some discussions on the dependence of the parameter $\eta$.

\item[(2)]
The numerical results of the branching ratios exhibited the same conclusion with the charm meson decays: the non-factorizable long-distance contributions 
play an important role in doubly charmed baryon decays.

\item[(3)]
We obtained the same counting rules as the analysis in SCET for the topological amplitudes in charm decays, 
that is ${|C|\over |T|}\sim{|C^\prime|\over |C|}\sim{|E_1|\over |C|}\sim{|E_2|\over |C|}\sim O({{\Lambda^h_{\text{QCD}}\over m_c}})\sim O(1)$, 
which will be significant guidance for the following studies to charmed baryon decays.

\item[(4)]
Large $SU(3)$ symmetry breaking effects are obtained in our method. It requires more studies on the $SU(3)$ breaking effects of the doubly charmed baryon decays in the future. 
\end{enumerate}

\section{Acknowledgement}
We are grateful to all the collaborators in the series of studies on the doubly heavy baryons. Especially, we are grateful to Cai-Dian L\"u, Run-Hui Li, Wei Wang, Zhen-Xing Zhao and Zhi-Tian Zou for the theoretical collaborations, and to Yuan-Ning Gao, Ji-Bo He and Yan-Xi Zhang for the discussions on the experimental searches. 
This work was supported by the National Natural Science Foundation of China under the Grant No.~11775117,~U1732101 and 11975112, and National Key Research and Development Program of China under Contracts No. 2020YFA0406400.


\appendix
		
\section{ Effective Lagrangians}\label{app:lag}

The effective Lagrangians used in the rescattering mechanism are those as given in Refs.~\cite{Aliev:2010yx,Yan:1992gz,Casalbuoni:1996pg,Meissner:1987ge,Li:2012bt,Aliev:2010nh}:
\begin{align}
\label{eq:LVPP}\mathcal{L}_{VPP}&=\frac{ig_{VPP}}{\sqrt{2}}Tr[V^\mu[P,\partial_\mu P] ], \\
\mathcal{L}_{VVV}&=\frac{ig_{VVV}}{\sqrt{2}}Tr[(\partial_\nu V_\mu V^\mu-V^\mu\partial_\nu V_\mu)V^\nu],\\
		\mathcal{L}_{PB_6B_6}&=g_{PB_6B_6}Tr[\bar{B}_6i\gamma_5PB_6],\\
		\mathcal{L}_{PB_{\bar{3}}B_{\bar{3}}}&=g_{PB_{\bar{3}}B_{\bar{3}}}Tr[\bar{B}_{\bar{3}}i\gamma_5PB_{\bar{3}}],\\
		\mathcal{L}_{PB_6B_{\bar{3}}}&=g_{PB_6B_{\bar{3}}}Tr[\bar{B}_6i\gamma_5PB_{\bar{3}}]+h.c.,\\
		\mathcal{L}_{VB_6B_6}&=f_{1VB_6B_6}Tr[\bar{B}_6\gamma_\mu V^\mu B_6]+\frac{f_{2PB_6B_6}}{2m_6}Tr[\bar{B}_6\sigma_{\mu\nu}\partial^\mu V^\nu B_6],
\end{align}
\begin{align}
		\mathcal{L}_{VB_{\bar{3}}B_{\bar{3}}}&=f_{1PB_{\bar{3}}B_{\bar{3}}}Tr[\bar{B}_{\bar{3}}\gamma_\mu V^\mu B_{\bar{3}}]
+\frac{f_{2PB_{\bar{3}}B_{\bar{3}}}}{2m_3}Tr[\bar{B}_{\bar{3}}\sigma_{\mu\nu}\partial^\mu V^\nu B_{\bar{3}}],\\		
\mathcal{L}_{VB_6B_{\bar{3}}}&= \{f_{1VB_6B_{\bar{3}}}Tr[\bar{B}_6\gamma_\mu V^\mu B_{\bar{3}}]
+\frac{f_{2VB_6B_{\bar{3}}}}{m_6+m_3}Tr[\bar{B}_6\sigma_{\mu\nu}\partial^\mu V^\nu B_{\bar{3}}]\}+h.c.,
\end{align}
\begin{align}
P(J^P=0^-)=\left(
\begin{array}{ccc}
	\frac{\pi^0}{\sqrt{2}}+\frac{\eta_8}{\sqrt{6}}&\pi^+&K^+\\
	\pi^-&-\frac{\pi^0}{\sqrt{2}}+\frac{\eta_8}{\sqrt{6}}&K^0\\
	K^-&\overline{K}^0&-\sqrt{\frac{2}{3}}\eta_8\\
\end{array}\right)+\frac{1}{\sqrt{3}}\left(
\begin{array}{ccc}
	\eta_1&0&0\\
	0&\eta_1&0\\
	0&0&\eta_1\\
\end{array}\right),
\end{align}
\begin{align}
		V(J^P=1^-) =\left(		\begin{array}{ccc}
		\frac{\rho^0}{\sqrt{2}}+\frac{\omega}{\sqrt{2}}&\rho^+&K^{\ast +}\\
		\rho^-&-\frac{\rho^0}{\sqrt{2}}+\frac{\omega}{\sqrt{2}}&K^{\ast 0}\\ 		K^{\ast -}&\overline{K}^{\ast 0}&\phi\\
		\end{array}\right) ,
\end{align}
\begin{align}
B_6(J^P=\frac{1}{2}^+)=\left(		\begin{array}{ccc}
		\Sigma_c^{++}&\frac{\Sigma_c^+}{\sqrt{2}}&\frac{\Xi_c^{\prime +}}{\sqrt{2}}\\
		\frac{\Sigma_c^+}{\sqrt{2}}&\Sigma_c^0&\frac{\Xi_c^{\prime 0}}{\sqrt{2}}\\
		\frac{\Xi_c^{\prime +}}{\sqrt{2}}&\frac{\Xi_c^{\prime 0}}{\sqrt{2}}&\Omega_c^0\\
		\end{array}\right), \quad
B_{\bar{3}}(J^P=\frac{1}{2}^+)=\left( 		\begin{array}{ccc}
		0&\Lambda_c^+&\Xi_c^+\\ 		-\Lambda_c^+&0&\Xi_c^0\\ 		-\Xi_c^+&-\Xi_c^0&0\\
		\end{array}\right).
\end{align}

Strong coupling constants are collected in Table.\ref{ap:VPP}, \ref{ap:PBB} and \ref{ap:VBB}.

\begin{table}[h]
	\centering
	\caption{Strong coupling constants of VPP and VVV vertices.}
	\label{ap:VPP}
	\begin{tabular}{|c|c|c|c|c|c|c|c|c|c|}
		\hline
		\hline
		\rule{0pt}{12pt}Vertex&g&Vertex&g&Vertex&g&Vertex&g&Vertex&g\\
		\hline
		\hline
		\rule{0pt}{12pt}$\rho^+\to\pi^0\pi^+$&6.05&$\rho^0\to\pi^+\pi^-$&6.05&$\rho^+\to K^+\overline{K}^{0}$&4.60&$\rho^0\to K^0\overline{K}^{0}$&-3.25&$\rho^0\to K^+K^-$&3.25\\
		\hline
		\rule{0pt}{12pt}$\phi\to K^-K^+$&4.60&$\overline{K}^{\ast 0}\to\eta_8\overline{K}^{0}$&5.63&$\overline{K}^{\ast 0}\to K^-\pi^+$&4.60&$\overline{K}^{\ast 0}\to \overline{K}^{0}\pi^0$&-3.25&$K^{\ast +}\to K^+\pi^0$&3.25\\
		\hline
		\rule{0pt}{12pt}$K^{\ast +}\to\eta_8 K^+$&5.63&$K^{\ast +}\to\pi^+K^0$&4.60&$K^{\ast 0}\to\pi^-K^+$&4.60&$K^{\ast 0}\to K^0\eta_8$&5.63&$K^{\ast 0}\to\pi^0K^0$&-3.25\\
		\hline
		\rule{0pt}{12pt}$\omega\to K^+K^-$&3.25&$\phi\to\overline{K}^{0}K^0$&4.60&$\omega\to K^0\overline{K}^{0}$&3.25&&&&\\
		\hline
		\hline
		\rule{0pt}{12pt}$\rho^+\to\rho^0\rho^+$&7.38&$\rho^0\to\rho^-\rho^+$&7.38&$\rho^+\to K^{\ast +}\overline{K}^{\ast 0}$&5.22&$\rho^0\to K^{\ast +}K^{\ast -}$&3.69&$\omega\to K^{\ast +}K^{\ast -}$&3.69\\
		\hline
		\rule{0pt}{12pt}$\overline{K}^{\ast 0}\to\phi\overline{K}^{\ast 0}$&5.22&$\overline{K}^{\ast 0}\to\overline{K}^{\ast 0}\rho^0$&-3.69&$\overline{K}^{\ast 0}\to\overline{K}^{\ast 0}\omega$&3.69&$K^{\ast +}\to\rho^+K^{\ast 0}$&5.22&$K^{\ast +}\to\phi K^{\ast +}$&5.22\\
		\hline
		\rule{0pt}{12pt}$K^{\ast +}\to\omega K^{\ast +}$&3.69&$K^{\ast 0}\to \rho^0 K^{\ast 0}$&-3.69&$K^{\ast 0}\to \omega K^{\ast 0}$&3.69&$K^{\ast 0}\to K^{\ast 0}\phi$&5.22&$\phi\to K^{\ast -}K^{\ast +}$&5.22\\
		\hline
		\rule{0pt}{12pt}$\omega\to K^{\ast 0}\overline{K}^{\ast 0}$&3.69&$\phi\to\overline{K}^{\ast 0}K^{\ast 0}$&5.22&$\rho^0\to K^{\ast 0}\overline{K}^{\ast 0}$&-3.69&$K^{\ast +}\to K^{\ast +}\rho^0$&3.69&$\overline{K}^{\ast 0}\to K^{\ast -} \rho^+$&5.22\\
		\hline
		\hline
		\end{tabular}
		\end{table}
		
		\begin{table}[h]
			\centering
			\caption{Strong coupling constants of $P\mathcal{B}_{\bar{3}}\mathcal{B}_{\bar{3}}$, $P\mathcal{B}_{\bar{3}}\mathcal{B}_6$ and $P\mathcal{B}_6\mathcal{B}_6$ vertex.}
			\label{ap:PBB}
			\begin{tabular}{|c|c|c|c|c|c|c|c|c|c|}
				\hline
				\hline
				\rule{0pt}{12pt}Vertex&g&Vertex&g&Vertex&g&Vertex&g&Vertex&g\\
				\hline
				\hline
				\rule{0pt}{12pt}$\Xi_c^+\to\Lambda_c^+\overline{K}^{0}$&0.9&$\Lambda_c^+\to\Xi_c^+K^0$&0.9&$\Xi_c^+\to\Xi_c^+\eta_8$&-0.7&$\Lambda_c^+\to\Lambda_c^+\eta_8$&0.81&$\Xi_c^0\to\Lambda_c^+K^-$&-0.9\\
				\hline
				\rule{0pt}{12pt}$\Xi_c^0\to\Xi_c^0\eta_8$&-0.7&$\Xi_c^0\to\Xi_c^+\pi^-$&0.99&$\Xi_c^+\to\Xi_c^0\pi^+$&0.99&$\Xi_c^0\to\Xi_c^0\pi^0$&-0.7&$\Xi_c^+\to\Xi_c^+\pi^0$&0.7\\
				\hline
				\rule{0pt}{12pt}$\Sigma_c^+\to\Xi_c^+ K^0$&-5.0&$\Xi_c^+\to\Sigma_c^{++} K^-$&-7.1&$\Sigma_c^{++}\to\Xi_c^+ K^+$&-7.1&$\Xi_c^+\to\Xi_c^{\prime +}\eta_8$&5.4&$\Xi_c^{\prime +}\to\Xi_c^+\eta_8$&5.4\\
				\hline
				\rule{0pt}{12pt}$\Sigma_c^+\to\Lambda_c^+\pi^0$&6.5&$\Lambda_c^+\to\Sigma_c^{++}\pi^-$&-6.5&$\Sigma_c^{++}\to\Lambda_c^+\pi^+$&-6.5&$\Lambda_c^+\to\Sigma_c^0\pi^+$&6.5&$\Sigma_c^0\to\Lambda_c^+\pi^-$&6.5\\
				\hline
				\rule{0pt}{12pt}$\Xi_c^{\prime +}\to\Lambda_c^+\overline{K}^{0}$&-4.6&$\Lambda_c^+\to\Xi_c^{\prime 0}K^+$&4.6&$\Xi_c^{\prime 0}\to\Lambda_c^+K^-$&4.4&$\Xi_c^0\to\Sigma_c^+K^-$&-5.0&$\Sigma_c^+\to\Xi_c^0K^+$&-5.0\\
				\hline
				\rule{0pt}{12pt}$\Sigma_c^0\to\Xi_c^0K^0$&-7.1&$\Xi_c^0\to\Xi_c^{\prime 0}\eta_8$&5.4&$\Xi_c^{\prime 0}\to\Xi_c^0\eta_8$&5.4&$\Xi_c^0\to\Omega_c^0K^0$&6.5&$\Omega_c^0\to\Xi_c^0\overline{K}^{0}$&6.5\\
				\hline
				\rule{0pt}{12pt}$\Xi_c^{\prime +}\to\Xi_c^0\pi^+$&4.4&$\Xi_c^0\to\Xi_c^{\prime 0}\pi^0$&3.1&$\Xi_c^{\prime 0}\to\Xi_c^0\pi^0$&3.1&$\Xi_c^+\to\Omega_c^0K^+$&6.5&$\Omega_c^0\to\Xi_c^+K^-$&6.5\\
				\hline
				\rule{0pt}{12pt}$\Xi_c^{\prime +}\to\Xi_c^+\pi^0$&3.1&$\Xi_c^+\to\Xi_c^{\prime 0}\pi^+$&4.4&$\Xi_c^{\prime 0}\to\Xi_c^+\pi^-$&4.4&$\Xi_c^{\prime +}\to\Sigma_c^+\overline{K}^{0}$&6.4&$\Sigma_c^+\to\Xi_c^{\prime +}K^0$&6.4\\
				\hline
				\rule{0pt}{12pt}$\Sigma_c^{++}\to\Xi_c^{\prime +}K^+$&9.0&$\Xi_c^{\prime +}\to\Xi_c^{\prime +}\eta_8$&-2.3&$\Sigma_c^+\to\Sigma_c^+\eta_8$&4.6&$\Sigma_c^+\to\Sigma_c^{++}\pi^-$&8.0&$\Sigma_c^{++}\to\Sigma_c^+\pi^+$&8.0\\
				\hline
				\rule{0pt}{12pt}$\Sigma_c^{++}\to\Sigma_c^{++}\pi^0$&8.0&$\Xi_c^{\prime 0}\to\Sigma_c^+K^-$&6.4&$\Sigma_c^+\to\Xi_c^{\prime 0}K^+$&6.4&$\Xi_c^{\prime 0}\to\Sigma_c^0\overline{K}^{0}$&9.0&$\Sigma_c^0\to\Xi_c^{\prime 0}K^0$&9.0\\
				\hline
				\rule{0pt}{12pt}$\Omega_c^0\to\Omega_c^0\eta_8$&-10.4&$\Omega_c^0\to\Xi_c^{\prime +}K^-$&9.0&$\Xi_c^{\prime +}\to\Omega_c^0K^+$&9.0&$\Omega_c^0\to\Xi_c^{\prime 0}\overline{K}^{0}$&9&$\Xi_b^{\prime-}\to\Omega_c^0K^0$&9\\
				\hline
				\rule{0pt}{12pt}$\Sigma_c^+\to\Sigma_c^0\pi^+$&8.0&$\Sigma_c^0\to\Sigma_c^0\eta_8$&4.6&$\Sigma_c^0\to\Sigma_c^0\pi^0$&-8.0&$\Xi_c^{\prime 0}\to\Xi_c^{\prime +}\pi^-$&5.7&$\Xi_c^{\prime +}\to\Xi_c^{\prime 0}\pi^+$&5.7\\
				\hline
				\rule{0pt}{12pt}$\Xi_c^{\prime +}\to\Xi_c^{\prime +}\pi^0$&4.0&$\Lambda_c^+\to\Xi_c^0K^+$&-0.9&$\Xi_c^+\to\Sigma_c^+\overline{K}^{0}$&-5.0&$\Lambda_c^+\to\Sigma_c^+\pi^0$&6.5&$\Lambda_c^+\to\Xi_c^{\prime +}K^0$&-4.6\\
				\hline
				\rule{0pt}{12pt}$\Xi_c^0\to\Sigma_c^0\overline{K}^{0}$&-6.5&$\Xi_c^0\to\Xi_c^{\prime +}\pi^-$&4.4&$\Xi_c^+\to\Xi_c^{\prime +}\pi^0$&3.1&$\Xi_c^{\prime +}\to\Sigma_c^{++}K^-$&9.0&$\Sigma_c^{++}\to\Sigma_c^{++}\eta_8$&4.6\\
				\hline
				\rule{0pt}{12pt}$\Xi_c^{\prime 0}\to\Xi_c^{\prime 0}\eta_8$&-2.3&$\Sigma_c^0\to\Sigma_c^+\pi^-$&6.5&$\Xi_c^{\prime 0}\to\Xi_c^{\prime 0}\pi^0$&-4.0&$\Xi_c^+\to\Xi_c^+\eta_1$&0.07&$\Lambda_c^+\to\Lambda_c^+\eta_1$&0.75\\
				\hline
				\rule{0pt}{12pt}$\Xi_c^0\to\Xi_c^0\eta_1$&0.07&$\Xi_c^{\prime +}\to\Xi_c^{\prime +}\eta_1$&2.6&$\Sigma_c^+\to\Sigma_c^+\eta_1$&2.6&$\Sigma_c^{++}\to\Sigma_c^{++}\eta_1$&2.6&$\Xi_c^{\prime 0}\to\Xi_c^{\prime 0}\eta_1$&2.6\\
				\hline
				\rule{0pt}{12pt}$\Omega_c^0\to\Omega_c^0\eta_1$&11.0&$\Sigma_c^0\to\Sigma_c^0\eta_1$&2.6&&&&&&\\
				\hline
				\hline
			\end{tabular}
		\end{table}
				
				\begin{table}[h]
					\centering
					\caption{Strong coupling constants of $V\mathcal{B}_{\bar{3}}\mathcal{B}_{\bar{3}}$, $V\mathcal{B}_{\bar{3}}\mathcal{B}_6$ and $V\mathcal{B}_6\mathcal{B}_6$ vertex.}
					\label{ap:VBB}
					\begin{tabular}{|c|c|c|c|c|c|c|c|c|c|c|c|}
						\hline
						\hline
						\rule{0pt}{12pt}Vertex&$f_1$&$f_2$&Vertex&$f_1$&$f_2$&Vertex&$f_1$&$f_2$&Vertex&$f_1$&$f_2$\\
						\hline
						\hline
						\rule{0pt}{12pt}$\Lambda_c^+\to\Lambda_c^+\omega$&4.9&6&$\Lambda_c^+\to\Xi_c^+K^{\ast 0}$&4.6&6&$\Xi_c^+\to\Lambda_c^+\overline{K}^{\ast 0}$&4.6&6&$\Xi_c^+\to\Xi_c^+\phi$&4.6&6\\
						\hline
						\rule{0pt}{12pt}$\Xi_c^0\to\Lambda_c^+K^{\ast -}$&-4.6&-6&$\Xi_c^0\to\Xi_c^0\phi$&4.6&16&$\Xi_c^0\to\Xi_c^+\rho^-$&8.5&10.6&$\Xi_c^+\to\Xi_c^0\rho^+$&8.5&10.6\\
						\hline
						\rule{0pt}{12pt}$\Xi_c^0\to\Xi_c^0\rho^0$&-6&-7.5&$\Xi_c^+\to\Xi_c^+\omega$&5.5&7.5&$\Xi_c^+\to\Xi_c^+\rho^0$&6&7.5&$\Lambda_c^+\to\Sigma_c^+\rho^0$&2.6&16\\
						\hline
						\rule{0pt}{12pt}$\Lambda_c^+\to\Sigma_c^{++}\rho^-$&-2.6&-16&$\Sigma_c^{++}\to\Lambda_c^+\rho^+$&-2.6&-16&$\Lambda_c^+\to\Xi_c^{\prime +}K^{\ast 0}$&-2.3&-14&$\Xi_c^{\prime +}\to\Lambda_c^+\overline{K}^{\ast 0}$&-2.3&-14\\
						\hline
						\rule{0pt}{12pt}$\Sigma_c^+\to\Xi_c^+K^{\ast 0}$&-2.2&-13&$\Xi_c^+\to\Sigma_c^{++}K^{\ast -}$&-3.1&-18.4&$\Sigma_c^{++}\to\Xi_c^+K^{\ast +}$&-3.1&-18.4&$\Xi_c^+\to\Xi_c^{\prime +}\phi$&-2.1&-13\\
						\hline
						\rule{0pt}{12pt}$\Lambda_c^+\to\Sigma_c^0\rho^+$&2.6&16&$\Sigma_c^0\to\Lambda_c^+\rho^-$&2.6&16&$\Lambda_c^+\to\Xi_c^{\prime 0}K^{\ast +}$&2.3&14.1&$\Xi_c^{\prime 0}\to\Lambda_c^+K^{\ast -}$&2.3&14.1\\
						\hline
						\rule{0pt}{12pt}$\Sigma_c^+\to\Xi_c^0K^{\ast +}$&-2.2&-13&$\Xi_c^0\to\Sigma_c^0\overline{K}^{\ast 0}$&-2.2&-13&$\Sigma_c^0\to\Xi_c^0K^{\ast 0}$&-2.2&-13&$\Xi_c^0\to\Xi_c^{\prime 0}\phi$&-2.1&-13\\
						\hline
						\rule{0pt}{12pt}$\Xi_c^0\to\Omega_c^0K^{\ast 0}$&3.3&20&$\Omega_c^0\to\Xi_c^0\overline{K}^{\ast 0}$&3.3&20&$\Xi_c^0\to\Xi_c^{\prime +}\rho^-$&2.1&115.6&$\Xi_c^{\prime +}\to\Xi_c^0\rho^+$&2.1&115.6\\
						\hline
						\rule{0pt}{12pt}$\Xi_c^{\prime 0}\to\Xi_c^0\omega$&1.2&8&$\Xi_c^0\to\Xi_c^{\prime 0}\rho^0$&-1.5&-11&$\Xi_c^{\prime 0}\to\Xi_c^0\rho^0$&-1.5&-11&$\Xi_c^+\to\Omega_c^0K^{\ast +}$&3.3&20\\
						\hline
						\rule{0pt}{12pt}$\Xi_c^+\to\Xi_c^{\prime 0}\rho^+$&2.1&15.6&$\Xi_b^{\prime-}\to\Xi_c^+\rho^-$&2.1&15.6&$\Xi_c^+\to\Xi_c^{\prime +}\omega$&1.5&11&$\Xi_c^{\prime +}\to\Xi_c^+\omega$&1.5&11\\
						\hline
						\rule{0pt}{12pt}$\Xi_c^{\prime +}\to\Xi_c^+\rho^0$&1.5&11.0&$\Sigma_c^+\to\Sigma_c^+\omega$&3.5&24&$\Sigma_c^+\to\Sigma_b^{\prime +}\rho^-$&4.0&27.0&$\Sigma_c^{++}\to\Sigma_c^+\rho^+$&4&27\\
						\hline
						\rule{0pt}{12pt}$\Xi_c^{\prime +}\to\Sigma_c^+\overline{K}^{\ast 0}$&3.5&21.2&$\Sigma_c^{++}\to\Sigma_c^{++}\omega$&3.5&24&$\Sigma_c^{++}\to\Sigma_c^{++}\rho^0$&4&27&$\Sigma_c^{++}\to\Xi_c^{\prime +}K^{\ast +}$&5&30\\
						\hline
						\rule{0pt}{12pt}$\Xi_c^{\prime +}\to\Xi_c^{\prime +}\phi$&4&21&$\Omega_c^0\to\Omega_c^0\phi$&11&52&$\Omega_c^0\to\Xi_c^{\prime +}K^{\ast -}$&7&35&$\Xi_c^{\prime +}\to\Omega_c^0K^{\ast +}$&7&35\\
						\hline
						\rule{0pt}{12pt}$\Xi_c^{\prime 0}\to\Omega_c^0K^{\ast 0}$&7&35&$\Sigma_c^0\to\Sigma_c^+\rho^-$&4&27&$\Sigma_c^+\to\Sigma_c^0\rho^+$&4&27&$\Sigma_c^0\to\Sigma_c^0\omega$&3.5&24\\
						\hline
						\rule{0pt}{12pt}$\Sigma_c^0\to\Xi_c^{\prime 0}K^{\ast 0}$&5&30&$\Xi_c^{\prime 0}\to\Sigma_c^0\overline{K}^{\ast 0}$&5&30&$\Sigma_c^+\to\Xi_c^{\prime 0}K^{\ast +}$&3.5&21.2&$\Xi_c^{\prime 0}\to\Sigma_c^+K^{\ast -}$&3.5&21.2\\
						\hline
						\rule{0pt}{12pt}$\Xi_c^{\prime 0}\Xi_c^{\prime +}\rho^-$&3.5&22.6&$\Xi_c^{\prime +}\to\Xi_c^{\prime 0}\rho^+$&3.5&22.6&$\Xi_c^{\prime 0}\to\Xi_c^{\prime 0}\omega$&2.4&15&$\Xi_c^{\prime 0}\to\Xi_c^{\prime 0}\rho^0$&-2.5&-16\\
						\hline
						\rule{0pt}{12pt}$\Xi_c^{\prime +}\to\Xi_c^{\prime +}\rho^0$&2.5&16&$\Lambda_c^+\to\Xi_c^0K^{\ast +}$&-4.6&-6&$\Xi_c^0\to\Xi_c^0\omega$&5.5&7.5&$\Sigma_c^+\to\Lambda_c^+\rho^0$&2.6&16\\
						\hline
						\rule{0pt}{12pt}$\Xi_c^+\to\Sigma_c^+\overline{K}^{\ast 0}$&-2.2&-13&$\Xi_c^{\prime +}\to\Xi_c^+\phi$&-2.1&-13&$\Xi_c^0\to\Sigma_c^+K^{\ast -}$&-2.2&-13&$\Xi_c^{\prime 0}\to\Xi_c^0\phi$&-2.1&-13\\
						\hline
						\rule{0pt}{12pt}$\Xi_c^0\to\Xi_c^{\prime 0}\omega$&1.2&8&$\Omega_c^0\to\Xi_c^+K^{\ast -}$&3.5&20&$\Xi_c^+\to\Xi_c^{\prime +}\rho^0$&1.2&8&$\Sigma_c^+\to\Xi_c^{\prime +}K^{\ast 0}$&3.5&21.2\\
						\hline
						\rule{0pt}{12pt}$\Xi_c^{\prime +}\to\Sigma_c^{++}K^{\ast -}$&5.0&30.0&$\Omega_c^0\to\Xi_c^{\prime 0}\overline{K}^{\ast 0}$&5&30&$\Sigma_c^0\to\Sigma_c^0\rho^0$&-4&-27&$\Xi_c^{\prime 0}\to\Xi_c^{\prime 0}\phi$&4&21\\
						\hline
						\rule{0pt}{12pt}$\Xi_c^{\prime +}\to\Xi_c^{\prime +}\omega$&2.4&15&&&&&&&&&\\
						\hline
						\hline
					\end{tabular}
				\end{table}

\section{ Expressions of amplitudes}\label{app:amp}

The expressions of amplitudes for all the forty-seven $\mathcal{B}_{cc}\to\mathcal{B}_cP$ decays cosidered in this paper  are collected in this Appendix.

{ \scriptsize
\begin{align}
\mathcal{A}(\Xi_{cc}^{++}\to\Lambda_c^+K^+)&=\mathcal{T}(\Xi_{cc}^{++}\to\Lambda_c^+K^+)+\mathcal{M}(K^+,\Lambda_c^+;\omega)+\mathcal{M}(K^+,\Sigma_c^+;\rho^0)+\mathcal{M}(K^{\ast +},\Lambda_c^+;\eta_8)+\mathcal{M}(K^{\ast +},\Sigma_c^+;\pi^0)\non
&+\mathcal{M}(K^+,\Lambda_c^+;\Xi_c^0)+\mathcal{M}(K^+,\Lambda_c^+;\Xi_c^{\prime 0})+\mathcal{M}(K^+,\Sigma_c^+;\Xi_c^0)+\mathcal{M}(K^+,\Sigma_c^+;\Xi_c^{\prime 0})+\mathcal{M}(K^{\ast +},\Lambda_c^+;\Xi_c^0)\non
&+\mathcal{M}(K^{\ast +},\Lambda_c^+;\Xi_c^{\prime 0})+\mathcal{M}(K^{\ast +},\Sigma_c^+;\Xi_c^0)+\mathcal{M}(K^{\ast +},\Sigma_c^+;\Xi_c^{\prime 0}),\\
\mathcal{A}(\Xi_{cc}^{++}\to\Lambda_c^+\pi^+)&=\mathcal{T}(\Xi_{cc}^{++}\to\Lambda_c^+\pi^+)+\mathcal{M}(\pi^+,\Sigma_c^+;\rho^0)+\mathcal{M}(\rho^+,\Sigma_c^+;\pi^0)+\mathcal{M}(K^+,\Xi_c^+;K^{\ast 0})+\mathcal{M}(K^+,\Xi_c^{\prime +};K^{\ast 0})\non
&+\mathcal{M}(K^{\ast +},\Xi_c^+;K^0)+\mathcal{M}(K^{\ast +},\Xi_c^{\prime +};K^0)+\mathcal{M}(\pi^+,\Lambda_c^+;\Sigma_c^0)+\mathcal{M}(\pi^+,\Sigma_c^+;\Sigma_c^0)+\mathcal{M}(\rho^+,\Lambda_c^+;\Sigma_c^0)\non
&+\mathcal{M}(\rho^+,\Sigma_c^+;\Sigma_c^0)+\mathcal{M}(K^+,\Xi_c^+;\Xi_c^0)+\mathcal{M}(K^+,\Xi_c^+;\Xi_c^{\prime 0})+\mathcal{M}(K^+,\Xi_c^{\prime +};\Xi_c^0)+\mathcal{M}(K^+,\Xi_c^{\prime +};\Xi_c^{\prime 0})\non
&+\mathcal{M}(K^{\ast +},\Xi_c^+;\Xi_c^0)+\mathcal{M}(K^{\ast +},\Xi_c^+;\Xi_c^{\prime 0})+\mathcal{M}(K^{\ast +},\Xi_c^{\prime +};\Xi_c^0)+\mathcal{M}(K^{\ast +},\Xi_c^{\prime +};\Xi_c^{\prime 0}),\\
\mathcal{A}(\Xi_{cc}^{++}\to\Xi_c^+K^+)&=\mathcal{T}(\Xi_{cc}^{++}\to\Xi_c^+K^+)+\mathcal{M}(K^+,\Xi_c^+;\rho^0)+\mathcal{M}(K^+,\Xi_c^+;\omega)+\mathcal{M}(K^+,\Xi_c^{\prime +};\rho^0)+\mathcal{M}(K^+,\Xi_c^{\prime +};\omega)\non
&+\mathcal{M}(K^{\ast +},\Xi_c^+;\pi^0)+\mathcal{M}(K^{\ast +},\Xi_c^+;\eta_8)+\mathcal{M}(K^{\ast +},\Xi_c^{\prime +};\pi^0)+\mathcal{M}(K^{\ast +},\Xi_c^{\prime +};\eta_8)+\mathcal{M}(\pi^+,\Lambda_c^+;\overline{K}^{\ast 0})\non
&+\mathcal{M}(\pi^+,\Sigma_c^+;\overline{K}^{\ast 0})+\mathcal{M}(\rho^+,\Lambda_c^+;\overline{K}^{0})+\mathcal{M}(\rho^+,\Sigma_c^+;\overline{K}^{0})+\mathcal{M}(K^+,\Xi_c^+;\phi)+\mathcal{M}(K^+,\Xi_c^{\prime +};\phi)\non
&+\mathcal{M}(K^{\ast +},\Xi_c^+;\eta_8)+\mathcal{M}(K^{\ast +},\Xi_c^{\prime +};\eta_8)+\mathcal{M}(K^{+},\Xi_c^+;\Omega_c^0)+\mathcal{M}(K^{+},\Xi_c^{\prime +};\Omega_c^0)+\mathcal{M}(K^{\ast +},\Xi_c^+;\Omega_c^0)\non
&+\mathcal{M}(K^{\ast +},\Xi_c^{\prime +};\Omega_c^0)+\mathcal{M}(\pi^{+},\Lambda_c^+;\Xi_c^0)+\mathcal{M}(\pi^{+},\Lambda_c^+;\Xi_c^{\prime 0})+\mathcal{M}(\pi^{+},\Sigma_c^+;\Xi_c^0)+\mathcal{M}(\pi^{+},\Sigma_c^+;\Xi_c^{\prime 0})\non
&+\mathcal{M}(\rho^{+},\Lambda_c^+;\Xi_c^0)+\mathcal{M}(\rho^{+},\Lambda_c^+;\Xi_c^{\prime 0})+\mathcal{M}(\rho^{+},\Sigma_c^+;\Xi_c^0)+\mathcal{M}(\rho^{+},\Sigma_c^+;\Xi_c^{\prime 0}),\\
\mathcal{A}(\Xi_{cc}^{++}\to\Xi_c^+\pi^+)&=\mathcal{T}(\Xi_{cc}^{++}\to\Xi_c^+\pi^+)+\mathcal{M}(\pi^+,\Xi_c^+;\rho^0)+\mathcal{M}(\pi^+,\Xi_c^{\prime +};\rho^0)+\mathcal{M}(\rho^+,\Xi_c^+;\pi^0)+\mathcal{M}(\rho^+,\Xi_c^{\prime +};\pi^0)\non
&+\mathcal{M}(\pi^+,\Xi_c^+;\Xi_c^0)+\mathcal{M}(\pi^+,\Xi_c^+;\Xi_c^{\prime 0})+\mathcal{M}(\pi^+,\Xi_c^{\prime +};\Xi_c^0)+\mathcal{M}(\pi^+,\Xi_c^{\prime +};\Xi_c^{\prime 0})+\mathcal{M}(\rho^+,\Xi_c^+;\Xi_c^0)\non
&+\mathcal{M}(\rho^+,\Xi_c^+;\Xi_c^{\prime 0})+\mathcal{M}(\rho^+,\Xi_c^{\prime +};\Xi_c^0)+\mathcal{M}(\rho^+,\Xi_c^{\prime +};\Xi_c^{\prime 0}),  \\
\mathcal{A}(\Xi_{cc}^{+}\to\Lambda_c^+\overline{K}^0		)&=\mathcal{C}_{SD}(\Xi_{cc}^{+}\to\Lambda_c^+\overline{K}^0)+\mathcal{M}(\pi^+,\Xi_c^0;K^{\ast +})+\mathcal{M}(\pi^+,\Xi_c^{\prime 0};K^{\ast +})+\mathcal{M}(\rho^+,\Xi_c^0;K^+)+\mathcal{M}(\rho^+,\Xi_c^{\prime 0};K^+)\non
&+\mathcal{M}(\pi^+,\Xi_c^0;\Sigma_c^0)+\mathcal{M}(\pi^+,\Xi_c^{\prime 0};\Sigma_c^0)+\mathcal{M}(\rho^+,\Xi_c^0;\Sigma_c^0)+\mathcal{M}(\rho^+,\Xi_c^{\prime 0};\Sigma_c^0),\\
\mathcal{A}(\Xi_{cc}^{+}\to\Lambda_c^+K^0)&=\mathcal{C}_{SD}(\Xi_{cc}^{+}\to\Lambda_c^+K^0)+\mathcal{M}(K^+,\Sigma_c^0;\rho^+)+\mathcal{M}(K^{\ast +},\Sigma_c^0;\pi^+)+\mathcal{M}(K+,\Sigma_c^0;\Xi_c^0)+\mathcal{M}(K^+,\Sigma_c^0;\Xi_c^{\prime 0})\non
&+\mathcal{M}(K^{\ast +},\Sigma_c^0;\Xi_c^0)+\mathcal{M}(K^{\ast +},\Sigma_c^0;\Xi_c^{\prime 0}),\\
\mathcal{A}(\Xi_{cc}^{+}\to\Lambda_c^+\pi^0)&=\mathcal{C}_{SD}(\Xi_{cc}^{+}\to\Lambda_c^+\pi^0)+\mathcal{M}(\pi^+,\Sigma_c^0;\rho^+)+\mathcal{M}(\rho^+,\Sigma_c^0;\pi^+)+\mathcal{M}(K^{+},\Xi_c^0;K^{\ast +})+\mathcal{M}(K^{+},\Xi_c^{\prime 0};K^{\ast +})\non
&+\mathcal{M}(K^{\ast +},\Xi_c^0;K^{+})+\mathcal{M}(K^{\ast +},\Xi_c^{\prime 0};K^{+})+\mathcal{M}(\pi^+,\Sigma_c^0;\Sigma_c^0)+\mathcal{M}(\rho^+,\Sigma_c^0;\Sigma_c^0)+\mathcal{M}(K^+,\Xi_c^0;\Xi_c^0)\non
&+\mathcal{M}(K^+,\Xi_c^0;\Xi_c^{\prime 0})+\mathcal{M}(K^+,\Xi_c^{\prime 0};\Xi_c^0)+\mathcal{M}(K^+,\Xi_c^{\prime 0};\Xi_c^{\prime 0}),\\
\mathcal{A}(\Xi_{cc}^{+}\to\Lambda_c^+\eta_1)&=\mathcal{C}_{SD}(\Xi_{cc}^{+}\to\Lambda_c^+\eta_1)+\mathcal{M}(K^+,\Xi_c^0;\Xi_c^0)+\mathcal{M}(K^+,\Xi_c^{\prime 0};\Xi_c^{\prime 0})+\mathcal{M}(K^{\ast +},\Xi_c^0;\Xi_c^0)+\mathcal{M}(K^{\ast +},\Xi_c^{\prime 0};\Xi_c^{\prime 0})\non
&+\mathcal{M}(\pi^+,\Sigma_c^0;\Sigma_c^0)+\mathcal{M}(\rho^+,\Sigma_c^0;\Sigma_c^0),\\
\mathcal{A}(\Xi_{cc}^{+}\to\Lambda_c^+\eta_8)&=\mathcal{C}_{SD}(\Xi_{cc}^{+}\to\Lambda_c^+\eta_8)+\mathcal{M}(K^+,\Xi_c^0;K^{\ast +})+\mathcal{M}(K^+,\Xi_c^{\prime 0};K^{\ast +})+\mathcal{M}(K^{\ast +},\Xi_c^0;K^{+})+\mathcal{M}(K^{\ast +},\Xi_c^{\prime 0};K^{+})\non
&+\mathcal{M}(K^+,\Xi_c^0;\Xi_c^0)+\mathcal{M}(K^+,\Xi_c^0;\Xi_c^{\prime 0})+\mathcal{M}(K^+,\Xi_c^{\prime 0};\Xi_c^0)+\mathcal{M}(K^+,\Xi_c^{\prime 0};\Xi_c^{\prime 0})+\mathcal{M}(K^{\ast +},\Xi_c^0;\Xi_c^0)\non
&+\mathcal{M}(K^{\ast +},\Xi_c^0;\Xi_c^{\prime 0})+\mathcal{M}(K^{\ast +},\Xi_c^{\prime 0};\Xi_c^0)+\mathcal{M}(K^{\ast +},\Xi_c^{\prime 0};\Xi_c^{\prime 0})+\mathcal{M}(\pi^+,\Sigma_c^0;\Sigma_c^0)+\mathcal{M}(\rho^+,\Sigma_c^0;\Sigma_c^0),\\
\mathcal{A}(\Xi_{cc}^{+}\to\Xi_c^+K^0)&=\mathcal{C}_{SD}(\Xi_{cc}^{+}\to\Xi_c^+K^0)+\mathcal{M}(K^+,\Xi_c^0;\rho^+)+\mathcal{M}(K^+,\Xi_c^{\prime 0};\rho^+)+\mathcal{M}(K^{\ast +},\Xi_c^0;\rho^+)+\mathcal{M}(K^{\ast +},\Xi_c^{\prime 0};\rho^+)\non
&+\mathcal{M}(K^+,\Xi_c^0;\Omega_c^0)+\mathcal{M}(K^+,\Xi_c^{\prime 0};\Omega_c^0)+\mathcal{M}(K^{\ast +},\Xi_c^0;\Omega_c^0)+\mathcal{M}(K^{\ast +},\Xi_c^{\prime 0};\Omega_c^0)+\mathcal{M}(\pi^+,\Sigma_c^0;\Xi_c^0)\non
&+\mathcal{M}(\pi^+,\Sigma_c^0;\Xi_c^{\prime 0})+\mathcal{M}(\rho^+,\Sigma_c^0;\Xi_c^0)+\mathcal{M}(\rho^+,\Sigma_c^0;\Xi_c^{\prime 0}),\\
\mathcal{A}(\Xi_{cc}^{+}\to\Xi_c^0K^+)&=\mathcal{T}(\Xi_{cc}^{+}\to\Xi_c^0K^+)+\mathcal{M}(\pi^+,\Sigma_c^0;\overline{K}^{\ast 0})+\mathcal{M}(\rho^+,\Sigma_c^0;\overline{K}^{0})+\mathcal{M}(K^+,\Xi_c^0;\phi)+\mathcal{M}(K^+,\Xi_c^{\prime 0};\phi)\non
&+\mathcal{M}(K^{\ast +},\Xi_c^0;\eta_8)+\mathcal{M}(K^{\ast +},\Xi_c^{\prime 0};\eta_8),\\
\mathcal{A}(\Xi_{cc}^{+}\to\Xi_c^0\pi^+)&=\mathcal{T}(\Xi_{cc}^{+}\to\Xi_c^0\pi^+)+\mathcal{M}(\pi^+,\Xi_c^0;\rho^0)+\mathcal{M}(\pi^+,\Xi_c^{\prime 0};\rho^0)+\mathcal{M}(\rho^+,\Xi_c^0;\pi^0)+\mathcal{M}(\rho^+,\Xi_c^{\prime 0};\pi^0),\\
\mathcal{A}(\Xi_{cc}^{+}\to\Xi_c^+\pi^0)&=\mathcal{C}_{SD}(\Xi_{cc}^{+}\to\Xi_c^+\pi^0)+\mathcal{M}(\pi^{+},\Xi_c^0;\rho^+)+\mathcal{M}(\pi^{+},\Xi_c^{\prime 0};\rho^+)+\mathcal{M}(\rho^{+},\Xi_c^0;\pi^+)+\mathcal{M}(\rho^{+},\Xi_c^{\prime 0};\pi^+)\non
&+\mathcal{M}(\pi^{+},\Xi_c^0;\Xi_c^0)+\mathcal{M}(\pi^{+},\Xi_c^0;\Xi_c^{\prime 0})+\mathcal{M}(\pi^{+},\Xi_c^{\prime 0};\Xi_c^0)+\mathcal{M}(\pi^{+},\Xi_c^{\prime 0};\Xi_c^{\prime 0})+\mathcal{M}(\rho^{+},\Xi_c^0;\Xi_c^0)\non
&+\mathcal{M}(\rho^{+},\Xi_c^0;\Xi_c^{\prime 0})+\mathcal{M}(\rho^{+},\Xi_c^{\prime 0};\Xi_c^0)+\mathcal{M}(\rho^{+},\Xi_c^{\prime 0};\Xi_c^{\prime 0}),\\
\mathcal{A}(\Xi_{cc}^{+}\to\Xi_c^+\eta_1)&=\mathcal{M}(\pi^+,\Xi_c^0;\Xi_c^0)+\mathcal{M}(\pi^+,\Xi_c^{\prime 0};\Xi_c^{\prime 0})+\mathcal{M}(\rho^+,\Xi_c^0;\Xi_c^0)+\mathcal{M}(\rho^+,\Xi_c^{\prime 0};\Xi_c^{\prime 0}),\\
\mathcal{A}(\Xi_{cc}^{+}\to\Xi_c^+\eta_8)&=\mathcal{M}(\pi^+,\Xi_c^0;\Xi_c^0)+\mathcal{M}(\pi^+,\Xi_c^{0};\Xi_c^{\prime 0})+\mathcal{M}(\pi^+,\Xi_c^{\prime 0};\Xi_c^0)+\mathcal{M}(\pi^+,\Xi_c^{\prime 0};\Xi_c^{\prime 0})+\mathcal{M}(\rho^+,\Xi_c^0;\Xi_c^0)\non
&+\mathcal{M}(\rho^+,\Xi_c^{0};\Xi_c^{\prime 0})+\mathcal{M}(\rho^+,\Xi_c^{\prime 0};\Xi_c^0)+\mathcal{M}(\rho^+,\Xi_c^{\prime 0};\Xi_c^{\prime 0}),\\
\mathcal{A}(\Xi_{cc}^{+}\to\Xi_c^{\prime +}\eta_1)&=\mathcal{M}(\pi^+,\Xi_c^0;\Xi_c^0)+\mathcal{M}(\pi^+,\Xi_c^{\prime 0};\Xi_c^{\prime 0})+\mathcal{M}(\rho^+,\Xi_c^0;\Xi_c^0)+\mathcal{M}(\rho^+,\Xi_c^{\prime 0};\Xi_c^{\prime 0}),\\
\mathcal{A}(\Xi_{cc}^{+}\to\Xi_c^{\prime +}\eta_8)&=\mathcal{M}(\pi^+,\Xi_c^0;\Xi_c^0)+\mathcal{M}(\pi^+,\Xi_c^{0};\Xi_c^{\prime 0})+\mathcal{M}(\pi^+,\Xi_c^{\prime 0};\Xi_c^0)+\mathcal{M}(\pi^+,\Xi_c^{\prime 0};\Xi_c^{\prime 0})+\mathcal{M}(\rho^+,\Xi_c^0;\Xi_c^0)\non
&+\mathcal{M}(\rho^+,\Xi_c^{0};\Xi_c^{\prime 0})+\mathcal{M}(\rho^+,\Xi_c^{\prime 0};\Xi_c^0)+\mathcal{M}(\rho^+,\Xi_c^{\prime 0};\Xi_c^{\prime 0}),\\
\mathcal{A}(\Xi_{cc}^{+}\to\Sigma_c^0K^+)&=\mathcal{T}(\Xi_{cc}^{+}\to\Sigma_c^0K^+),\\
\mathcal{A}(\Omega_{cc}^{+}\to\Omega_c^0\pi^+)&=\mathcal{T}(\Omega_{cc}^{+}\to\Omega_c^0\pi^+),\\
\mathcal{A}(\Omega_{cc}^{+}\to\Lambda_c^+\eta_1)&=\mathcal{C}_{SD}(\Omega_{cc}^{+}\to\Lambda_c^+\eta_1)+\mathcal{M}(K^+,\Xi_c^0;\Xi_c^0)+\mathcal{M}(K^+,\Xi_c^{\prime 0};\Xi_c^{\prime 0})+\mathcal{M}(K^{\ast +},\Xi_c^0;\Xi_c^0)+\mathcal{M}(K^{\ast +},\Xi_c^{\prime 0};\Xi_c^{\prime 0}),\\
\mathcal{A}(\Omega_{cc}^{+}\to\Lambda_c^+\eta_8)&=\mathcal{C}_{SD}(\Omega_{cc}^{+}\to\Lambda_c^+\eta_8)+\mathcal{M}(K^{+},\Xi_c^0;K^{\ast +})+\mathcal{M}(K^{+},\Xi_c^{\prime 0};K^{\ast +})+\mathcal{M}(K^{\ast +},\Xi_c^0;K^{+})\non
&+\mathcal{M}(K^{\ast +},\Xi_c^{\prime 0};K^{+})+\mathcal{M}(K^{+},\Xi_c^0;\Xi_c^0)+\mathcal{M}(K^{+},\Xi_c^0;\Xi_c^{\prime 0})+\mathcal{M}(K^{+},\Xi_c^{\prime 0};\Xi_c^0)+\mathcal{M}(K^{+},\Xi_c^{\prime 0};\Xi_c^{\prime 0})\non
&+\mathcal{M}(K^{\ast +},\Xi_c^0;\Xi_c^0)+\mathcal{M}(K^{\ast +},\Xi_c^0;\Xi_c^{\prime 0})+\mathcal{M}(K^{\ast +},\Xi_c^{\prime 0};\Xi_c^0)+\mathcal{M}(K^{\ast +},\Xi_c^{\prime 0};\Xi_c^{\prime 0}),\\
\mathcal{A}(\Omega_{cc}^{+}\to\Xi_c^{\prime +}\eta_1)&=\mathcal{C}_{SD}(\Omega_{cc}^{+}\to\Xi_c^{\prime +}\eta_1)+\mathcal{M}(K^+,\Omega_c^0;\Omega_c^0)+\mathcal{M}(K^{\ast +},\Omega_c^0;\Omega_c^0)+\mathcal{M}(\pi^+,\Xi_c^0;\Xi_c^0)+\mathcal{M}(\pi^+,\Xi_c^{\prime 0};\Xi_c^{\prime 0})\non
&+\mathcal{M}(\rho^+,\Xi_c^0;\Xi_c^0)+\mathcal{M}(\rho^+,\Xi_c^{\prime 0};\Xi_c^{\prime 0}),\\
\mathcal{A}(\Omega_{cc}^{+}\to\Xi_c^{\prime +}\eta_8)&=\mathcal{C}_{SD}(\Omega_{cc}^{+}\to\Xi_c^{\prime +}\eta_8)+\mathcal{M}(K^+,\Omega_c^0;K^{\ast +})+\mathcal{M}(K^{\ast +},\Omega_c^0;K^+)+\mathcal{M}(K^+,\Omega_c^0;\Omega_c^0)+\mathcal{M}(K^{\ast +},\Omega_c^0;\Omega_c^0)\non
&+\mathcal{M}(\pi^+,\Xi_c^0;\Xi_c^0)+\mathcal{M}(\pi^+,\Xi_c^{\prime};\Xi_c^0)+\mathcal{M}(\pi^+,\Xi_c^0;\Xi_c^{\prime})+\mathcal{M}(\pi^+,\Xi_c^{\prime 0};\Xi_c^{\prime 0})\non
&+\mathcal{M}(\rho^+,\Xi_c^0;\Xi_c^0)+\mathcal{M}(\rho^+,\Xi_c^{\prime};\Xi_c^0)+\mathcal{M}(\rho^+,\Xi_c^0;\Xi_c^{\prime})+\mathcal{M}(\rho^+,\Xi_c^{\prime 0};\Xi_c^{\prime 0}),\\
\mathcal{A}(\Omega_{cc}^{+}\to\Xi_c^{+}\eta_1)&=\mathcal{C}_{SD}(\Omega_{cc}^{+}\to\Xi_c^{+}\eta_1)+\mathcal{M}(K^+,\Omega_c^0;\Omega_c^0)+\mathcal{M}(K^{\ast +},\Omega_c^0;\Omega_c^0)+\mathcal{M}(\pi^+,\Xi_c^0;\Xi_c^0)+\mathcal{M}(\pi^+,\Xi_c^{\prime 0};\Xi_c^{\prime 0})\non
&+\mathcal{M}(\rho^+,\Xi_c^0;\Xi_c^0)+\mathcal{M}(\rho^+,\Xi_c^{\prime 0};\Xi_c^{\prime 0}),\\
\mathcal{A}(\Omega_{cc}^{+}\to\Xi_c^{+}\eta_8)&=\mathcal{C}_{SD}(\Omega_{cc}^{+}\to\Xi_c^{+}\eta_8)+\mathcal{M}(K^+,\Omega_c^0;K^{\ast +})+\mathcal{M}(K^{\ast +},\Omega_c^0;K^+)+\mathcal{M}(K^+,\Omega_c^0;\Omega_c^0)+\mathcal{M}(K^{\ast +},\Omega_c^0;\Omega_c^0)\non
&+\mathcal{M}(\pi^+,\Xi_c^0;\Xi_c^0)+\mathcal{M}(\pi^+,\Xi_c^{\prime};\Xi_c^0)+\mathcal{M}(\pi^+,\Xi_c^0;\Xi_c^{\prime})+\mathcal{M}(\pi^+,\Xi_c^{\prime 0};\Xi_c^{\prime 0})\non
&+\mathcal{M}(\rho^+,\Xi_c^0;\Xi_c^0)+\mathcal{M}(\rho^+,\Xi_c^{\prime};\Xi_c^0)+\mathcal{M}(\rho^+,\Xi_c^0;\Xi_c^{\prime})+\mathcal{M}(\rho^+,\Xi_c^{\prime 0};\Xi_c^{\prime 0}),\\
\mathcal{A}(\Omega_{cc}^{+}\to\Sigma_c^+\eta_1)&=\mathcal{C}_{SD}(\Omega_{cc}^{+}\to\Sigma_c^+\eta_1)+\mathcal{M}(K^+,\Xi_c^0;\Xi_c^0)+\mathcal{M}(K^+,\Xi_c^{\prime 0};\Xi_c^{\prime 0})+\mathcal{M}(K^{\ast +},\Xi_c^0;\Xi_c^0)+\mathcal{M}(K^{\ast +},\Xi_c^{\prime 0};\Xi_c^{\prime 0}),\\
\mathcal{A}(\Omega_{cc}^{+}\to\Sigma_c^+\eta_8)&=\mathcal{C}_{SD}(\Omega_{cc}^{+}\to\Sigma_c^+\eta_8)+\mathcal{M}(K^{+},\Xi_c^0;K^{\ast +})+\mathcal{M}(K^{+},\Xi_c^{\prime 0};K^{\ast +})+\mathcal{M}(K^{\ast +},\Xi_c^0;K^{+})\non
&+\mathcal{M}(K^{\ast +},\Xi_c^{\prime 0};K^{+})+\mathcal{M}(K^{+},\Xi_c^0;\Xi_c^0)+\mathcal{M}(K^{+},\Xi_c^0;\Xi_c^{\prime 0})+\mathcal{M}(K^{+},\Xi_c^{\prime 0};\Xi_c^0)+\mathcal{M}(K^{+},\Xi_c^{\prime 0};\Xi_c^{\prime 0})\non
&+\mathcal{M}(K^{\ast +},\Xi_c^0;\Xi_c^0)+\mathcal{M}(K^{\ast +},\Xi_c^0;\Xi_c^{\prime 0})+\mathcal{M}(K^{\ast +},\Xi_c^{\prime 0};\Xi_c^0)+\mathcal{M}(K^{\ast +},\Xi_c^{\prime 0};\Xi_c^{\prime 0}),\\
\mathcal{A}(\Omega_{cc}^{+}\to\Lambda_c^+\overline{K}^0)&=\mathcal{C}_{SD}(\Omega_{cc}^{+}\to\Lambda_c^+\overline{K}^0)+\mathcal{M}(\pi^+,\Xi_c^0;K^{\ast +})+\mathcal{M}(\pi^+,\Xi_c^{\prime 0};K^{\ast +})+\mathcal{M}(\rho^+,\Xi_c^0;K^{+})+\mathcal{M}(\rho^+,\Xi_c^{\prime 0};K^{+})\non
&+\mathcal{M}(\pi^+,\Xi_c^0;\Sigma_c^0)+\mathcal{M}(\pi^+,\Xi_c^{\prime 0};\Sigma_c^0)+\mathcal{M}(\rho^+,\Xi_c^0;\Sigma_c^0)+\mathcal{M}(\rho^+,\Xi_c^{\prime 0};\Sigma_c^0)+\mathcal{M}(K^+,\Omega_c^0;\Xi_c^0)\non
&+\mathcal{M}(K^+,\Omega_c^0;\Xi_c^{\prime 0})+\mathcal{M}(K^{\ast +},\Omega_c^0;\Xi_c^0)+\mathcal{M}(K^{\ast +},\Omega_c^0;\Xi_c^{\prime 0}),\\
\mathcal{A}(\Omega_{cc}^{+}\to\Lambda_c^+\pi^0)&=\mathcal{M}(K^{+},\Xi_c^0;K^{\ast +})+\mathcal{M}(K^{+},\Xi_c^{\prime 0};K^{\ast +})+\mathcal{M}(K^{\ast +},\Xi_c^0;K^{+})+\mathcal{M}(K^{\ast +},\Xi_c^{\prime 0};K^{+})\non
&+\mathcal{M}(K^{+},\Xi_c^0;\Xi_c^0)+\mathcal{M}(K^{+},\Xi_c^0;\Xi_c^{\prime 0})+\mathcal{M}(K^{+},\Xi_c^{\prime 0};\Xi_c^0)+\mathcal{M}(K^{+},\Xi_c^{\prime 0};\Xi_c^{\prime 0})\non
&+\mathcal{M}(K^{\ast +},\Xi_c^0;\Xi_c^0)+\mathcal{M}(K^{\ast +},\Xi_c^0;\Xi_c^{\prime 0})+\mathcal{M}(K^{\ast +},\Xi_c^{\prime 0};\Xi_c^0)+\mathcal{M}(K^{\ast +},\Xi_c^{\prime 0};\Xi_c^{\prime 0}),\\
\mathcal{A}(\Omega_{cc}^{+}\to\Xi_c^0K^+)&=\mathcal{T}(\Omega_{cc}^{+}\to\Xi_c^0K^+)+\mathcal{M}(K^{+},\Xi_c^0;\phi)+\mathcal{M}(K^{+},\Xi_c^{\prime 0};\phi)+\mathcal{M}(K^{\ast +},\Xi_c^0;\eta_8)+\mathcal{M}(K^{\ast +},\Xi_c^{\prime 0};\eta_8),\\
\mathcal{A}(\Omega_{cc}^{+}\to\Xi_c^0\pi^+)&=\mathcal{T}(\Omega_{cc}^{+}\to\Xi_c^0\pi^+)+\mathcal{M}(\pi^+,\Xi_c^0;\rho^0)+\mathcal{M}(\pi^+,\Xi_c^{\prime 0};\rho^0)+\mathcal{M}(\rho^+,\Xi_c^0;\pi^0)+\mathcal{M}(\rho^+,\Xi_c^{\prime 0};\pi^0)\non
&+\mathcal{M}(K^+,\Omega_c^0;K^{\ast 0})+\mathcal{M}(K^{\ast +},\Omega_c^0;K^{0}), \\
\mathcal{A}(\Omega_{cc}^{+}\to\Xi_c^+\overline{K}^0)&=\mathcal{C}_{SD}(\Omega_{cc}^{+}\to\Xi_c^+\overline{K}^0)+\mathcal{M}(\pi^+,\Omega_c^0;K^{\ast +})+\mathcal{M}(\rho^+,\Omega_c^0;K^+)+\mathcal{M}(\pi^+,\Omega_c^0;\Xi_c^0)+\mathcal{M}(\pi^+,\Omega_c^0;\Xi_c^{\prime 0})\non
&+\mathcal{M}(\rho^+,\Omega_c^0;\Xi_c^0)+\mathcal{M}(\rho^+,\Omega_c^0;\Xi_c^{\prime 0}),\\
\mathcal{A}(\Omega_{cc}^{+}\to\Xi_c^+K^0)&=\mathcal{C}_{SD}(\Omega_{cc}^{+}\to\Xi_c^+K^0)+\mathcal{M}(K^+,\Xi_c^0;\rho^+)+\mathcal{M}(K^+,\Xi_c^{\prime 0};\rho^+)+\mathcal{M}(K^{\ast +},\Xi_c^0;\pi^+)+\mathcal{M}(K^{\ast +},\Xi_c^{\prime 0};\pi^+)\non
&+\mathcal{M}(K^{+},\Xi_c^0;\Omega_c^0)+\mathcal{M}(K^{+},\Xi_c^0{\prime -};\Omega_c^0)+\mathcal{M}(K^{\ast +},\Xi_c^0;\Omega_c^0)+\mathcal{M}(K^{\ast +},\Xi_c^{\prime 0};\Omega_c^0),\\
\mathcal{A}(\Omega_{cc}^{+}\to\Xi_c^+\pi^0)&=\mathcal{C}_{SD}(\Omega_{cc}^{+}\to\Xi_c^+\pi^0)+\mathcal{M}(\pi^{+},\Xi_c^0;\rho^+)+\mathcal{M}(\pi^{+},\Xi_c^{\prime 0};\rho^+)+\mathcal{M}(\rho^{+},\Xi_c^0;\pi^+)+\mathcal{M}(\rho^{+},\Xi_c^{\prime 0};\pi^+)\non
&+\mathcal{M}(K^{+},\Omega_c^0;K^{\ast +})+\mathcal{M}(K^{\ast +},\Omega_c^0;K^{+})+\mathcal{M}(\pi^+,\Xi_c^0;\Xi_c^0)+\mathcal{M}(\pi^+,\Xi_c^0;\Xi_c^{\prime 0})+\mathcal{M}(\pi^+,\Xi_c^{\prime 0};\Xi_c^0)\non
&+\mathcal{M}(\pi^+,\Xi_c^{\prime 0};\Xi_c^{\prime 0})+\mathcal{M}(\rho^+,\Xi_c^0;\Xi_c^0)+\mathcal{M}(\rho^+,\Xi_c^0;\Xi_c^{\prime 0})+\mathcal{M}(\rho^+,\Xi_c^{\prime 0};\Xi_c^0)+\mathcal{M}(\rho^+,\Xi_c^{\prime 0};\Xi_c^{\prime 0}), \\
\mathcal{A}(\Xi_{cc}^{++}\to\Sigma_c^+K^+)&=\mathcal{T}(\Xi_{cc}^{++}\to\Sigma_c^+K^+)+\mathcal{M}(K^{+},\Lambda_c^+;\rho^0)+\mathcal{M}(K^{+},\Sigma_c^+;\omega)+\mathcal{M}(K^{\ast +},\Lambda_c^+;\pi^0)+\mathcal{M}(K^{\ast +},\Sigma_c^+;\eta_8)\non
&+\mathcal{M}(K^{+},\Lambda_c^+;\Xi_c^0)+\mathcal{M}(K^{+},\Lambda_c^+;\Xi_c^{\prime 0})+\mathcal{M}(K^{+},\Sigma_c^+;\Xi_c^0)+\mathcal{M}(K^{+},\Sigma_c^+;\Xi_c^{\prime 0})+\mathcal{M}(K^{\ast +},\Lambda_c^+;\Xi_c^0)\non
&+\mathcal{M}(K^{\ast +},\Lambda_c^+;\Xi_b^{\prime})+\mathcal{M}(K^{\ast +},\Sigma_c^+;\Xi_c^0)+\mathcal{M}(K^{\ast +},\Sigma_c^+;\Xi_c^{\prime 0}),\\
\mathcal{A}(\Xi_{cc}^{++}\to\Sigma_c^+\pi^+)&=\mathcal{T}(\Xi_{cc}^{++}\to\Sigma_c^+\pi^+)+\mathcal{M}(\pi^{+},\Lambda_c^+;\rho^0)+\mathcal{M}(\rho^{+},\Lambda_c^+;\pi^0)+\mathcal{M}(K^{+},\Xi_c^+;K^{\ast 0})+\mathcal{M}(K^{+},\Xi_c^{\prime +};K^{\ast 0})\non
&+\mathcal{M}(K^{\ast +},\Xi_c^+;K^{0})+\mathcal{M}(K^{\ast +},\Xi_c^{\prime +};K^{0})+\mathcal{M}(\pi^+,\Lambda_c^+;\Sigma_c^0)+\mathcal{M}(\pi^+,\Sigma_c^+;\Sigma_c^0)+\mathcal{M}(\rho^+,\Lambda_c^+;\Sigma_c^0)\non
&+\mathcal{M}(\rho^+,\Sigma_c^+;\Sigma_c^0)+\mathcal{M}(K^+,\Xi_c^+;\Xi_c^0)+\mathcal{M}(K^+,\Xi_c^+;\Xi_c^{\prime 0})+\mathcal{M}(K^+,\Xi_c^{\prime +};\Xi_c^0)+\mathcal{M}(K^+,\Xi_c^{\prime +};\Xi_c^{\prime 0})\non
&+\mathcal{M}(K^{\ast +},\Xi_c^+;\Xi_c^0)+\mathcal{M}(K^{\ast +},\Xi_c^+;\Xi_c^{\prime 0})+\mathcal{M}(K^{\ast +},\Xi_c^{\prime +};\Xi_c^0)+\mathcal{M}(K^{\ast +},\Xi_c^{\prime +};\Xi_c^{\prime 0}),\\
\mathcal{A}(\Xi_{cc}^{++}\to\Sigma_c^{++}\overline{K}^0)&=\mathcal{C}_{SD}(\Xi_{cc}^{++}\to\Sigma_c^{++}\overline{K}^0)+\mathcal{M}(\pi^+,\Xi_c^+;K^{\ast +})+\mathcal{M}(\pi^+,\Xi_c^{\prime +};K^{\ast +})+\mathcal{M}(\rho^+,\Xi_c^+;K^{+})\non
&+\mathcal{M}(\rho^+,\Xi_c^{\prime +};K^{+})+\mathcal{M}(\pi^+,\Xi_c^+;\Sigma_c^+)+\mathcal{M}(\pi^+,\Xi_c^+;\Lambda_c^+)+\mathcal{M}(\pi^+,\Xi_c^{\prime +};\Sigma_c^+)+\mathcal{M}(\pi^+,\Xi_c^{\prime +};\Lambda_c^+)\non
&+\mathcal{M}(\rho^+,\Xi_c^+;\Sigma_c^+)+\mathcal{M}(\rho^+,\Xi_c^+;\Lambda_c^+)+\mathcal{M}(\rho^+,\Xi_c^{\prime +};\Sigma_c^+)+\mathcal{M}(\rho^+,\Xi_c^{\prime +};\Lambda_c^+),\\
\mathcal{A}(\Xi_{cc}^{++}\to\Sigma_c^{++}K^0)&=\mathcal{C}_{SD}(\Xi_{cc}^{++}\to\Sigma_c^{++}K^0)+\mathcal{M}(K^{\ast +},\Lambda_c^+;\pi^+)+\mathcal{M}(K^{\ast +},\Sigma_c^+;\pi^+)+\mathcal{M}(K^{+},\Lambda_c^+;\rho^+)\non
&+\mathcal{M}(K^{+},\Sigma_c^+;\rho^+)+\mathcal{M}(K^{+},\Lambda_c^+;\Xi_c^+)+\mathcal{M}(K^{+},\Lambda_c^+;\Xi_c^{\prime +})+\mathcal{M}(K^{+},\Sigma_c^+;\Xi_c^+)+\mathcal{M}(K^{+},\Sigma_c^+;\Xi_c^{\prime +})\non
&+\mathcal{M}(K^{\ast +},\Lambda_c^+;\Xi_c^+)+\mathcal{M}(K^{\ast +},\Lambda_c^+;\Xi_c^{\prime +})+\mathcal{M}(K^{\ast +},\Sigma_c^+;\Xi_c^+)+\mathcal{M}(K^{\ast +},\Sigma_c^+;\Xi_c^{\prime +}), \\
\mathcal{A}(\Xi_{cc}^{++}\to\Sigma_c^{++}\pi^0)&=\mathcal{C}_{SD}(\Xi_{cc}^{++}\to\Sigma_c^{++}\pi^0)+\mathcal{M}(\pi^+,\Lambda_c^+;\rho^+)+\mathcal{M}(\pi^+,\Sigma_c^+;\rho^+)+\mathcal{M}(\rho^+,\Lambda_c^+;\pi^+)+\mathcal{M}(\rho^+,\Sigma_c^+;\pi^+)\non
&+\mathcal{M}(\pi^+,\Lambda_c^+;\Sigma^+)+\mathcal{M}(\pi^+,\Sigma_c^+;\Lambda^+)+\mathcal{M}(\rho^+,\Lambda_c^+;\Sigma^+)+\mathcal{M}(\rho^+,\Sigma_c^+;\Lambda^+)+\mathcal{M}(K^+,\Xi_c^+;K^{\ast +})\non
&+\mathcal{M}(K^+,\Xi_c^{\prime +};K^{\ast +})+\mathcal{M}(K^{\ast +},\Xi_c^+;K^{+})+\mathcal{M}(K^{\ast +},\Xi_c^{\prime +};K^{+})+\mathcal{M}(K^{+},\Xi_c^+;\Xi_c^+)\non
&+\mathcal{M}(K^{+},\Xi_c^+;\Xi_c^{\prime +})+\mathcal{M}(K^{+},\Xi_c^{\prime +};\Xi_c^+)+\mathcal{M}(K^{+},\Xi_c^{\prime +};\Xi_c^{\prime +})+\mathcal{M}(K^{\ast +},\Xi_c^+;\Xi_c^+)\non
&+\mathcal{M}(K^{\ast +},\Xi_c^+;\Xi_c^{\prime +})+\mathcal{M}(K^{\ast +},\Xi_c^{\prime +};\Xi_c^+)+\mathcal{M}(K^{\ast +},\Xi_c^{\prime +};\Xi_c^{\prime +}),\\
\mathcal{A}(\Xi_{cc}^{++}\to\Xi_c^{\prime +}K^+)&=\mathcal{T}(\Xi_{cc}^{++}\to\Xi_c^{\prime +}K^+)+\mathcal{M}(K^{+},\Xi_c^+;\rho^0)+\mathcal{M}(K^{+},\Xi_c^+;\omega)+\mathcal{M}(K^{+},\Xi_c^{\prime +};\rho^0)+\mathcal{M}(K^{+},\Xi_c^{\prime +};\omega)\non
&+\mathcal{M}(K^{\ast +},\Xi_c^+;\pi^0)+\mathcal{M}(K^{\ast +},\Xi_c^+;\eta_8)+\mathcal{M}(K^{\ast +},\Xi_c^{\prime +};\pi^0)+\mathcal{M}(K^{\ast +},\Xi_c^{\prime +};\eta_8)+\mathcal{M}(\pi^+,\Lambda_c^+;\overline{K}^{\ast 0})\non
&+\mathcal{M}(\pi^+,\Sigma_c^+;\overline{K}^{\ast 0})+\mathcal{M}(\rho^+,\Lambda_c^+;\overline{K}^{0})+\mathcal{M}(\rho^+,\Sigma_c^+;\overline{K}^{0})+\mathcal{M}(K^+,\Xi_c^+;\phi)+\mathcal{M}(K^+,\Xi_c^{\prime +};\phi)\non
&+\mathcal{M}(K^+,\Xi_c^+;\Omega_c^0)+\mathcal{M}(K^+,\Xi_c^{\prime +};\Omega_c^0)+\mathcal{M}(K^{\ast +},\Xi_c^+;\Omega_c^0)+\mathcal{M}(K^{\ast +},\Xi_c^{\prime +};\Omega_c^0)+\mathcal{M}(\pi^+,\Lambda_c^+;\Xi_c^0)\non
&+\mathcal{M}(\pi^+,\Lambda_c^+;\Xi_c^{\prime 0})+\mathcal{M}(\pi^+,\Sigma_c^+;\Xi_c^0)+\mathcal{M}(\pi^+,\Sigma_c^+;\Xi_c^{\prime 0})+\mathcal{M}(\rho^+,\Lambda_c^+;\Xi_c^0)+\mathcal{M}(\rho^+,\Lambda_c^+;\Xi_c^{\prime 0})\non
&+\mathcal{M}(\rho^+,\Sigma_c^+;\Xi_c^0)+\mathcal{M}(\rho^+,\Sigma_c^+;\Xi_c^{\prime 0}),\\
\mathcal{A}(\Xi_{cc}^{++}\to\Xi_c^{\prime +}\pi^+)&=\mathcal{T}(\Xi_{cc}^{++}\to\Xi_c^{\prime +}\pi^+)+\mathcal{M}(\pi^+,\Xi_c^+;\rho^0)+\mathcal{M}(\pi^+,\Xi_c^{\prime +};\rho^0)+\mathcal{M}(\rho^+,\Xi_c^+;\pi^0)+\mathcal{M}(\rho^+,\Xi_c^{\prime +};\pi^0)\non
&+\mathcal{M}(\pi^+,\Xi_c^+;\Xi_c^0)+\mathcal{M}(\pi^+,\Xi_c^+;\Xi_c^{\prime 0})+\mathcal{M}(\pi^+,\Xi_c^{\prime +};\Xi_c^0)+\mathcal{M}(\pi^+,\Xi_c^{\prime +};\Xi_c^{\prime 0})+\mathcal{M}(\rho^+,\Xi_c^+;\Xi_c^0)\non
&+\mathcal{M}(\rho^+,\Xi_c^+;\Xi_c^{\prime 0})+\mathcal{M}(\rho^+,\Xi_c^{\prime +};\Xi_c^0)+\mathcal{M}(\rho^+,\Xi_c^{\prime +};\Xi_c^{\prime 0}) ,\\
\mathcal{A}(\Xi_{cc}^{++}\to\Sigma_c^{++}\eta_1)&=\mathcal{C}_{SD}(\Xi_{cc}^{++}\to\Sigma_c^{++}\eta_1)+\mathcal{M}(\pi^+,\Sigma_c^+;\Sigma_c^+)+\mathcal{M}(\pi^+,\Lambda_c^+;\Lambda_c^+)+\mathcal{M}(\rho^+,\Sigma_c^+;\Sigma_c^+)+\mathcal{M}(\rho^+,\Lambda_c^+;\Lambda_c^+)\non&+\mathcal{M}(K^+,\Xi_c^+;\Xi_c^+)+\mathcal{M}(K^+,\Xi_c^{\prime +};\Xi_c^{\prime +})+\mathcal{M}(K^{\ast +},\Xi_c^+;\Xi_c^+)+\mathcal{M}(K^{\ast +},\Xi_c^{\prime +};\Xi_c^{\prime +}),\\
\mathcal{A}(\Xi_{cc}^{++}\to\Sigma_c^{++}\eta_8)&=\mathcal{C}_{SD}(\Xi_{cc}^{++}\to\Sigma_c^{++}\eta_8)+\mathcal{M}(\pi^+,\Sigma_c^+;\Sigma_c^+)+\mathcal{M}(\pi^+,\Lambda_c^+;\Lambda_c^+)+\mathcal{M}(\rho^+,\Sigma_c^+;\Sigma_c^+)+\mathcal{M}(\rho^+,\Lambda_c^+;\Lambda_c^+)\non&+\mathcal{M}(K^+,\Xi_c^+;K^{\ast +})+\mathcal{M}(K^+,\Xi_c^{\prime +};K^{\ast +})+\mathcal{M}(K^{\ast +},\Xi_c^+;K^+)+\mathcal{M}(K^{\ast +},\Xi_c^{\prime +};K^+)+\mathcal{M}(K^+,\Xi_c^+;\Xi_c^+)\non
&+\mathcal{M}(K^+,\Xi_c^{+};\Xi_c^{\prime +})+\mathcal{M}(K^+,\Xi_c^{\prime +};\Xi_c^+)+\mathcal{M}(K^+,\Xi_c^{\prime +};\Xi_c^{\prime +})+\mathcal{M}(K^{\ast +},\Xi_c^+;\Xi_c^+)\non
&+\mathcal{M}(K^{\ast +},\Xi_c^{+};\Xi_c^{\prime +})+\mathcal{M}(K^{\ast +},\Xi_c^{\prime +};\Xi_c^+)+\mathcal{M}(K^{\ast +},\Xi_c^{\prime +};\Xi_c^{\prime +}),\\
\mathcal{A}(\Xi_{cc}^{+}\to\Omega_c^0K^+)&=\mathcal{M}(\pi^+,\Xi_c^0;\overline{K}^{\ast 0})+\mathcal{M}(\pi^+,\Xi_c^{\prime 0};\overline{K}^{\ast 0})+\mathcal{M}(\rho^+,\Xi_c^0;\overline{K}^{0})+\mathcal{M}(\rho^+,\Xi_c^{\prime 0};\overline{K}^{0}), \\
\mathcal{A}(\Xi_{cc}^{+}\to\Sigma_c^+\overline{K}^0)&=\mathcal{C}_{SD}(\Xi_{cc}^{+}\to\Sigma_c^+\overline{K}^0)+\mathcal{M}(\pi^+,\Xi_c^0;K^{\ast +})+\mathcal{M}(\pi^+,\Xi_c^{\prime 0};K^{\ast +})+\mathcal{M}(\rho^+,\Xi_c^0;K^{+})\non
&+\mathcal{M}(\rho^+,\Xi_c^{\prime 0};K^{+})+\mathcal{M}(\pi^+,\Xi_c^0;\Sigma_c^0)+\mathcal{M}(\pi^+,\Xi_c^{\prime 0};\Sigma_c^0)+\mathcal{M}(\rho^+,\Xi_c^0;\Sigma_c^0)+\mathcal{M}(\rho^+,\Xi_c^{\prime 0};\Sigma_c^0),\\
\mathcal{A}(\Xi_{cc}^{+}\to\Sigma_c^+K^0)&=\mathcal{C}_{SD}(\Xi_{cc}^{+}\to\Sigma_c^+K^0)+\mathcal{M}(K^+,\Sigma_c^0;\rho^+)+\mathcal{M}(K^{\ast +},\Sigma_c^0;\pi^+)+\mathcal{M}(K^+,\Sigma_c^0;\Xi_c^0)+\mathcal{M}(K^+,\Sigma_c^0;\Xi_c^{\prime 0})\non
&+\mathcal{M}(K^{\ast +},\Sigma_c^0;\Xi_c^0)+\mathcal{M}(K^{\ast +},\Sigma_c^0;\Xi_c^{\prime 0}),\\
\mathcal{A}(\Xi_{cc}^{+}\to\Sigma_c^+\pi^0)&=\mathcal{C}_{SD}(\Xi_{cc}^{+}\to\Sigma_c^+\pi^0)+\mathcal{M}(\pi^+,\Sigma_c^0;\rho^+)+\mathcal{M}(\rho^+,\Sigma_c^0;\pi^+)+\mathcal{M}(K^+,\Xi_c^0;K^{\ast +})+\mathcal{M}(K^+,\Xi_c^{\prime 0};K^{\ast +})\non
&+\mathcal{M}(K^{\ast +},\Xi_c^0;K^{+})+\mathcal{M}(K^{\ast +},\Xi_c^{\prime 0};K^{+})+\mathcal{M}(\pi^+,\Sigma_c^0;\Sigma_c^0)+\mathcal{M}(\rho^+,\Sigma_c^0;\Sigma_c^0)+\mathcal{M}(K^+,\Xi_c^0;\Xi_c^0)\non
&+\mathcal{M}(K^+,\Xi_c^0;\Xi_c^{\prime 0})+\mathcal{M}(K^+,\Xi_c^{\prime 0};\Xi_c^0)+\mathcal{M}(K^+,\Xi_c^{\prime 0};\Xi_c^{\prime 0})+\mathcal{M}(K^{\ast +},\Xi_c^0;\Xi_c^0)+\mathcal{M}(K^{\ast +},\Xi_c^0;\Xi_c^{\prime 0})\non
&+\mathcal{M}(K^{\ast +},\Xi_c^{\prime 0};\Xi_c^0)+\mathcal{M}(K^{\ast +},\Xi_c^{\prime 0};\Xi_c^{\prime 0}),\\
\mathcal{A}(\Xi_{cc}^{+}\to\Sigma_c^+\eta_1)&=\mathcal{C}_{SD}(\Xi_{cc}^{+}\to\Sigma_c^+\eta_1)+\mathcal{M}(K^+,\Xi_c^0;\Xi_c^0)+\mathcal{M}(K^+,\Xi_c^{\prime 0};\Xi_c^{\prime 0})+\mathcal{M}(K^{\ast +},\Xi_c^0;\Xi_c^0)\non
&+\mathcal{M}(K^{\ast +},\Xi_c^{\prime 0};\Xi_c^{\prime 0})+\mathcal{M}(\pi^+,\Sigma_c^0;\Sigma_c^0)+\mathcal{M}(\rho^+,\Sigma_c^0;\Sigma_c^0),\\
\mathcal{A}(\Xi_{cc}^{+}\to\Sigma_c^+\eta_8)&=\mathcal{C}_{SD}(\Xi_{cc}^{+}\to\Sigma_c^+\eta_8)+\mathcal{M}(K^+,\Xi_c^0;K^{\ast +})+\mathcal{M}(K^+,\Xi_c^{\prime 0};K^{\ast +})+\mathcal{M}(K^{\ast +},\Xi_c^0;K^{+})+\mathcal{M}(K^{\ast +},\Xi_c^{\prime 0};K^{+})\non
&+\mathcal{M}(K^{+},\Xi_c^0;\Xi_c^0)+\mathcal{M}((K^{+},\Xi_c^0;\Xi_c^{\prime 0})+\mathcal{M}((K^{+},\Xi_c^{\prime 0};\Xi_c^0)+\mathcal{M}((K^{+},\Xi_c^{\prime 0};\Xi_c^{\prime 0})+\mathcal{M}(K^{\ast +},\Xi_c^0;\Xi_c^0)\non
&+\mathcal{M}((K^{\ast +},\Xi_c^0;\Xi_c^{\prime 0})+\mathcal{M}((K^{\ast +},\Xi_c^{\prime 0};\Xi_c^0)+\mathcal{M}((K^{\ast +},\Xi_c^{\prime 0};\Xi_c^{\prime 0})\mathcal{M}(\pi^+,\Sigma_c^0;\Sigma_c^0)+\mathcal{M}(\rho^+,\Sigma_c^0;\Sigma_c^0),\\
\mathcal{A}(\Xi_{cc}^{+}\to\Sigma_c^0\pi^+)&=\mathcal{T}(\Xi_{cc}^{+}\to\Sigma_c^0\pi^+)+\mathcal{M}(\pi^+,\Sigma_c^0;\rho^0)+\mathcal{M}(\rho^+,\Sigma_c^0;\pi^0)+\mathcal{M}(K^+,\Xi_c^0;K^{\ast 0})+\mathcal{M}(K^+,\Xi_c^{\prime 0};K^{\ast 0})\non
&+\mathcal{M}(K^{\ast +},\Xi_c^0;K^{0})+\mathcal{M}(K^{\ast +},\Xi_c^{\prime 0};K^{0}),\\
\mathcal{A}(\Xi_{cc}^{+}\to\Sigma_c^{++}K^-)&=\mathcal{M}(\pi^+,\Xi_c^0;\Sigma_c^+)+\mathcal{M}(\pi^+,\Xi_c^0;\Lambda_c^+)+\mathcal{M}(\pi^+,\Xi_c^{\prime 0};\Sigma_c^+)+\mathcal{M}(\pi^+,\Xi_c^{\prime 0};\Lambda_c^+)+\mathcal{M}(\rho^+,\Xi_c^0;\Sigma_c^+)\non
&+\mathcal{M}(\rho^+,\Xi_c^0;\Lambda_c^+)+\mathcal{M}(\rho^+,\Xi_c^{\prime 0};\Sigma_c^+)+\mathcal{M}(\rho^+,\Xi_c^{\prime 0};\Lambda_c^+),\\
\mathcal{A}(\Xi_{cc}^{+}\to\Sigma_c^{++}\pi^-)&=\mathcal{M}(\pi^+,\Sigma_b^{-};\Sigma_c^+)+\mathcal{M}(\pi^+,\Sigma_b^{-};\Lambda_c^+)+\mathcal{M}(\rho^+,\Sigma_b^{-};\Sigma_c^+)+\mathcal{M}(\rho^+,\Sigma_b^{-};\Lambda_c^+)+\mathcal{M}(K^+,\Xi_c^{0};\Xi_c^+)\non
&+\mathcal{M}(K^+,\Xi_c^{0};\Xi_c^{\prime +})+\mathcal{M}(K^+,\Xi_c^{\prime 0};\Xi_c^+)+\mathcal{M}(K^+,\Xi_c^{\prime 0};\Xi_c^{\prime +})+\mathcal{M}(K^{\ast +},\Xi_c^{0};\Xi_c^+)\non
&+\mathcal{M}(K^{\ast +},\Xi_c^{0};\Xi_c^{\prime +})+\mathcal{M}(K^{\ast +},\Xi_c^{\prime 0};\Xi_c^+)+\mathcal{M}(K^{\ast +},\Xi_c^{\prime 0};\Xi_c^{\prime +}), \\
\mathcal{A}(\Xi_{cc}^{+}\to\Xi_c^{\prime 0}K^+)&=\mathcal{T}(\Xi_{cc}^{+}\to\Xi_c^{\prime 0}K^+)+\mathcal{M}(\pi^+,\Sigma_b^{-};\overline{K}^{\ast 0})+\mathcal{M}(\rho^+,\Sigma_b^{-};\overline{K}^{0})+\mathcal{M}(K^+,\Xi_c^{0};\phi)+\mathcal{M}(K^+,\Xi_c^{\prime 0};\phi)\non
&+\mathcal{M}(K^{\ast +},\Xi_c^{0};\eta_8)+\mathcal{M}(K^{\ast +},\Xi_c^{\prime 0};\eta_8),\\
\mathcal{A}(\Xi_{cc}^{+}\to\Xi_c^{\prime 0}\pi^+)&=\mathcal{T}(\Xi_{cc}^{+}\to\Xi_c^{\prime 0}\pi^+)+\mathcal{M}(\pi^+,\Xi_c^{0};\rho^0)+\mathcal{M}(\pi^+,\Xi_c^{\prime 0};\rho^0)+\mathcal{M}(\rho^+,\Xi_c^{0};\pi^0)+\mathcal{M}(\rho^+,\Xi_c^{\prime 0};\pi^0),\\
\mathcal{A}(\Xi_{cc}^{+}\to\Xi_c^{\prime +}K^0)&=\mathcal{C}_{SD}(\Xi_{cc}^{+}\to\Xi_c^{\prime +}K^0)+\mathcal{M}(K^+,\Xi_c^{0};\rho^+)+\mathcal{M}(K^+,\Xi_c^{\prime 0};\rho^+)+\mathcal{M}(K^{\ast +},\Xi_c^{0};\pi^+)\non
&+\mathcal{M}(K^{\ast +},\Xi_c^{\prime 0};\pi^+)+\mathcal{M}(K^+,\Xi_c^{0};\Omega_c^0)+\mathcal{M}(K^+,\Xi_c^{\prime 0};\Omega_c^0)+\mathcal{M}(K^{\ast +},\Xi_c^{0};\Omega_c^0)+\mathcal{M}(K^{\ast +},\Xi_c^{\prime 0};\Omega_c^0)\non
&+\mathcal{M}(\pi^+,\Sigma_b^{-};\Xi_c^0)+\mathcal{M}(\pi^+,\Sigma_b^{-};\Xi_c^{\prime 0})+\mathcal{M}(\rho^+,\Sigma_b^{-};\Xi_c^0)+\mathcal{M}(\rho^+,\Sigma_b^{-};\Xi_c^{\prime 0}), \\
\mathcal{A}(\Xi_{cc}^{+}\to\Xi_c^{\prime +}\pi^0)&=\mathcal{C}_{SD}(\Xi_{cc}^{+}\to\Xi_c^{\prime +}\pi^0)+\mathcal{M}(\pi^+,\Xi_c^{0};\rho^+)+\mathcal{M}(\pi^+,\Xi_c^{\prime 0};\rho^+)+\mathcal{M}(\rho^+,\Xi_c^{0};\pi^+)+\mathcal{M}(\rho^+,\Xi_c^{\prime 0};\pi^+)\non
&+\mathcal{M}(\pi^+,\Xi_c^{0};\Xi_c^0)+\mathcal{M}(\pi^+,\Xi_c^{0};\Xi_c^{\prime 0})+\mathcal{M}(\pi^+,\Xi_c^{\prime 0};\Xi_c^0)+\mathcal{M}(\pi^+,\Xi_c^{\prime 0};\Xi_c^{\prime 0})+\mathcal{M}(\rho^+,\Xi_c^{0};\Xi_c^0)\non
&+\mathcal{M}(\rho^+,\Xi_c^{0};\Xi_c^{\prime 0})+\mathcal{M}(\rho^+,\Xi_c^{\prime 0};\Xi_c^0)+\mathcal{M}(\rho^+,\Xi_c^{\prime 0};\Xi_c^{\prime 0}),\\
\mathcal{A}(\Omega_{cc}^{+}\to\Omega_c^0K^+)&=\mathcal{T}(\Omega_{cc}^{+}\to\Omega_c^0K^+)+\mathcal{M}(\pi^+,\Xi_c^{0};\overline{K}^{\ast 0})+\mathcal{M}(\pi^+,\Xi_c^{\prime 0};\overline{K}^{\ast 0})+\mathcal{M}(\rho^+,\Xi_c^{0};\overline{K}^{0})+\mathcal{M}(\rho^+,\Xi_c^{\prime 0};\overline{K}^{0})\non
&+\mathcal{M}(K^+,\Omega_c^{0};\phi)+\mathcal{M}(K^{\ast +},\Omega_c^{0};\eta_8),\\
\mathcal{A}(\Omega_{cc}^{+}\to\Sigma_c^+\overline{K}^0)&=\mathcal{C}_{SD}(\Omega_{cc}^{+}\to\Sigma_c^+\overline{K}^0)+\mathcal{M}(\pi^+,\Xi_c^{0};K^{\ast +})+\mathcal{M}(\pi^+,\Xi_c^{\prime 0};K^{\ast +})+\mathcal{M}(\rho^+,\Xi_c^{0};K^{+})\non
&+\mathcal{M}(\rho^+,\Xi_c^{\prime 0};K^{+})+\mathcal{M}(\pi^+,\Xi_c^{0};\Sigma_c^0)+\mathcal{M}(\pi^+,\Xi_c^{\prime 0};\Sigma_c^0)+\mathcal{M}(\rho^+,\Xi_c^{0};\Sigma_c^0)+\mathcal{M}(\rho^+,\Xi_c^{\prime 0};\Sigma_c^0)\non
&+\mathcal{M}(K^+,\Omega_c^{0};\Xi_c^0)+\mathcal{M}(K^+,\Omega_c^{0};\Xi_c^{\prime 0})+\mathcal{M}(K^{\ast +},\Omega_c^{0};\Xi_c^0)+\mathcal{M}(K^{\ast +},\Omega_c^{0};\Xi_c^{\prime 0}),\\
\mathcal{A}(\Omega_{cc}^{+}\to\Sigma_c^+\pi^0)&=\mathcal{M}(K^{+},\Xi_c^{0};K^{\ast +})+\mathcal{M}(K^{+},\Xi_c^{\prime 0};K^{\ast +})+\mathcal{M}(K^{\ast +},\Xi_c^{0};K^{+})+\mathcal{M}(K^{\ast +},\Xi_c^{\prime 0};K^{+})\non
&+\mathcal{M}(K^{+},\Xi_c^{0};\Xi_c^0)+\mathcal{M}(K^{+},\Xi_c^{0};\Xi_c^{\prime 0})+\mathcal{M}(K^{+},\Xi_c^{\prime 0};\Xi_c^0)+\mathcal{M}(K^{+},\Xi_c^{\prime 0};\Xi_c^{\prime 0})+\mathcal{M}(K^{\ast +},\Xi_c^{0};\Xi_c^0)\non
&+\mathcal{M}(K^{\ast +},\Xi_c^{0};\Xi_c^{\prime 0})+\mathcal{M}(K^{\ast +},\Xi_c^{\prime 0};\Xi_c^0)+\mathcal{M}(K^{\ast +},\Xi_c^{\prime 0};\Xi_c^{\prime 0}), \\
\mathcal{A}(\Omega_{cc}^{+}\to\Sigma_c^0\pi^+)&=\mathcal{M}(K^{+},\Xi_c^{0};K^{\ast 0})+\mathcal{M}(K^{+},\Xi_c^{\prime 0};K^{\ast 0})+\mathcal{M}(K^{\ast +},\Xi_c^{0};K^{0})+\mathcal{M}(K^{\ast +},\Xi_c^{\prime 0};K^{0}),\\
\mathcal{A}(\Omega_{cc}^{+}\to\Sigma_c^{++}K^-)&=\mathcal{M}(\pi^+,\Xi_c^{0};\Sigma_c^+)+\mathcal{M}(\pi^+,\Xi_c^{0};\Lambda_c^+)+\mathcal{M}(\pi^+,\Xi_c^{\prime 0};\Sigma_c^+)+\mathcal{M}(\pi^+,\Xi_c^{\prime 0};\Lambda_c^+)+\mathcal{M}(\rho^+,\Xi_c^{0};\Sigma_c^+)\non
&+\mathcal{M}(\rho^+,\Xi_c^{0};\Lambda_c^+)+\mathcal{M}(\rho^+,\Xi_c^{\prime 0};\Sigma_c^+)+\mathcal{M}(\rho^+,\Xi_c^{\prime 0};\Lambda_c^+)+\mathcal{M}(K^+,\Omega_c^{0};\Xi_c^+)+\mathcal{M}(K^+,\Omega_c^{0};\Xi_c^{\prime +})\non
&+\mathcal{M}(K^{\ast +},\Omega_c^{0};\Xi_c^+)+\mathcal{M}(K^{\ast +},\Omega_c^{0};\Xi_c^{\prime +}),\\
\mathcal{A}(\Omega_{cc}^{+}\to\Sigma_c^{++}\pi^-)&=\mathcal{M}(K^{+},\Xi_c^{0};\Xi_c^+)+\mathcal{M}(K^{+},\Xi_c^{0};\Xi_c^{\prime +})+\mathcal{M}(K^{+},\Xi_c^{\prime 0};\Xi_c^+)+\mathcal{M}(K^{+},\Xi_c^{\prime 0};\Xi_c^{\prime +})+\mathcal{M}(K^{\ast +},\Xi_c^{0};\Xi_c^+)\non
&+\mathcal{M}(K^{\ast +},\Xi_c^{0};\Xi_c^{\prime +})+\mathcal{M}(K^{\ast +},\Xi_c^{\prime 0};\Xi_c^+)+\mathcal{M}(K^{\ast +},\Xi_c^{\prime 0};\Xi_c^{\prime +}),\\
\mathcal{A}(\Omega_{cc}^{+}\to\Xi_c^{\prime 0}K^+)&=\mathcal{T}(\Omega_{cc}^{+}\to\Xi_c^{\prime 0}K^+)+\mathcal{M}(K^{+},\Xi_c^{0};\phi)+\mathcal{M}(K^{+},\Xi_c^{\prime 0};\phi)+\mathcal{M}(K^{\ast +},\Xi_c^{0};\eta_8)+\mathcal{M}(K^{\ast +},\Xi_c^{\prime 0};\eta_8), \\
\mathcal{A}(\Omega_{cc}^{+}\to\Xi_c^{\prime 0}\pi^+)&=\mathcal{T}(\Omega_{cc}^{+}\to\Xi_c^{\prime 0}\pi^+)+\mathcal{M}(\pi^+,\Xi_c^{0};\rho^0)+\mathcal{M}(\pi^+,\Xi_c^{\prime 0};\rho^0)+\mathcal{M}(\rho^+,\Xi_c^{0};\pi^0)+\mathcal{M}(\rho^+,\Xi_c^{\prime 0};\pi^0)\non
&+\mathcal{M}(K^+,\Omega_c^{0};K^{\ast 0})+\mathcal{M}(K^{\ast +},\Omega_c^{0};K^{0}),\\
\mathcal{A}(\Omega_{cc}^{+}\to\Xi_c^{\prime +}\overline{K}^0)&=\mathcal{C}_{SD}(\Omega_{cc}^{+}\to\Xi_c^{\prime +}\overline{K}^0)+\mathcal{M}(\pi^+,\Omega_c^{0};K^{\ast +})+\mathcal{M}(\rho^+,\Omega_c^{0};K^{+})+\mathcal{M}(\pi^+,\Omega_c^{0};\Xi_c^0)+\mathcal{M}(\pi^+,\Omega_c^{0};\Xi_c^{\prime 0})\non
&+\mathcal{M}(\rho^+,\Omega_c^{0};\Xi_c^0)\mathcal{M}(\rho^+,\Omega_c^{0};\Xi_c^{\prime 0}),\\
\mathcal{A}(\Omega_{cc}^{+}\to\Xi_c^{\prime +}K^0)&=\mathcal{C}_{SD}(\Omega_{cc}^{+}\to\Xi_c^{\prime +}K^0)+\mathcal{M}(K^+,\Xi_c^{0};\rho^+)+\mathcal{M}(K^+,\Xi_c^{\prime 0};\rho^+)+\mathcal{M}(K^{\ast +},\Xi_c^{0};\pi^+)+\mathcal{M}(K^{\ast +},\Xi_c^{\prime 0};\pi^+)\non
&+\mathcal{M}(K^{+},\Xi_c^{0};\Omega_c^0)+\mathcal{M}(K^{+},\Xi_c^{\prime 0};\Omega_c^0)+\mathcal{M}(K^{\ast +},\Xi_c^{0};\Omega_c^0)+\mathcal{M}(K^{\ast +},\Xi_c^{\prime 0};\Omega_c^0),\\
\mathcal{A}(\Omega_{cc}^{+}\to\Xi_c^{\prime +}\pi^0)&=\mathcal{C}_{SD}(\Omega_{cc}^{+}\to\Xi_c^{\prime +}\pi^0)+\mathcal{M}(\pi^+,\Xi_c^{0};\rho^+)+\mathcal{M}(\pi^+,\Xi_c^{\prime 0};\rho^+)+\mathcal{M}(\rho^+,\Xi_c^{0};\pi^+)+\mathcal{M}(\rho^+,\Xi_c^{\prime 0};\pi^+)\non
&+\mathcal{M}(K^+,\Omega_c^{0};K^{\ast +})+\mathcal{M}(K^{\ast +},\Omega_c^{0};K^{+})+\mathcal{M}(\pi^+,\Xi_c^{0};\Xi_c^0)+\mathcal{M}(\pi^+,\Xi_c^{0};\Xi_c^{\prime 0})+\mathcal{M}(\pi^+,\Xi_c^{\prime 0};\Xi_c^0)\non
&+\mathcal{M}(\pi^+,\Xi_c^{\prime 0};\Xi_c^{\prime 0})+\mathcal{M}(\rho^+,\Xi_c^{0};\Xi_c^0)+\mathcal{M}(\rho^+,\Xi_c^{0};\Xi_c^{\prime 0})+\mathcal{M}(\rho^+,\Xi_c^{\prime 0};\Xi_c^0)+\mathcal{M}(\rho^+,\Xi_c^{\prime 0};\Xi_c^{\prime 0}).
\end{align}	}



\end{document}